# Bayes linear covariance matrix adjustment

A thesis presented for the degree

of Doctor of Philosophy at the

University of Durham.

Darren James Wilkinson

*Department of Mathematical Sciences,*

*University of Durham,*

*DURHAM, DH1 3LE.*

September 1995

# Abstract

In this thesis, a Bayes linear methodology for the adjustment of covariance matrices is presented and discussed. A geometric framework for quantifying uncertainties about covariance matrices is set up, and an inner-product for spaces of random matrices is motivated and constructed. The inner-product on this space captures aspects of belief about the relationships between covariance matrices of interest, providing a structure rich enough to adjust beliefs about unknown matrices in the light of data such as sample covariance matrices, exploiting second-order exchangeability and related specifications to obtain representations allowing analysis.

Adjustment is associated with orthogonal projection, and illustrated by examples for some common problems. The difficulties of adjusting the covariance matrices underlying exchangeable random vectors is tackled and discussed. Learning about the covariance matrices associated with multivariate time series dynamic linear models is shown to be amenable to a similar approach. Diagnostics for matrix adjustments are also discussed.

# Preface

For a great many reasons, the three years I have spent working on my Ph.D. have been the most interesting and enjoyable of my life. I can but hope that at least a small fraction of the interest and enthusiasm I have for the work contained in this thesis will infect the reader. There are a great many people who have supported and encouraged me over the last few years. Unfortunately I have space to thank just a few of them here.

Thanks, first and foremost must go to my supervisor, Michael Goldstein. I am of course, very grateful to him for the many fascinating technical discussions relating to the theory and applications of Bayes linear methods we have had since I began my doctoral research. Mainly however, I am indebted to him for reinforcing my belief that inference only makes sense in the context of subjective revision of personal belief, that such revision must necessarily revolve around limited aspects of belief, and that there can be no entirely prescriptive methodology for carrying out this revision of belief. This is truly the greatest insight any statistician (indeed, any scientist) can have, and it is this insight that I shall carry with me for the rest of my life.

I would also like to thank David Wooff, not only for writing [B/D], the Bayes linear computer programming language, but also for the maintenance and development of the program which he has carried out over the last three years; much of which has been at my request, and for my benefit.

This thesis would not exist, were it not for the financial support provided to me by the EPSRC. I am grateful to Positive Concepts Ltd. for supplying the data for the time series example discussed in Chapter 5. There are many people in the Department of Mathematical Sciences, in Durham, who have supported me in less tangible, but no less important, ways. In particular, the graduate students have formed a very close community and I shall miss them all, when we finally disperse to every corner of the globe.

Thanks are also due to Herman Rubin, not just for enlightening me with respect to the general moment structure of the multivariate normal distribution, but also for the regular source of great wisdom he provides to the statistical community, through his postings to the `sci.stat.math` USENET News group.

Finally, I would like to thank my family, for always supporting me in everything that I do, and in particular, my wife, Alison, who could not possibly have known the extent to which the work for my Ph.D. would take over my life.

# Statement of originality

Some of the ideas behind Chapters 4 and 5 were the result of joint work with Michael Goldstein. However, the formulation of the matrix space, and the form of the natural matrix inner-product are my own, as are all of the algebraic results. The rest of the work for this thesis is entirely my own.

I declare that no part of this thesis has been submitted for any part of any degree at this or any other university.

# Contents

# List of Tables

# List of Figures

# Chapter 1

# Introduction

> . . . There is no way, however, in which the individual can avoid the burden of responsibility for his own evaluations. The key cannot be found that will unlock the enchanted garden wherein, among the fairy-rings and the shrubs of magic wands, beneath the trees laden with monads and noumena, blossom forth the flowers of *probabilitas realis*. With these fabulous blooms safely in our button-holes we would be spared the necessity of forming opinions, and the heavy loads we bear upon our necks would be rendered superfluous once and for all.
>
> Bruno de Finetti
>
> Theory of Probability, Vol 2

## 1.1 Introduction

Often *random variables* (or unknown quantities) are not independent of one another. That is, knowledge of the outcome of a particular random variable effects beliefs about other random variables. Clearly the random variables *the amount of rain to fall this Saturday* and *the amount of rain to fall this Sunday* fall into this category, since knowledge of the amount of rain that fell on Saturday will be very helpful in assessing the amount of rain to fall on Sunday. A common way of measuring the amount of *linear* association between two random variables is the *covariance* between them. Given collections of random variables, one may extend the concept of covariance to that of the covariance matrix. This is a matrix of numbers containing information about the pairwise linear association between variables. This thesis is concerned with the modelling and revising of covariance matrices in the light of predictive data. A *Bayes linear* approach will be taken to the problem, and the theory will be illustrated with examples for some common statistical problems.

## 1.2 Prevision and expectation

In this thesis, a *random quantity* is any well-determined real number whose precise value is unknown. Beliefs about the "location" of a given random quantity, $X$, are quantified by making a statement about it's *prevision* (de Finetti 1974, Chapter 3), $\mathrm{P}(X)$. $\mathrm{P}(X)$ is the quantity, $x$ that *you* would choose in order to minimise the loss, or penalty, $L$, given by

$$L = K(x - X)^2 \tag{1.1}$$

for some unit of loss, $K$ (de Finetti 1974, Section 3.3.6). Such an approach to the quantification of uncertainty was advocated in de Finetti (1974), and is used as a starting point for the subjectivist theory of statistical inference developed by Goldstein and others (see for example, Goldstein (1981), Farrow and Goldstein (1993) and Goldstein (1994)).

Unfortunately there is a problem with such a definition of prevision, since one must have a linear preference for the incurred loss, $L$. In fact, the loss should be in units of *utility*. de Finetti was, of course, well aware of this, hence his *Digression on* *decisions and utilities* (de Finetti 1974, Section 3.2). However, his "resolution" is not really satisfactory, and in any case, one must be extremely careful to avoid circularity, since utility is usually defined in terms of probability, something which we wish the concept of prevision to supercede.

Frank Ramsey also recognised the subjective nature of probability, and the problem with the usual *betting rate* definition (non-linear preferences). Ramsey gives his own solution to the problem in Ramsey (1931) (or Ramsey (1990)). However, I prefer a solution which works with *probability currency*, essentially by working with tickets in a raffle for a single fixed prize. Such a solution is discussed for upper and lower probabilities in Walley (1991, Section 2.2). The idea of using the lottery analogy for ensuring linear preference is discussed in Savage (1954, Chapter 5). This idea is developed and used for a probability currency approach to belief elicitation in Smith (1961). I particularly like Smith's "small diamond in a block of beeswax" formulation, but as Savage (1971) points out, this would not be a truly "utility-free currency for exploring a subject's opinions about the future of a diamond market"!

Note that these definitions all rely in some way, on the concept of *equally likely events*. Indeed, I strongly suspect that any careful definition of subjective probability or prevision *must necessarily* make exactly such a recourse at some point.

To make precise the definition of prevision for unbounded quantities requires a limiting argument, and one must be careful to ensure that all limits exist. There are, of course, many random quantities for which there does not exist a prevision, but such quantities are all unbounded. These will not be considered further, and so attention will be restricted to strictly bounded quantities, since this thesis is not really the place to discus the origin and foundations of prevision.

For the rest of this thesis it will be assumed that there is a well-defined notion of prevision which obeys the conditions of *coherence*

$$\mathrm{P}(X+Y) = \mathrm{P}(X) + \mathrm{P}(Y) \tag{1.2}$$

$$\inf(X) \leq \mathrm{P}(X) \leq \sup(X) \tag{1.3}$$

discussed in Section 3.1.5 of de Finetti (1974). Note that the choice of $\mathrm{P}(X)$ is not viewed as some kind of decision problem, under a quadratic loss function, but as a one-dimensional summary of the random quantity $X$, with desirable linearity properties, such as

$$\mathrm{P}(aX + bY + cZ + \cdots) = a\mathrm{P}(X) + b\mathrm{P}(Y) + c\mathrm{P}(Z) + \cdots \tag{1.4}$$

and always of intrinsic interest, irrespective of any decision problem which may or may not be present.

Prevision is simply a primitively defined *expectation*, $\mathrm{E}(X)$, of a random quantity, $X$, and the two concepts usually coincide provided that you are *coherent*. However, the notation $\mathrm{E}(X)$ is reserved to mean something which is in practice, the same, but conceptually different (a precise definition will be given later). The prevision of a vector or matrix of random quantities is the vector, or matrix of previsions.

## 1.3 Covariance

Covariance is a measure of linear association between two random quantities. In this thesis, the covariance, $\mathrm{Cov}(X, Y)$, between the random quantities $X$ and $Y$ will be defined by

$$\mathrm{Cov}(X, Y) = \mathrm{P}(XY) - \mathrm{P}(X)\mathrm{P}(Y), \quad \forall X, Y \tag{1.5}$$

The notion of covariance easily extends to vectors of random quantities. For two vectors of random quantities, $\boldsymbol{X}$ and $\boldsymbol{Y}$, the covariance matrix between them, Cov($\boldsymbol{X}$, $\boldsymbol{Y}$), is defined by

$$\mathrm{Cov}(\boldsymbol{X}, \boldsymbol{Y}) = \mathrm{P}(\boldsymbol{X}\boldsymbol{Y}^{\mathrm{T}}) - \mathrm{P}(\boldsymbol{X})\mathrm{P}(\boldsymbol{Y})^{\mathrm{T}}, \quad \forall \boldsymbol{X}, \boldsymbol{Y} \tag{1.6}$$

The covariance of a vector of random quantities with itself, $\mathrm{Cov}(\boldsymbol{X}, \boldsymbol{X})$, is the covariance matrix for $\boldsymbol{X}$, and is denoted $\mathrm{Var}(\boldsymbol{X})$. This thesis is concerned with the quantification of uncertainty about such matrices, and methods for learning about such matrices.

## 1.4 Bayes linear methods

The Bayes linear approach to subjective statistical inference is founded upon de Finetti's theory of prevision. The idea that the foundations of statistical inference could be based upon concept of revision of prevision was given in Goldstein (1981). A more gentle overview of the methodology from a slightly more simplistic viewpoint is given in Farrow and Goldstein (1993). Analysis is carried out using only first and second order belief specifications. Linear Bayesian methods have been considered previously in the literature. Some of the basic ideas and key results can be found in Stone (1963) and Hartigan (1969).

Bayes linear methods require only a specification of prevision for every quantity under consideration, and also specifications for the covariance between every pair of quantities (in other words, the covariance matrix for the vector of all quantities under consideration). For example, if we are interested in a vector of random quantities, $\boldsymbol{B}$, and wish to learn about it using a vector of observable quantities, $\boldsymbol{D}$, we would form

the vector

$$\mathrm{P}(\boldsymbol{B}) + \mathrm{Cov}(\boldsymbol{B}, \boldsymbol{D})\mathrm{Var}(\boldsymbol{D})^{-1}[\boldsymbol{D} - \mathrm{P}(\boldsymbol{D})] \tag{1.7}$$

Note that this is a vector which depends only on our prior specifications and the data. In fact, it is the *Bayes linear rule* for $\boldsymbol{B}$, using the data, $\boldsymbol{D}$. In this thesis, it will be assumed that all the variance matrices being considered are strictly positive definite, and hence invertible. In fact, such a restriction can nearly always be dropped. All that one requires is that the specifications for the covariance matrix are coherent; namely that the covariance matrix is non-negative definite. The corresponding results can usually be obtained, simply by using any generalised inverse in place of an inverse. Moreover, the Moore-Penrose generalised inverse is a natural choice, with some desirable properties. More complete discussion of such issues can be found in Goldstein and Wooff (1995b).

It is worth pointing out at this point, that the Bayes linear rule is just a projection in the space spanned by the random quantities and the observables, with respect to expected quadratic loss. Moreover, when the quantities involved in the equations are all indicators for events, this projection gives the usual form of conditional probability

$$\mathrm{P}(B|D) = \frac{\mathrm{P}(B \cap D)}{\mathrm{P}(D)} \tag{1.8}$$

leading to the famous Theorem of Bayes (1763) (or Bayes (1958)),

$$\mathrm{P}(B|D) = \frac{\mathrm{P}(D|B)\mathrm{P}(B)}{\mathrm{P}(D)} \tag{1.9}$$

Thus, from a foundational perspective, the linear Bayes rule should be regarded as a *generalisation* of Bayes' Theorem, and not as some sort of approximation to it. This is the crux of the argument in Goldstein (1981). It is also worth noting that Bayes (1763) begins his essay by defining probability. Paraphrasing him, in more modern parlance, he essentially defines the probability of an event to be the expected value of it's indicator, as does de Finetti (1974), and as do we. Obviously, he considered expectation to be a more primitive concept than that of probability. Conventional treatments, which make probability primitive, and define expected values with respect to probability measures, have perhaps obscured the simplicity of the concept of expectation.

The use of second-order exchangeability as a fundamental modelling assumption was first discussed in Goldstein (1986a), but has since been discussed from a foundational perspective in Goldstein (1994).

Second order belief structures have many useful properties with respect to linear adjustment, and these are discussed in Goldstein (1988a). The particular properties associated with exchangeable belief structures are discussed in Goldstein and Wooff (1995a), together with the important notion of Bayes linear sufficiency, first outlined in Goldstein (1986b), and more recently discussed in the context of an example, in Goldstein and O'Hagan (1995). Comparisons of belief structures are discussed in Goldstein (1991), and diagnostics for adjustments are given in Goldstein (1988b). Graphical summaries of diagnostics are given in Farrow and Goldstein (1995).

Later, the geometric construction underlying second-order belief structures and linear belief adjustment will be given; for now it is sufficient to note that it is highly dependent on the covariance specifications made. Consequently, it is important that the specified covariances are appropriate. Bayes linear methods for estimating scale parameters were considered, both by Stone (1963) and Hartigan (1969), by fitting variance parameters linearly on quadratic data. Further, Stone points out that genuine variance estimation would require fourth order moment specifications. Modifying linear Bayes estimates due to information about variance was considered in Goldstein (1979) and Goldstein (1983). In some ways, this could be considered a precursor to the work contained in this thesis.

## 1.5 [B/D], the Bayes linear programming language

[B/D] is a fully functional interpreted programming language which implements most features of the Bayes linear methodology. It is freely available to the academic community over the World-Wide Web, via the URL `http://fourier.dur.ac.uk:8000/stats/bd/`. The program is outlined in Wooff (1992) and Goldstein and Wooff (1995b), and documented fully in Wooff (1995b). It provides a framework for belief specification and analysis, and facilities for carrying out adjustments using data, and producing diagnostic summaries of data adjustments. It also has facilities for producing Bayes linear diagnostic influence diagrams, such as those described in Goldstein, Farrow, and Spiropoulos (1993) or Goldstein and Wooff (1995b). A tutorial introduction to [B/D] can be obtained from the sequence of technical reports Goldstein (1995), Wooff (1995a) and Wooff and Goldstein (1995). All of the examples in this thesis were implemented in [B/D], and reference will be made to the package, where this is felt to be appropriate.

## 1.6 Revising beliefs about covariance structures

Quantifying relationships between variables is of fundamental importance in Bayesian analysis. However, there are many difficulties associated even with learning about covariances. For example, it is often difficult to make prior covariance specifications, but it is usually even harder to make the statements about the uncertainty in these covariance statements which are required in order to learn about the covariance statements from data. Further, a covariance structure is more than just a collection of random quantities, so we should aim to analyse such structures in a space where they live naturally. In this thesis, such an approach will be developed and illustrated, based around a geometric representation for variance matrices and exploiting second-order exchangeability specifications for them.

In Chapter 2, a methodology will be developed for the modelling and quantification of uncertainty about covariances between random variables. In Chapter 3, the geometric representation of covariance matrices is discussed. In Chapter 4, this representation is used to enable learning about the covariance structure underlying exchangeable random vectors.

## 1.7 Covariance estimation for dynamic linear models

In Chapter 5, the suggested approach to covariance estimation is applied to the development of a methodology for the revision of the underlying covariance structures for a dynamic linear model, free from any distributional restrictions, using Bayes linear estimators for the covariance matrices based on simple quadratic observables. This is done by constructing an inner-product space of random matrices containing both the underlying covariance matrices and observables predictive for them. Bayes linear estimates for the underlying matrices follow by orthogonal projection.

The method is illustrated with data derived from the weekly sales of six leading brands of shampoo from a medium sized cash-and-carry depot. The sales are modelled taking into account the underlying demand and competition effects and the covariance structure over the resulting dynamic linear model is adjusted using the weekly sales data.

## 1.8 Diagnostic analysis of matrix adjustments

In Chapter 6, methodology for the diagnostic analysis of matrix adjustments is discussed. A framework is outlined whereby the *a posteriori* observed changes in belief may be compared to the *a priori* expected changes in belief for general Bayes linear problems. This theory provides a general unified framework for the *a priori* and *a posteriori* analysis of Bayes linear statistical problems.

## 1.9 Distributional Bayesian approaches to covariance estimation

Until recently, most authors have followed a Wishart conjugate prior approach to covariance matrix estimation (see for example, Evans (1965), Chen (1979), Haff (1980) or Dickey, Lindley, and Press (1985)). This approach, whilst tractable, places severe restrictions on the form of the prior distribution (there is only one hyper-parameter which expresses uncertainty about the matrix), and requires a multivariate normal assumption for the distribution of the residuals.

More recently, a different approach has been proposed by Leonard and Hsu (1992). Essentially, they learn about the logarithm of the covariance matrix using the logarithm of sample covariance matrices. This solves the positivity problems associated with covariance revision, but imposes a tremendous specification burden, for parameters without an intuitive interpretation. Further, they require a joint multivariate normal assumption for the elements of the logarithm of the covariance matrix, and since they rely on sampling methodology, have serious computational problems for large matrices.

Brown, Le, and Zidek (1994), make further progress: working within a distributional Bayesian paradigm, they develop a reasonably flexible prior over the elements of a covariance structure, and offer interpretations for the parameters that one is required to specify. However, this work is still restricted to multivariate Normal likelihoods, and there is a weak restriction on the form of the mean structure for the data.

Covariance matrix adjustment for dynamic linear models is reviewed in West and Harrison (1989). For multivariate time series, the observational covariance matrix can be updated for a class of models known as *matrix normal models* using a simple conjugate prior approach. However, the distributional assumptions required are extremely restrictive, and there is no method which allows data-driven learning for the covariance matrix for the updating of the state vector.

It would seem that those authors who have considered the problem of covariance matrix revision have come to the conclusion that it is such a difficult problem that they are prepared to make whatever distributional assumptions necessary in order to make the analysis as simple as possible. In particular, the sole justification for the Wishart conjugate prior approach seems to be that it makes the problem simple and tractable. The distributional assumptions made are usually such that expectations and conditional expectations have desirable linearity properties which simplify the problem. In this thesis, no distributional assumptions are required, but exactly these sorts of linearity properties are used as a starting point.

# Chapter 2

# Partial belief specification and exchangeability

## 2.1 Partial belief specifications

Given a set of random quantities of interest, and a selection of observable random quantities predictive for them, a distributional Bayesian approach would require a full joint probability distribution to be specified over all of the variables of interest, before any analysis could take place. On the contrary, all that is required for a Bayes linear analysis is the first and second moments of that joint distribution, since many aspects of the relationships between variables are captured by such specifications.

The utility of working with first and second order characteristics has long been appreciated in other disciplines. For example, in classical mechanics, when summarising the properties of a heavy object, one describes it using the position of the centre of mass (its first order characteristics) and its moments of inertia (its second order characteristics). One could work exclusively with the mass distribution function for the object, but often there would be little utility to be gained from doing so. Further, working with first and second order characteristics tends to make analyses simple, tractable and robust. This analogy is taken much further in de Finetti (1974). Of course, the analogy only goes so far, and second-order characteristics can mean a lot more than just covariance specifications (for example, a full distributional Bayesian analysis of a statistical problem can be undertaken within this second-order framework). See Goldstein (1981) and Goldstein (1994) for a more complete discussion of the general foundational viewpoint.

## 2.2 Exchangeable representations for covariances

Let $\boldsymbol{X}_1, \boldsymbol{X}_2, \ldots$ be an infinite, second-order exchangeable sequence of random vectors, each of length $r$, namely a sequence for which $\boldsymbol{X}_k = (X_{1k}, \ldots, X_{rk})^{\mathrm{T}}$, $\boldsymbol{\mu} = \mathrm{P}(\boldsymbol{X}_i), \Sigma = \mathrm{Var}(\boldsymbol{X}_i)$ does not depend on $i$, and $\Delta = \mathrm{Cov}(\boldsymbol{X}_k, \boldsymbol{X}_l), k \neq l$ does not depend on $k, l$. In other words, the second-order beliefs are invariant under an arbitrary permutation of the index, $k$. Exchangeable sequences are the subjectivists generalisation of independent, identically distributed variables. Of course, they are not independent (or even uncorrelated), but they are very well behaved, as the following representation shows.

From the given specification, we may use the second-order exchangeability representation theorem (Goldstein 1986a) to decompose $\boldsymbol{X}_k$ as

$$\boldsymbol{X}_k = \boldsymbol{M} + \boldsymbol{R}_k \tag{2.1}$$

where $\boldsymbol{M}$ and $\boldsymbol{R}_k$ are vectors of random quantities, and $\mathrm{P}(\boldsymbol{R}_k) = 0$, $\mathrm{Cov}(\boldsymbol{M}, \boldsymbol{R}_k) = 0$, $\forall k$, $\mathrm{Cov}(\boldsymbol{R}_k, \boldsymbol{R}_l) = 0, \forall k \neq l$, and the vectors $\boldsymbol{R}_k = (R_{1k}, \ldots, R_{rk})^{\mathrm{T}}$ form a second order exchangeable sequence. Here, $\boldsymbol{M}$ may be thought of as representing underlying population behaviour, and $\boldsymbol{R}_k$ as representing individual variation. We can now see that whilst the quantities, $\boldsymbol{X}_k$, themselves are not necessarily uncorrelated, they would be uncorrelated if there were no uncertainty about the underlying mean, $\boldsymbol{M}$ of the quantities.

Bayes linear updating for such a representation would be informative for the means, and so $\text{Var}(\boldsymbol{M})$ would go to zero, given sufficient data. However, the data is not informative for future $\boldsymbol{R}_k$ under a second-order analysis, and so we do not learn in any way about the matrix $\text{Var}(\boldsymbol{R}_k)$. $\text{Var}(\boldsymbol{R}_k)$ is the underlying covariance matrix for the data, and has an important effect on the way in which we learn about the means. A method for quantifying uncertainty about the matrix, $\text{Var}(\boldsymbol{R}_k)$, is now presented, a necessary step on the way to providing a method for learning about such a matrix.

For the matrix $A = (\boldsymbol{a}_1, \boldsymbol{a}_2, \ldots, \boldsymbol{a}_n)$, define

$$\mathbf{vec}A = (\boldsymbol{a}_1{}^{\mathrm{T}}, \boldsymbol{a}_2{}^{\mathrm{T}}, \ldots, \boldsymbol{a}_n{}^{\mathrm{T}})^{\mathrm{T}} \tag{2.2}$$

The **vec** operator and it's properties are discussed in Searle (1982, Section 12.9).

Consider the sequence of $r^2$-dimensional vectors

$$\boldsymbol{Y}_k = \mathbf{vec}(\boldsymbol{R}_k \boldsymbol{R}_k{}^{\mathrm{T}}) \tag{2.3}$$

representing the quadratic products of the residuals. It is assumed that the $\boldsymbol{Y}_k$ are second-order exchangeable, and the additional specifications $\Sigma' = \text{Var}(\boldsymbol{Y}_k)$ and $\Delta' = \text{Cov}(\boldsymbol{Y}_k, \boldsymbol{Y}_l), k \neq l$ are expressed. Once again, $\Sigma'$ and $\Delta'$ should be specified directly. Note however, that if a multivariate normal assumption was felt appropriate for the $\boldsymbol{R}_k$, then $\Sigma' - \Delta'$ may be inferred from the fourth moments of that distribution. This is the subject of the next section. Of course, if one feels that some other distribution is more appropriate, then the moments of that distribution may be used. Indeed, if one feels up to the task, it is preferable to make the specifications directly. One is only constrained by the usual conditions of coherence.

The exchangeability representation theorem may be used to decompose the vector $\boldsymbol{Y}_k$ and then re-write the representation in matrix form in the following way.

$$\boldsymbol{R}_k \boldsymbol{R}_k^{\mathrm{T}} = V + U_k \tag{2.4}$$

where $V$ and $U_k$ are $r \times r$ random matrices such that $\mathrm{P}(U_k) = 0$, $\mathrm{Cov}(\mathbf{vec}V, \mathbf{vec}U_k) = \mathrm{Cov}(\mathbf{vec}U_k, \mathbf{vec}U_l) = 0$, $\forall k \neq l$, and the $\mathbf{vec}U_k$ form a second-order exchangeable sequence. In particular, $\mathrm{Var}(\mathbf{vec}V) = \Delta'$ and $\mathrm{Var}(\mathbf{vec}U_k) = \Sigma' - \Delta'$ is not dependent on $k$. Here, $V$ represents underlying covariance behaviour, and $U_k$ represents individual variation within the quadratic products of residuals (note that $\mathrm{P}(V) = \mathrm{P}(\boldsymbol{R}_k \boldsymbol{R}_k^{\mathrm{T}}) = \mathrm{Var}(\boldsymbol{R}_k)$).

If we observe a sample $\boldsymbol{X}_1, \ldots, \boldsymbol{X}_n$ of size $n$, then the sample covariance matrix takes the form

$$\begin{aligned} S &= \frac{1}{n-1} \sum_{w=1}^{n} (\boldsymbol{X}_w - \bar{\boldsymbol{X}})(\boldsymbol{X}_w - \bar{\boldsymbol{X}})^{\mathrm{T}} && (2.5) \\ &= \frac{1}{n-1} \sum_{w=1}^{n} (\boldsymbol{R}_w - \bar{\boldsymbol{R}})(\boldsymbol{R}_w - \bar{\boldsymbol{R}})^{\mathrm{T}} && (2.6) \end{aligned}$$

where $\bar{\boldsymbol{Z}} = (1/n) \sum_{i=1}^{n} \boldsymbol{Z}_i$, $\forall \boldsymbol{Z}_i$. Beliefs for the sample covariance matrix $S$ are, by (2.6), uniquely determined by representation (2.4). Imagine forming a sequence of sample covariance matrices, $S_1, S_2, \ldots$; each based on $n$ observations. Then the covariance structure for the sample covariance matrices takes on the following form.

$$\begin{aligned} \mathrm{Cov}(\mathbf{vec}V, \mathbf{vec}S_k) &= \mathrm{Var}(\mathbf{vec}V) \quad \forall k && (2.7) \\ \mathrm{Cov}(\mathbf{vec}S_k, \mathbf{vec}S_l) &= \mathrm{Var}(\mathbf{vec}V) \quad \forall k \neq l && (2.8) \\ \mathrm{Var}(\mathbf{vec}S_k) &= \mathrm{Var}(\mathbf{vec}V) + \frac{\mathrm{Var}(\mathbf{vec}U_k)}{n} \quad \forall k && (2.9) \end{aligned}$$

These results are derived in Appendix A.

Observing sample covariances from a sample of size $n$ reduces uncertainty for $V$, the underlying covariance matrix, by linear fitting, but is uninformative for $U_k$ for $k > n$.

## 2.3 Automated specification of the residual quadratic structure

Specification of the matrix $\mathrm{Var}(\mathbf{vec}U_k)$ is a very unfamiliar problem. However, if it is assumed that conditional upon knowledge of $V$ (*ie.* $\mathrm{Var}(\mathbf{vec}V) = 0$), the $\boldsymbol{R}_k$ are multivariate normally distributed, then $\mathrm{Var}(\mathbf{vec}U_k)$ may be inferred from the fourth moments of the multivariate normal distribution with mean vector zero, and covariance matrix $\mathrm{Var}(\boldsymbol{R}_k)$. The multivariate normal distribution is discussed in Searle (1982, Section 13.4), and quadratic forms thereof in Searle (1982, Section 13.5).

If $\boldsymbol{X} = (X_1, X_2, X_3, X_4)^{\mathrm{T}}$ is a multivariate normal vector such that $\mathrm{P}(\boldsymbol{X}) = 0$, then

$$\mathrm{Cov}(\boldsymbol{X}^{\mathrm{T}}P\boldsymbol{X}, \boldsymbol{X}^{\mathrm{T}}Q\boldsymbol{X}) = 2\mathrm{Tr}[P\mathrm{Var}(\boldsymbol{X})Q\mathrm{Var}(\boldsymbol{X})] \tag{2.10}$$

where $P$ and $Q$ are any constant conformable matrices. This is a well known result from normal theory, and is discussed in Searle (1971) and Rohde and Tallis (1969). From this, one may easily deduce that

$$\mathrm{P}(X_1X_2X_3X_4) = \mathrm{P}(X_1X_2)\mathrm{P}(X_3X_4) + \mathrm{P}(X_1X_3)\mathrm{P}(X_2X_4) + \mathrm{P}(X_1X_4)\mathrm{P}(X_2X_3) \tag{2.11}$$

This may also be deduced from the fact that the moment generating function is

$$m_X(\boldsymbol{t}) = exp\left\{\frac{1}{2}\boldsymbol{t}^{\mathrm{T}}\mathrm{Var}(\boldsymbol{X})\boldsymbol{t}\right\} \tag{2.12}$$

and that

$$\mathrm{P}(X_1X_2X_3X_4) = \frac{\partial^4}{\partial t_1 \partial t_2 \partial t_3 \partial t_4} m_X(\boldsymbol{t}) \tag{2.13}$$

The algebra is a little tedious, but leads to the more general result that previsions (expectations) of odd products are zero, and previsions of even products can be calculated by forming all possible pairs, then for each combination, form the product of the prevision of the pairs, then sum over pairs. A vectorised version of (2.11) is required. First some notation for two different "direct products" of matrices is needed. The direct product is discussed in Searle (1982, Section 10.7) and it's relationship with the **vec** operator is discussed in Searle (1982, Section 12.9).

**Definition 1** *For $r \times r$ matrices $A$ (having entry $a_{ij}$ in row $i$, column $j$) and $B$ (having entry $b_{ij}$ in row $i$, column $j$) the* (left) direct product, *$A \otimes B$ of $A$ and $B$ is defined to be the $r^2 \times r^2$ matrix with the element $a_{jk}b_{lm}$ in row $r(l-1)+j$, column $r(m-1)+k$.*

**Definition 2** *For $r \times r$ matrices $A$ (having entry $a_{ij}$ in row $i$, column $j$) and $B$ (having entry $b_{ij}$ in row $i$, column $j$) the* star product *$A \star B$ of $A$ and $B$ is defined to be the $r^2 \times r^2$ matrix with the element $a_{jk}b_{lm}$ in row $r(k-1)+m$, column $r(l-1)+j$.*

It is worth noting that

$$A \star B = I_{r,r}(A \otimes B) \tag{2.14}$$

where $I_{r,r}$ is the $(r,r)^{th}$ *vec-permutation matrix*. A full review of the definitions of, and the relationships between **vec**, vec-permutation matrices and direct products can be found in Henderson and Searle (1981).

Now, given (2.11), it trivially follows that for the mean zero, MVN vector $\boldsymbol{X}$, we have

$$\mathrm{Var}(\mathbf{vec}(\boldsymbol{X}\boldsymbol{X}^{\mathrm{T}})) = \mathrm{Var}(\boldsymbol{X}) \otimes \mathrm{Var}(\boldsymbol{X}) + \mathrm{Var}(\boldsymbol{X}) \star \mathrm{Var}(\boldsymbol{X}) \tag{2.15}$$

This is the dispersion matrix for the $p$-dimensional Wishart distribution with 1 degree of freedom, and (2.15) is a special case of the Wishart dispersion result in Henderson and Searle (1979). Equation (2.15) may be regarded as a primitive statement about the way in which the fourth-order moments depend on the second-order moments, thus weakening the requirement of a multivariate normality assumption. The distributional assumption was made only so that the fourth order moments may be deduced from the first and second-order moments. Whatever distributional assumption is made, if the fourth-order centred moments depend only on the second-order centred moments, then the function will be symmetric across covariances. Equation (2.15) represents one of the simplest symmetric functions on the covariance possible, and so the assumption of this form for the dependencies may be natural independently of any normality assumption.

An analogy may be useful at this point. The usual second-order Bayes linear adjustment gives results identical to those which would have been obtained using a distributional Bayesian approach, together with the assumption of multivariate normality. However, the Bayes linear adjustment is used as a natural method for updating moments without making any distributional assumptions at all. In the same way, I suspect that (2.15) represents a natural way to assign the fourth-order moments using only the second-order moments, irrespective of any distributional assumptions.

In all of the illustrative examples in this thesis, (2.15) was used in order to “deduce” the form of the fourth order residual structure. A [B/D] macro was written which automatically formed quadratic products of desired variables, and made the residual specifications accordingly.

It is appropriate to end this section with a major caveat. The *specifications* made for the quadratic (and indeed, any other) covariance structure must necessarily be *statements of belief*. Ultimately, any method which automatically assigns specifica-tions for the fourth-order moments using lower-order moments is nothing more than a crude approximation (although there will often be situations where the analysis is quite insensitive to the precise specifications at this level). I would say that specification of belief for quadratic products (and specification of belief, more generally) is an area that should be given more attention. However, the requirement that all such specifications *must* be made in order simply to revise beliefs about covariance structures is unreasonable, and is one of the central themes of this thesis. In the next chapter, a structure will be developed in such a way that the *minimum* specification required in order to carry out such analyses is vastly reduced, consequently diminishing the need to resort to techniques such as those outlined in this section.

## 2.4 Bayes linear adjustment for the quadratic structure

Let the sample covariance matrix $S$, have elements $S_{ij}$, and the underlying covariance matrix, $V$, have elements $V_{ij}$. Form the vector space, $\mathcal{V}$ of all (real) linear combinations of the these elements (together with the unit constant, 1).

$$\mathcal{V} = span\{1, V_{ij}, S_{ij} | \forall i, j\} \tag{2.16}$$

and then define the inner-product, $(\cdot, \cdot) : \mathcal{V} \times \mathcal{V} \rightarrow \mathbb{R}$ via

$$(X, Y) = \mathrm{P}(XY), \ \forall X, Y \in \mathcal{V} \tag{2.17}$$

Note that this inner-product induces the distance (or *loss*) function

$$d(X, Y) = \mathrm{P}([X - Y]^2), \ \forall X, \in \mathcal{V} \tag{2.18}$$

Now define the observable subspace $D \subseteq \mathcal{V}$ via

$$D = span\{1, S_{ij} | \forall i, j\} \tag{2.19}$$

Now for any subspace $G \subseteq \mathcal{V}$, define the adjusted expectation operator, $\mathrm{E}_G : \mathcal{V} \rightarrow G$ to be the bounded linear operator which orthogonally projects $Y \in \mathcal{V}$ into $G$. In particular, form $\mathrm{E}_D(V_{ij})$, $\forall i, j$. These depend only on the data, and represent the *Bayes linear rules* for the $V_{ij}$, given the data. Explicitly, using (1.7), (2.7) and (2.9),

$$\begin{aligned} \mathrm{E}_D(\mathbf{vec}V) &= \mathrm{P}(\mathbf{vec}V) + \mathrm{Cov}(\mathbf{vec}V, \mathbf{vec}S)\mathrm{Var}(\mathbf{vec}S)^{-1}[\mathbf{vec}S - \mathrm{P}(\mathbf{vec}V)] && (2.20) \\ &= \mathrm{P}(\mathbf{vec}V) + \mathrm{Var}(\mathbf{vec}V)\left\{\mathrm{Var}(\mathbf{vec}V) + \frac{\mathrm{Var}(\mathbf{vec}U_k)}{n}\right\}^{-1}\mathbf{vec}[S - \mathrm{P}(V)] && (2.21) \end{aligned}$$

Further, if beliefs over $U_k$ are assigned by MVN fitting, using (2.15), this becomes,

$$\begin{aligned} \mathrm{E}_D(\mathbf{vec}V) &= \mathrm{P}(\mathbf{vec}V) + \mathrm{Var}(\mathbf{vec}V) \\ &\quad \times \left\{\mathrm{Var}(\mathbf{vec}V) + \frac{\mathrm{Var}(R)\otimes\mathrm{Var}(R) + \mathrm{Var}(R)\star\mathrm{Var}(R)}{n}\right\}^{-1} \\ &\quad \times\mathbf{vec}[S - \mathrm{P}(V)] && (2.22) \end{aligned}$$

They are related to posterior beliefs about the $V_{ij}$ in a way made explicit in Goldstein (1994).

The specifications made, can therefore be used as a basis for a Bayes linear analysis of the covariance structures. However, for large matrices, the number of quantities involved in the adjustments will be prohibitively large (though simplifications could be made by focussing on small subsets of the problem). It would be desirable to analyse covariance matrices in a space where they live more naturally, exploiting their matrix structure.

# Chapter 3

# Bayes linear matrix spaces

## 3.1 Bayes linear inner-products

At the end of the last chapter, a special case of general construction of belief structures was demonstrated. In general, given a collection of quantities of interest, $B = [B_1, B_2, \ldots]$ and a collection of predictive observables, $D = [D_0 = 1, D_1, D_2, \ldots]$, form the real vector space, $\mathcal{V}$, of all linear combinations of these quantities

$$\mathcal{V} = span\{B \cup D\} \tag{3.1}$$

and then define an inner product, $(\cdot, \cdot) : \mathcal{V} \times \mathcal{V} \to \mathbb{R}$ on $\mathcal{V}$ via

$$(X, Y) = \mathrm{P}(XY), \ \forall X, Y \in \mathcal{V} \tag{3.2}$$

This induces the norm

$$\|X\|^2 = \mathrm{P}(X^2), \ \forall X \in \mathcal{V} \tag{3.3}$$

which in turn induces the metric

$$d(X,Y)^2 = \mathrm{P}([X-Y]^2),\ \forall X,Y \in \mathcal{V} \tag{3.4}$$

This is a natural metric to have on a space of random quantities. Note that if all quantities have zero prevision, then this inner-product simply corresponds to covariance. However, the inner-product has not been defined to be covariance, since two quantities should not be viewed as being "the same", simply because they have a correlation of one. Viewing them as being "the same" requires them to have the same prevision, and a correlation of one. Note that there is still a slight subtlety associated with the use of this inner-product, since strictly speaking, it is only an inner-product over equivalence classes of random quantities whose normed difference is zero. However, from a linear perspective, this inner-product captures exactly what is required. An introduction to the functional analytic ideas used in this chapter can be found in Kreyszig (1978). Where necessary, form the completion of the space $\mathcal{V}$, and denote the resulting Hilbert space by $\mathcal{H}$ (see Kreyszig 1978, Section 3.2-3).

For any closed subspace $G \subseteq \mathcal{H}$, define the *adjusted expectation operator*, $\mathrm{E}_G : \mathcal{H} \to G$, to be such that, for all $Y \in \mathcal{H}$, $\mathrm{E}_G(Y)$ is the orthogonal projection of $Y$ into $G$. Note then that $\mathrm{E}_{D_0}(Y) = \mathrm{P}(Y),\ \forall Y$, and so where there is no confusion, the subscript is dropped, so that $\mathrm{E}(Y) = \mathrm{P}(Y)$. Note that expectation and adjusted expectation are defined entirely in terms of the inner-product, $(\cdot,\cdot)$, and that the inner-product is defined in terms of prevision.

By keeping distinct notation for the two concepts of prevision and adjusted expectation, it is at all times clear whether or not we are dealing with a primitively defined prevision, or a projection in the relevant Hilbert space. Whilst this is not quite so critical in the case of scalar adjustments, it is very helpful for the spaces of random matrices which will be constructed in the next section.

Also note that for any $X \in B$, $\mathrm{E}_D(X)$ is the *Bayes linear rule* for $X$, given the data, (see Section 1.4) since the orthogonal projection of $X$ into $D$ is the linear combination of elements of $D$ which minimises expected quadratic loss. Consequently,

$$\mathrm{E}_D(X) = \mathrm{P}(X) + \mathrm{Cov}(X, \boldsymbol{D})\mathrm{Var}(\boldsymbol{D})^{-1}[\boldsymbol{D} - \mathrm{P}(\boldsymbol{D})] \tag{3.5}$$

where $\boldsymbol{D}$ is the vector of elements of $D$.

## 3.2 Spaces of random matrices

In Chapter 2, we saw how beliefs about a covariance matrix may be revised, first by forming quadratic products of the scalar quantities, giving rise to the covariance matrix, and then specifying covariance matrices over the second-order exchangeable decompositions of all of these quantities, and then carrying out adjustment in the usual Bayes linear way. However, it is immediately obvious that faced with problems where the covariance matrix to learn about is large, the magnitude of belief specification and computation required in order to carry out adjustment is going to be considerable. It would be desirable to create a framework whereby matrices may be analysed in a space where they may be either treated as a whole, or broken down into as many components as the belief analyst feels comfortable working with. For small problems, or in problems where a great deal of detailed knowledge about the interaction of variables is known, it may well be desirable to work with the components of the matrices directly. However, faced with large problems, or problems where it is infeasible to elicit such detailed specifications, one may simply wish to make simple scalar statements representing uncertainty in the prior or sample matrices, and the "interaction" between them. It is perfectly possible to set up a framework where a Bayes linear analysis may take place given such limited specifications.

The representation which will allow us to treat a covariance matrix as a single object is now developed. Let

$$B = [B_1, B_2, \ldots] \tag{3.6}$$

be a collection of random $r \times r$ non-negative definite, real symmetric matrices, representing unknown matrices of interest. These might, for example, represent population covariance matrices. Let

$$D = [D_1, D_2, \ldots] \tag{3.7}$$

be another such collection, representing observable matrices (such as sample covariance matrices). Finally, form a collection of $r^2$ linearly independent $r \times r$ constant matrices such that $C_{r(i-1)+j}$ is the matrix with a 1 in the $(i,j)^{th}$ position, and zeros elsewhere, where $i$ and $j$ range from 1 to $r$ and call this collection

$$C = [C_1, \ldots, C_{r^2}] \tag{3.8}$$

This collection of matrices is a basis for the space of constant $r \times r$ matrices. Next form a vector space

$$\mathcal{N} = span\{B \cup C \cup D\} \tag{3.9}$$

of all linear combinations of the elements of these collections, and define the inner-product (over equivalence classes) on $\mathcal{N}$ as

$$(P, Q) = \mathrm{P}(\mathrm{Tr}(PQ^{\mathrm{T}})) \quad \forall P, Q \in \mathcal{N} \tag{3.10}$$

which induces the norm

$$||P||^2 = \mathrm{P}(||P||_F^2), \ \forall P \in \mathcal{N} \tag{3.11}$$

which in turn induces the metric

$$d(P, Q)^2 = \mathrm{P}(\|P - Q\|_F^2) \quad \forall P, Q \in \mathcal{N}, \tag{3.12}$$

where $\|\cdot\|_F$ denotes the Frobenius norm of a matrix. This is the sum of the squares of the elements, or equivalently, the sum of the squares of the eigenvalues. Where necessary, form the completion of the space. The complete inner-product space, or Hilbert space, is denoted by $\mathcal{M}$.

Analogously with the revision of belief over scalar quantities (Goldstein 1981), we learn about the elements of the collection $B$, by orthogonal projection into closed subspaces of $\mathcal{M}$ spanned by elements of the collection $C \cup D$, in order to obtain the corresponding adjusted expectations, namely the linear combinations of sample covariance matrices which give our adjusted beliefs.

Note also that the projection into the constant space, $\mathrm{E}_C$, is such that

$$\mathrm{E}_C(Q) = \mathrm{P}(Q),\ \forall Q \in \mathcal{M} \tag{3.13}$$

and so where there is no confusion, we often drop the subscript to get $\mathrm{E}(Q) = \mathrm{P}(Q)$.

## 3.3 Matrix inner-product

Why choose the inner-product $(P, Q) = \mathrm{P}(\mathrm{Tr}(PQ^{\mathrm{T}}))$ for the matrix space? Are there other inner-products which would be equally appropriate? As for the Bayes linear theory for random scalars, all of the theory is developed for a general inner-product space, and so one is free to use any inner-product which one feels to be appropriate. However, there are foundational reasons for using this inner-product, since the induced norm is based on the expectation of a proper scoring rule for matrices (Goldstein 1996), and so an argument for using this inner-product can be given which is similar to that used in Goldstein (1986b). Further, there is a sense in which the inner-product chosen here is the natural extension of the inner-product $(X, Y) = \mathrm{P}(XY)$ used for scalar quantities.

Given a vector space, $\mathcal{N}$ of $n \times n$ random matrices, let $\phi_{ij} : \mathcal{N} \to \mathcal{P}_{ij}$ be the homomorphism which maps the matrix to it's $(i, j)^{th}$ element, for all $i$ and $j$. For example, for any $P \in \mathcal{N}$, $\phi_{ij}(P)$ is the $(i, j)^{th}$ element of $P$ — a random scalar. $\mathcal{P}_{ij}$ is a vector space of random scalars, $\phi_{ij}(\mathcal{N})$. Define an inner-product $(\cdot, \cdot)_{ij} : \mathcal{P}_{ij} \times \mathcal{P}_{ij} \to \mathbb{R}$ over each of the vector spaces $\mathcal{P}_{ij}$, and henceforth regard them as inner-product spaces. Now define a new space $\mathcal{Q}$ to be the direct sum of the spaces $\mathcal{P}_{ij}$ (direct sums are discussed in Section 3.3 of Kreyszig (1978)).

$$\mathcal{Q} = \bigoplus_{i=1}^{n} \bigoplus_{j=1}^{n} \mathcal{P}_{ij} \tag{3.14}$$

The inner-product, $[\cdot, \cdot]$ on $\mathcal{Q}$ is uniquely determined by the inner-products on the subspaces it is composed of.

$$[p, q] = \sum_{i=1}^{n} \sum_{j=1}^{n} (p_{ij}, q_{ij})_{ij}, \ \forall p = (p_{11}, p_{12}, \ldots, p_{nn}), q = (q_{11}, q_{12}, \ldots, q_{nn}) \in \mathcal{Q} \tag{3.15}$$

Ignoring the inner-products, the vector space, $\mathcal{Q}$ is isomorphic to the vector space $\mathcal{N}$, via the isomorphism, $\phi : \mathcal{Q} \to \mathcal{N}$ defined via

$$\phi(p_{11}, p_{12}, \ldots, p_{nn}) = \begin{pmatrix} p_{11} & p_{12} & \cdots & p_{1n} \\ p_{21} & p_{22} & \cdots & p_{2n} \\ \vdots & \vdots & \ddots & \vdots \\ p_{n1} & p_{n2} & \cdots & p_{nn} \end{pmatrix} \tag{3.16}$$

If one is prepared to accept $\phi$ as an inner-product space isomorphism from $\mathcal{Q}$, to $\mathcal{N}$, the inner-product $\{\cdot, \cdot\} : \mathcal{N} \times \mathcal{N} \to \mathbb{R}$ over $\mathcal{N}$ is induced. For example, if $\phi(p) = P$,

and $\phi(q) = Q$, then

$$\{P, Q\} = [p, q] = \sum_{i=1}^{n}\sum_{j=1}^{n}(p_{ij}, q_{ij})_{ij} \tag{3.17}$$

But if on each of the scalar spaces, the usual Bayes linear inner-product

$$(p_{ij}, q_{ij})_{ij} = \mathrm{P}(p_{ij}q_{ij}), \ \forall p_{ij}, q_{ij} \in \mathcal{P}_{ij} \tag{3.18}$$

is used, one may deduce that

$$\begin{aligned} \{P, Q\} &= [p, q] && (3.19) \\ &= \sum_{i=1}^{n}\sum_{j=1}^{n}\mathrm{P}(p_{ij}q_{ij}) && (3.20) \\ &= \mathrm{P}(\mathrm{Tr}[PQ^{\mathrm{T}}]), \ \forall P, Q \in \mathcal{N} && (3.21) \end{aligned}$$

It is important to note that adopting the usual Bayes linear inner-product on the scalar subspaces in no way *forces* us to adopt the inner-product advocated for the matrix space. The spaces $\mathcal{Q}$ and $\mathcal{N}$ are only *necessarily* isomorphic when considered purely as *vector spaces*. The matrix inner-product for $\mathcal{N}$ is only implied given the additional *specification* that the *inner-product spaces* $\mathcal{Q}$ and $\mathcal{N}$ are isomorphic. Vector and inner-product space isomorphisms are defined in Sections 2.8-8 and 3.2-2 of Kreyszig (1978), respectively.

Note that by viewing the matrix space in this way, many desirable properties of the matrix inner-product become apparent, which link matrix and scalar spaces. An important property has already been mentioned — projection of a random matrix into the constant space gives it's prevision. Also, note that the induced norms are *consistent*. In other words, for any matrix $P$,

$$\|P\|^2 = \{P, P\} = 0 \Leftrightarrow \|p_{ij}\|^2 = (p_{ij}, p_{ij}) = 0 \quad \forall i, j \tag{3.22}$$

This is important when it comes to matching up the completions of matrix and component scalar spaces.

If all matrices of interest contain only one non-zero component (all in the same position), the inner product becomes $(P, Q) = \mathrm{P}(P_{ij}Q_{ij})$, inducing the distance $d(P, Q)^2 = \mathrm{P}((P_{ij} - Q_{ij})^2)$, as for the usual Bayes linear theory for scalar quantities. Further, when matrices are decomposed, the different subspaces representing different parts of the matrices remain orthogonal, preventing different subspaces from influencing one another. The matrix structure is a generalisation of the scalar Bayes linear structure, and scalar Bayes linear adjustments can be recovered by decomposing all variance structures to the one component level. Viewing the matrix space as a direct sum of a large number of orthogonal subspaces of matrix components is analagous to viewing a probability distribution as an amalgamation of a partition of indicator functions for the component events.

The matrices we are considering do not have to be finite dimensional. All of the theory remains valid if we think in terms of representations of random bounded linear self-adjoint operators on a (possibly infinite-dimensional) vector space.

## 3.4 General Bayes linear representations

### 3.4.1 $n$-step exchangeable collections

There is a common form of symmetry which often arises amongst ordered vectors of random quantities. It is essentially just a slightly weaker concept than that of (second-order) exchangeability. The covariance structure is invariant under arbitrary translations and reflections of the ordering, and the auto-correlation function becomes constant after some distance, $n$. We will call ordered vectors with this property, *second-order $n$-step exchangeable*. As we have seen, covariance may be interpreted as an inner-product on a space of random quantities. This same symmetry also occurs, under the same sorts of circumstances, for collections of random matrices in a random matrix inner-product space. Hence, a concept of $n$-step exchangeability which is sufficiently general that it is also valid for spaces of matrices is required, and so the concept is formalised as follows.

Let $\{Y_{jk} | \forall j, k\}$ be a collection of random entities of interest. Also form a maximal linearly independent collection of constant entities of the same type, and call this collection $C = [C_1, C_2, \ldots]$. When dealing with random scalars, $C$ will consist of the single scalar, $C_1 = 1$. In Section 3.2, the constant space for a collection of random matrices was described. Form the vector space

$$\mathcal{V} = span\{C_i, Y_{jk} | \forall i, j, k\} \tag{3.23}$$

so that the random entities are now vectors within this space. Define an inner-product $(\cdot, \cdot) : \mathcal{V} \times \mathcal{V} \longrightarrow \mathbb{R}$ on $\mathcal{V}$. The inner-product should capture certain aspects of our beliefs about the relationships between the elements of $\mathcal{V}$. Form the completion of the space $\mathcal{V}$, and denote this Hilbert space by $\mathcal{H}$. In such a general Bayes linear space, a bounded linear *expectation* function $\mathrm{E}(\cdot) : \mathcal{H} \longrightarrow span\{C\}$, is defined such that $\forall Y \in \mathcal{H}$, $\mathrm{E}(Y)$ is the orthogonal projection of $Y$ into $span\{C\}$, with respect to the inner-product $(\cdot, \cdot)$.

**Definition 3** *If* $\exists n \in \mathbb{N}$ *such that* $\mathrm{E}(Y_{jk}) = e_j \quad \forall j, k,$ *and*

$$\begin{array}{rcll}
(Y_{ik}, Y_{jl}) & = & d_{0ij} & \forall i, j, |k-l| = 0 \\
(Y_{ik}, Y_{jl}) & = & d_{1ij} & \forall i, j, |k-l| = 1 \\
\vdots & \vdots & \vdots & \vdots \\
(Y_{ik}, Y_{jl}) & = & d_{(n-1)ij} & \forall i, j, |k-l| = n-1
\end{array}$$

$$(Y_{ik}, Y_{jl}) = c_{ij} \qquad \forall i, j, |k - l| \geq n \tag{3.24}$$

*the collection* $\{Y_{jk} | \forall j, k\}$ *is said to be* generalised (second-order) $n$-step exchangeable over $k$. *If* $n = 1$, *the collection is said to be* generalised (second-order) exchangeable over $k$.

□

### 3.4.2 Representation for $n$-step exchangeable collections

Goldstein (1986a) constructs a general representation for second-order exchangeable collections. There is an analagous representation for collections with the weaker property of $n$-step exchangeability, constructed in a similar way.

**Theorem 1** *Let* $\{Y_{jk} | \forall j, k\}$ *be generalised second-order* $n$*-step exchangeable over* $k$ *with respect to the inner-product* $(\cdot, \cdot)$. *Then the* $Y_{jk}$ *may be represented as*

$$Y_{jk} = M_j + R_{jk} \quad \forall j, k \tag{3.25}$$

*where the* $M_j$ *and* $R_{jk}$ *have the following properties:*

$$\mathrm{E}(Y_{jk}) = \mathrm{E}(M_j), \ \mathrm{E}(R_{jk}) = 0 \quad \forall j, k \tag{3.26}$$

$$(M_i, M_j) = (M_i, Y_{jk}) = c_{ij}, \quad (M_i, R_{jk}) = 0 \quad \forall i, j, k \tag{3.27}$$

$$(R_{ik}, R_{jl}) = (Y_{ik}, R_{jl}) = (Y_{ik}, Y_{jl}) - c_{ij} \quad \forall i, j, k, l \tag{3.28}$$

*Further, the* $\{R_{jk} | \forall j, k\}$ *are generalised second-order* $n$*-step exchangeable over* $k$, *with* $(R_{ik}, R_{jl}) = 0 \quad \forall i, j, |k - l| \geq n$.

*Proof*

Let

$$M_{im} = \frac{1}{m}\sum_{k=1}^{m} Y_{ik} \quad \forall i, m \tag{3.29}$$

Observe that the sequence $M_{i1}, M_{i2}, \ldots$ is *Cauchy* $\forall i$ ie. that $(M_{ik}-M_{il}, M_{ik}-M_{il}) \longrightarrow 0$ as $k, l \longrightarrow \infty$, which follows directly from the properties of $n$-step exchangeable sequences. Construct the quantity $M_i$ to be the Cauchy limit of this sequence so that

$$\lim_{m\to\infty}(M_{im}, Y) = (M_i, Y) \ \forall i, \forall Y \in \mathcal{H} \tag{3.30}$$

Continuity of the inner-product is given in 3.2-2 of Kreyszig (1978). Linearity of $\mathrm{E}(\cdot)$ gives $\mathrm{E}(M_{im}) = e_i \quad \forall i, m$, and hence applying (3.30) for $Y \in C$ we deduce $\mathrm{E}(M_i) = e_i$. Define $R_{ik}$ via $R_{ik} = X_{ik} - M_i \ \forall i$, so that $\mathrm{E}(R_{im}) = 0 \quad \forall i, m$. The other properties of the representation follow directly from (3.30). □

As for the case of second-order exchangeability, the mean components of the representation, $M_j$, represent the quantities which may be learned about by linear fitting on the data. It is possible to resolve as much uncertainty as is wished about these quantities given a sufficient number of observations, by such linear fitting. Therefore, the $n$-step exchangeable collection $\{Y_{jk} | \forall j, k\}$ with representation $Y_{jk} = M_j + R_{jk} \quad \forall j, k$ will be said to *identify* the random quantities $M_j, \ \forall j$.

### 3.4.3 Example

Consider the following simple time series model for a sequence of observations $\{X_1, X_2, \ldots\}$.

$$X_t = M_t + R_t, \quad \forall t \geq 1 \tag{3.31}$$

$$M_t = M_{t-1} + S_t, \quad \forall t \geq 2 \tag{3.32}$$

$$M_1 = S_1 \tag{3.33}$$

Where the random quantities $\{R_i, S_i | \forall i \geq 1\}$ have zero prevision, are mutually uncorrelated, and are such that $\text{Var}(R_i) = r$ and $\text{Var}(S_i) = s \ \forall i \geq 1$. Now form the one step differences of the observations.

$$X'_t = X_t - X_{t-1} = R_t - R_{t-1} + S_t \tag{3.34}$$

Applying Definition 3 to the space $\mathcal{V} = span\{1, X'_i | \forall i \geq 2\}$ with the inner-product $(X, Y) = \text{P}(XY), \ \forall X, Y \in \mathcal{V}$, we see that the $\{X'_i | \forall i \geq 2\}$ are (second-order) 2-step exchangeable. Applying Theorem 1, we see that the $\{X'_i | \forall i \geq 2\}$ identify zero.

### 3.4.4 Matrix example

Reconsider the exchangeable vector example developed in Chapter 2. Imagine forming a sequence of sample covariance matrices, $\{S_1, S_2, \ldots\}$, each based upon $n$ observations. Then form the matrix vector space

$$\mathcal{V} = span\{\boldsymbol{R}_1\boldsymbol{R}_1{}^{\text{T}}, \boldsymbol{R}_2\boldsymbol{R}_2{}^{\text{T}}, \ldots, S_1, S_2, \ldots\} \tag{3.35}$$

(each $S_i$ is based upon $n$ of the $\boldsymbol{R}_j\boldsymbol{R}_j{}^{\text{T}}$) and impose the inner-product

$$(A, B) = \text{P}(\text{Tr}(AB^{\text{T}})), \ \forall A, B \in \mathcal{V} \tag{3.36}$$

Complete this space into a Hilbert space, $\mathcal{M}$. Limit points such as $V$, the underlying covariance matrix, will be added to the space upon completion. It is clear that both the residual and sample covariance matrices are generalised (second-order) exchangeable, and that they have exchangeable representations

$$\boldsymbol{R}_k \boldsymbol{R}_k^{\mathrm{T}} = V + U_k \tag{3.37}$$

and

$$S_k = V + T_k \tag{3.38}$$

where $(T_k, T_k) = (U_k, U_k)/n$. This is deduced from equations (2.7), (2.8) and (2.9) using the consistency of the inner-products on the scalar and matrix spaces. It is clear that the sample covariance matrices identify the underlying covariance matrix, $V$. Hence we may learn as much as is desired about this matrix by linear fitting on sufficiently many sample covariance matrices.

## 3.5 Primitive specification of the matrix inner-product

This thesis is primarily concerned with making specifications for the random matrix inner-product by building it up from specifications made for the quadratic scalars of which the matrices are composed. In a sense, this is analogous to specifying an expectation of a random quantity by breaking it up over a partition of events and specifying probabilities over the partition. However, just as there are many advantages to making expectation primitive, and specifying expectations directly, so there are with matrix inner-products. Initially, it may seem difficult to make primitive specifications for the matrix object inner-product, simply because it is a very unfamiliar problem. Nevertheless a scheme for elicitation based upon graphical modelling of the relationships between matrices, followed by quantifications of uncertainty, and proportions of uncertainty resolved due to knowledge of parent nodes, could be used in a way very similar to that often used for random scalars. In this way, the specification burden will be vastly reduced. Given a problem involving just a few (possibly large) matrices, all that will be required in order to carry out a basic analysis is a specification for the inner-product between every pair of matrices, rather than between every pair of scalars of which they are comprised.

# Chapter 4

# Covariance matrix adjustment for exchangeable vectors

## 4.1 Introduction

Consider the problem of learning about the covariance matrix for the $r$-dimensional exchangeable random vectors, $\{\boldsymbol{X}_k | \forall k \in \mathbb{N}\}$. As described in Chapter 2, form the exchangeable decomposition of the quadratic products of the residual vectors.

$$\boldsymbol{R}_k {\boldsymbol{R}_k}^{\mathrm{T}} = V + U_k \tag{4.1}$$

Let $C$ be a basis for the constant observable matrices, and let $D = [D_1, D_2, \ldots]$ be a collection of observable matrices predictive for the matrix, $V$ in (4.1). Then form a vector space of random matrices, as described in Chapter 3,

$$\mathcal{V} = span\{C \cup V \cup D\} \tag{4.2}$$

and impose the usual matrix inner-product

$$(A, B) = \mathrm{P}(\mathrm{Tr}(AB^{\mathrm{T}})) \tag{4.3}$$

If necessary, form the completion of the space. Denote the resulting Hilbert space by $\mathcal{M}$.

## 4.2 Decomposing the variance structure

As a simple example, $D$ might consist only of the sample covariance matrix, $S$, based on $n$ observations. In this case, the adjusted expectation for the "population" matrix would be a weighted linear combination of the prior and sample covariance matrices. However, by breaking down the sample covariance matrix into its component sub-matrices, one may resolve a greater proportion of our uncertainty about the "population" covariance matrix.

For simplicity, consider the problem of learning about the covariance structure induced by representation (2.4) for 2-dimensional vectors. The covariance matrices will be $2 \times 2$. Consider the sample covariance matrix

$$S = \begin{pmatrix} S_{11} & S_{12} \\ S_{12} & S_{22} \end{pmatrix} \tag{4.4}$$

and the corresponding "population" covariance matrix

$$V = \begin{pmatrix} V_{11} & V_{12} \\ V_{12} & V_{22} \end{pmatrix} \tag{4.5}$$

In the notation of the previous chapter, attention could be restricted to

$$B = [V], D_S = [S], C = \left[ \begin{pmatrix} 1 & 0 \\ 0 & 0 \end{pmatrix}, \begin{pmatrix} 0 & 1 \\ 1 & 0 \end{pmatrix}, \begin{pmatrix} 0 & 0 \\ 0 & 1 \end{pmatrix} \right], \tag{4.6}$$

where all $2 \times 2$ *symmetric* matrices can be constructed as linear combinations of the elements of $C^*$. Using these collections, our adjusted expectation for $V$ given $D_S$ (and $C$) would take the form

$$\mathrm{E}_{C+D_S}(V) = (1-\alpha)\mathrm{P}(V) + \alpha S \tag{4.7}$$

where $\alpha$ is the coefficient of the orthogonal projection determined by the inner-product (4.3). This simple form arises because from (P6) of Goldstein (1988a),

$$\mathrm{E}_{C+D}(V) = \mathrm{E}_C(V) + \mathrm{E}_{D-\mathrm{E}_C(D)}(V) \tag{4.8}$$

Also, by (3.38), $\mathrm{P}(D) = \mathrm{P}(S) = \mathrm{P}(V)$, and for the constant space, $C$, $\mathrm{E}_C(\cdot) = \mathrm{P}(\cdot)$. Explicitly, the coefficient $\alpha$ takes the form:

$$\alpha = \frac{(V - \mathrm{P}(V), V - \mathrm{P}(V))}{(S - \mathrm{P}(S), S - \mathrm{P}(S))} \tag{4.9}$$

$$= \frac{\sum_{i=1}^{2}\sum_{j=1}^{2} n\mathrm{Var}(V_{ij})}{\sum_{i=1}^{2}\sum_{j=1}^{2}\{n\mathrm{Var}(V_{ij}) + \mathrm{Var}(U_{ij})\}} \tag{4.10}$$

However, to improve the precision of the estimates, the projection space could be enlarged by constructing

$$D_I = \left[ \begin{pmatrix} S_{11} & 0 \\ 0 & 0 \end{pmatrix}, \begin{pmatrix} 0 & S_{12} \\ S_{12} & 0 \end{pmatrix}, \begin{pmatrix} 0 & 0 \\ 0 & S_{22} \end{pmatrix} \right] \tag{4.11}$$

Such a space is termed the *individual variance collection*. This allows different sample covariances to have different weights, if for example, there is higher prior uncertainty about some of the variances. Indeed, this may be taken a stage further, by construct-

*In the last chapter, a basis for all real matrices was suggested. However, if all of the other matrices in the problem are symmetric (as will usually be the case), then it is sufficient to work with a basis for only the symmetric matrices. All we require is that for all matrices, $M$, which are linear combinations of matrices in the problem, the constant matrix, $\mathrm{P}(M) \in span\{C\}$.

ing

$$
\begin{aligned}
D_F \;=\; & \left[\begin{pmatrix} S_{11} & 0 \\ 0 & 0 \end{pmatrix}, \begin{pmatrix} 0 & S_{11} \\ S_{11} & 0 \end{pmatrix}, \begin{pmatrix} 0 & 0 \\ 0 & S_{11} \end{pmatrix},\right. \\
& \begin{pmatrix} S_{12} & 0 \\ 0 & 0 \end{pmatrix}, \begin{pmatrix} 0 & S_{12} \\ S_{12} & 0 \end{pmatrix}, \begin{pmatrix} 0 & 0 \\ 0 & S_{12} \end{pmatrix}, \\
& \left.\begin{pmatrix} S_{22} & 0 \\ 0 & 0 \end{pmatrix}, \begin{pmatrix} 0 & S_{22} \\ S_{22} & 0 \end{pmatrix}, \begin{pmatrix} 0 & 0 \\ 0 & S_{22} \end{pmatrix}\right] \qquad (4.12)
\end{aligned}
$$

This last collection is called the *full variance collection*. This not only allows the different covariances to have different weights, but also allows relationships between covariances to have an effect on the adjustment. If $V$ is projected into the span of $D_F$ and $C$, then the adjusted expectation for $V$ will correspond precisely with the adjustment which would have been obtained using Bayes linear estimation on the quadratic products of the residuals in the scalar space.

Breaking down the population matrix in the same way, we let

$$
V_I = \left[\begin{pmatrix} V_{11} & 0 \\ 0 & 0 \end{pmatrix}, \begin{pmatrix} 0 & V_{12} \\ V_{12} & 0 \end{pmatrix}, \begin{pmatrix} 0 & 0 \\ 0 & V_{22} \end{pmatrix}\right]. \qquad (4.13)
$$

As the projection space is enlarged, more of the uncertainty about the variance structures is resolved, at the expense of doing more work. Generally projection should be carried out in as rich a space as is practicable, but for large variance matrices, the difference both in computational effort and in effort required for prior specification, between adjusting by $D_S$, $D_I$ and $D_F$ is substantial, so that a subjective assessment of the relative benefits of each adjustment must be made.

# 4.3 Example

## 4.3.1 Examination performance

A simple example, based on data relating to the examination performance of first year mathematics undergraduate students at Durham university, is presented. Those students who have only one A Level in mathematics are of particular interest, and so attention is restricted to these in this account. For illustrative purposes, focus on a few key variables, namely a summary of A Level performance ($A$), performance in the Christmas exams ($X$), and the end of year exam average ($E$).

For the exchangeable decomposition of (say) $A_k$, write

$$A_k = M_A + R_{A_k} \tag{4.14}$$

and for the exchangeable decomposition of (say) $R_{A_k}R_{X_k}$, write

$$R_{A_k}R_{X_k} = V_{AX} + U_{AX_k} \tag{4.15}$$

so that, for example, $V_{AX}$ represents the underlying covariance between the $A$ and $X$ variables, and $U_{AX_k}$ represents the residual for the $k^{th}$ observation. Construct the "population" and sample covariance matrices:

$$V = \begin{pmatrix} V_{AA} & V_{AE} & V_{AX} \\ V_{AE} & V_{EE} & V_{EX} \\ V_{AX} & V_{EX} & V_{XX} \end{pmatrix}, \quad S = \begin{pmatrix} S_{AA} & S_{AE} & S_{AX} \\ S_{AE} & S_{EE} & S_{EX} \\ S_{AX} & S_{EX} & S_{XX} \end{pmatrix} \tag{4.16}$$

A Bayes linear belief net was formed to represent beliefs about the relationships between the quadratic products of the residuals (Figure 4.1). Such graphs are discussed in Goldstein (1990), and from a more general perspective in Smith (1990). In general, the graphs have the property that nodes are *generalised conditionally inde-*

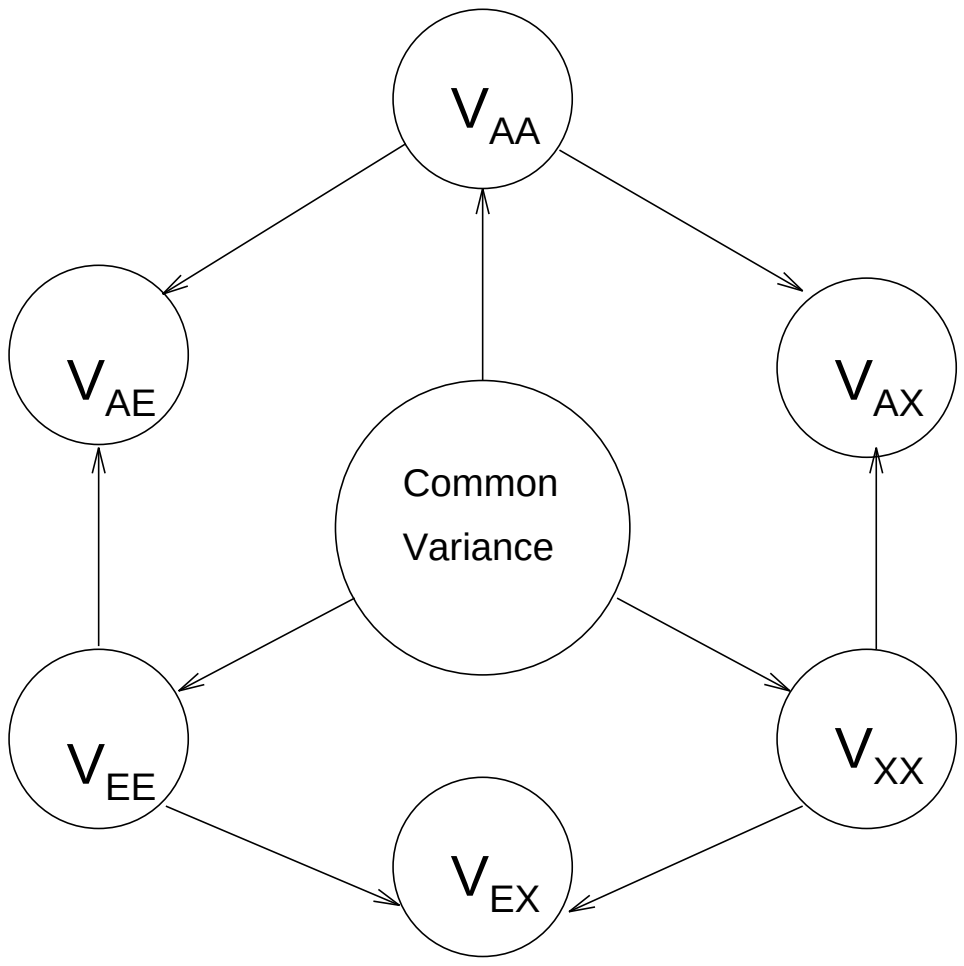

Figure 4.1: A conditional linear independence graph for the mean components of the quadratic products of the residuals

*pendent* given their parents, where generalised conditional independence is determined by a tertiary operator, $\cdot \perp\!\!\!\perp \cdot / \cdot$ which obeys the following conditions.

- $B \perp\!\!\!\perp C/(C+D)$
- $B \perp\!\!\!\perp C/D \Leftrightarrow C \perp\!\!\!\perp B/D$
- $B \perp\!\!\!\perp (C+D)/E \Leftrightarrow [\{B \perp\!\!\!\perp C/(D+E)\} \cap \{B \perp\!\!\!\perp D/E\}]$

We say that $A$ is generalised conditionally independent of $B$ given $C$ if and only if $A \perp\!\!\!\perp B/C$. Goldstein (1990) shows that adjusted orthogonality is a generalised conditional independence property. Explicity,

$$B \perp\!\!\!\perp C/D \Leftrightarrow \mathrm{E}_{C+D}(B) = \mathrm{E}_D(B) \tag{4.17}$$

defines a generalised conditional independence relation, which Smith (1990) refers to as *weak conditional independence*. This is the generalised conditional independence used to define the graph in Figure 4.1. The common variance node of Figure 4.1 reflects beliefs about the positive correlation between variances. This node does not represent an observable quantity, but is modelled conceptually in order to simplify the graph. Covariances are influenced by the corresponding variances. This graph was used to help structure the belief specification over the mean components of the variance structure. First a list ordering on the nodes is chosen: $\{CV, V_{AA}, V_{XX}, V_{EE}, V_{AX}, V_{AE}, V_{EX}\}$ ($CV$ denotes the node representing common variance).

The nodes are modelled as linear combinations of an orthonormal basis for the random variables. The coefficients of the combinations obviously form a lower triangular matrix with respect to a list ordering on the nodes, so that

$$\boldsymbol{V} = \Lambda \boldsymbol{E} \tag{4.18}$$

where $\boldsymbol{V}$ is the vector of list ordered nodes, $\boldsymbol{E}$ is a vector of orthonormal random variables, and $\Lambda$ is the lower triangular matrix of coefficients. Clearly we have

$$\mathrm{Var}(\boldsymbol{V}) = \Lambda\Lambda^{\mathrm{T}} \tag{4.19}$$

and so $\Lambda$ is the Choelesky triangle for the variance matrix.

By thinking about the uncertainty of nodes, and the contribution to that uncertainty by the parents of that node, the Choelesky triangle of the covariance matrix for the vector of list ordered nodes was specified as follows:

$$\Lambda = \begin{pmatrix} 1 & 0 & 0 & 0 & 0 & 0 & 0 \\ 0.71 & 0.71 & 0 & 0 & 0 & 0 & 0 \\ 13.26 & 0 & 13.26 & 0 & 0 & 0 & 0 \\ 7.07 & 0 & 0 & 7.07 & 0 & 0 & 0 \\ 2.17 & 1.08 & 1.08 & 0 & 1.25 & 0 & 0 \\ 3.25 & 1.62 & 0 & 1.62 & 0 & 1.86 & 0 \\ 6.50 & 0 & 3.25 & 3.25 & 0 & 0 & 3.75 \end{pmatrix} \tag{4.20}$$

For example, the common variance node, $CV$ was assigned a variance of 1, arbitrarily. Next, $V_{AA}$ was assessed to have a variance of 1, and to be such that one half of the standard deviation of $V_{AA}$ would be resolved due to knowledge of the unobservable common variance node. That determines the second row of the matrix $\Lambda$, since the coefficients for the modelled orthonormal variables contributing $CV$ and $V_{AA}$ must be the same, and the sum of their squares must be 1. Next, the quantity $V_{XX}$ was assessed to have a standard deviation of 18.75, and to be such that one half of it's variance would be resolved due to knowledge of the common variance node, thus determining the third row. Similar specifications were made in order to determine the rest of the matrix.

This implies the covariance matrix, $M = \Lambda\Lambda^{\mathrm{T}}$:

$$M = \begin{pmatrix} 1 & 0.71 & 13.26 & 7.07 & 2.17 & 3.25 & 6.50 \\ 0.71 & 1 & 9.38 & 5.00 & 2.30 & 3.44 & 4.59 \\ 13.26 & 9.38 & 351.56 & 93.75 & 43.06 & 43.06 & 129.17 \\ 7.07 & 5.00 & 93.75 & 100.00 & 15.31 & 34.44 & 68.89 \\ 2.17 & 2.30 & 43.06 & 15.31 & 8.59 & 8.79 & 17.58 \\ 3.25 & 3.44 & 43.06 & 34.44 & 8.79 & 19.34 & 26.37 \\ 6.50 & 4.59 & 129.17 & 68.89 & 17.58 & 26.37 & 77.34 \end{pmatrix} \tag{4.21}$$

An alternative approach to the specification of a covariance matrix is given in Garthwaite and Dickey (1992). It may be instructive to look at the induced correlation matrix, $\widehat{M}$.

$$\widehat{M} = \begin{pmatrix} 1 & 0.71 & 0.71 & 0.71 & 0.74 & 0.74 & 0.74 \\ 0.71 & 1 & 0.5 & 0.5 & 0.78 & 0.78 & 0.52 \\ 0.71 & 0.5 & 1 & 0.5 & 0.78 & 0.52 & 0.78 \\ 0.71 & 0.5 & 0.5 & 1 & 0.52 & 0.78 & 0.78 \\ 0.74 & 0.78 & 0.78 & 0.52 & 1 & 0.68 & 0.68 \\ 0.74 & 0.78 & 0.52 & 0.78 & 0.68 & 1 & 0.68 \\ 0.74 & 0.52 & 0.78 & 0.78 & 0.68 & 0.68 & 1 \end{pmatrix} \tag{4.22}$$

Also consider the inverse of the covariance matrix:

$$M^{-1} = \begin{pmatrix} 4.00 & -1.41 & -0.08 & -0.14 & 0 & 0 & 0 \\ -1.41 & 5.00 & 0.08 & 0.15 & -0.98 & -0.65 & 0 \\ -0.08 & 0.08 & 0.01 & 0.01 & -0.05 & 0 & -0.02 \\ -0.14 & 0.15 & 0.01 & 0.05 & 0 & -0.07 & -0.03 \\ 0 & -0.98 & -0.05 & 0 & 0.64 & 0 & 0 \\ 0 & -0.65 & 0 & -0.07 & 0 & 0.28 & 0 \\ 0 & 0 & -0.02 & -0.03 & 0 & 0 & 0.07 \end{pmatrix} \quad (4.23)$$

Table 4.1 shows the zero and non-zero elements for the lower triangle of $M^{-1}$ ($\square$ represents a non-zero element). It can be seen that the zeros in the inverse covariance matrix correspond to the adjusted orthogonalities represented by the graph. For example, the zero in row $V_{EX}$, column $V_{AA}$, means that $V_{EX}$ and $V_{AA}$ are orthogonal after adjusting by the other variables. Such properties of graphical models and covariance matrices are discussed in Whittaker (1989).

| | $CV$ | $V_{AA}$ | $V_{XX}$ | $V_{EE}$ | $V_{AX}$ | $V_{AE}$ | $V_{EX}$ |
|---|---|---|---|---|---|---|---|
| $CV$ | | | | | | | |
| $V_{AA}$ | $\square$ | | | | | | |
| $V_{XX}$ | $\square$ | $\square$ | | | | | |
| $V_{EE}$ | $\square$ | $\square$ | $\square$ | | | | |
| $V_{AX}$ | 0 | $\square$ | $\square$ | 0 | | | |
| $V_{AE}$ | 0 | $\square$ | 0 | $\square$ | 0 | | |
| $V_{EX}$ | 0 | 0 | $\square$ | $\square$ | 0 | 0 | |

Table 4.1: The conditional independences implied by Figure 4.1

Specifications are also required over the residual components of the variance structure. These specifications are more difficult to make, since we are not used to thinking about such quantities. In this example, for simplicity, belief specifications over the residual structure were chosen to be consistent with those imposed under a multivariate normal specification corresponding to the prior specifications over the elements $R_{ik}$, as discussed in Section 2.3. With respect to the ordering

$\{V_{AA}, V_{XX}, V_{EE}, V_{AX}, V_{AE}, V_{EX}\}$, the residual covariance matrix, $N$, took the form

$$N = \begin{pmatrix} 127 & 496 & 248 & 251 & 178 & 351 \\ 496 & 19997 & 5626 & 3150 & 1670 & 10607 \\ 248 & 5626 & 6332 & 1182 & 1254 & 5969 \\ 251 & 3150 & 1182 & 1046 & 599 & 1949 \\ 178 & 1671 & 1254 & 599 & 573 & 1477 \\ 351 & 10607 & 5969 & 1949 & 1477 & 8439 \end{pmatrix} \tag{4.24}$$

Having made specifications over the quadratic products of residuals, beliefs over all relevant covariance matrices are determined.

From the sample covariance matrix, $S = D_S$, construct the individual variance collection, $D_I$ (6 objects) and the full variance collection, $D_F$ (36 objects), as well as the individual collection for the mean structure, $V_I$ (6 objects). Form the random matrix space, $\mathcal{M}$ over all these objects, and investigate adjustments in this space.

### 4.3.2 Quantitative analysis

The prior covariance matrix was specified directly as follows:

$$\mathrm{E}(V) = \begin{pmatrix} 7.98 & 11.14 & 15.75 \\ 11.14 & 56.26 & 53.04 \\ 15.75 & 53.04 & 100.00 \end{pmatrix} \tag{4.25}$$

This matrix was specified using a graphical model, and a variance component approach, as discussed for the quadratic structure in the last section. The sample covariance matrix (34 cases) is:

$$S = \begin{pmatrix} 8.28 & 20.15 & 24.75 \\ 20.15 & 178.30 & 160.74 \\ 24.75 & 160.74 & 258.26 \end{pmatrix} \tag{4.26}$$

Note that the sample covariance matrix is not too far from the prior specification. The adjusted matrices were formed as the appropriate linear combinations of the observables, as described in Section 3.2, and derived explicitly for the simplest case in Section 4.2.

$$\mathrm{E}_{D_S}(V) = \begin{pmatrix} 8.08 & 14.08 & 18.69 \\ 14.08 & 96.08 & 88.18 \\ 18.69 & 88.18 & 151.65 \end{pmatrix} \tag{4.27}$$

$$\mathrm{E}_{D_I}(V) = \begin{pmatrix} 8.04 & 15.96 & 17.72 \\ 15.96 & 98.90 & 78.63 \\ 17.72 & 78.63 & 159.21 \end{pmatrix} \tag{4.28}$$

$$\mathrm{E}_{D_F}(V) = \begin{pmatrix} 8.30 & 15.43 & 20.06 \\ 15.43 & 92.04 & 80.66 \\ 20.06 & 80.66 & 156.79 \end{pmatrix} \tag{4.29}$$

In fact, these matrices are the observed values of $\mathrm{E}_{C+D_S}(V)$, $\mathrm{E}_{C+D_I}(V)$ and $\mathrm{E}_{C+D_F}(V)$ respectively, but the $C$ is dropped from the notation, and assumed implicitly to be included in the projection space. These adjusted matrices may be used as a basis for assessing our posterior beliefs about the matrix object (see Goldstein 1994 and Goldstein 1996). They represent prior inferences for posterior judgements.

Note that the last matrix (4.29) represents the adjusted matrix which would have been obtained using a standard Bayes linear analysis on the quadratic products of the residuals. In this particular example, all adjusted matrices are positive definite. In general, we view negative eigenvalues in the revised structure as providing diagnostic warnings of possible conflicts between prior beliefs and the data, or as warning of inappropriate model choice or selection of projection space.

It is desirable to be able to compare the estimates of $V$: $\mathrm{E}_{D_F}(V)$, $\mathrm{E}_{D_S}(V)$, and $\mathrm{E}_{D_I}(V)$. Thus, the standard interpretive and diagnostic features of the Bayes linear methodology are used to assess the model and understand the adjustments taking place.

Goldstein (1991) develops a formal framework for the comparison of covariance structures. Essentially, one focusses attention on the eigenstructure of the transformation which maps one covariance structure to the other. The transformation which accomplishes this is known as the *belief transform*, and it's eigenstructure contains all necessary information about the adjustment. Further details are given in Goldstein and Wooff (1994).

Quantitatively, given any two covariance matrices, $R$ and $S$, the belief transform is the linear transformation, $T$ such that

$$TR = S \tag{4.30}$$

Since all of the matrices are strictly positive definite, we can compute $T$ as

$$T = SR^{-1} \tag{4.31}$$

The eigenvalues of this matrix are those quoted in Tables 4.2, 4.3, 4.4 and 4.5. However, the eigenvectors quoted are normalised with respect to the second matrix, in order to make interpretation easier. A discussion of this, and other issues which arise in the case where the matrices are not of full rank is given in Goldstein and Wooff (1995b).

| Variable | Primary eigenvector | Secondary eigenvector |
|---|---|---|
| A | 0.12 | 0.08 |
| E | -0.11 | 0.11 |
| X | -0.01 | -0.12 |
| Eigenvalue ratio | 1.81 | 1.44 |

Table 4.2: Eigenstructure of the belief transform for the mapping from $\mathrm{E}(V)$ to $\mathrm{E}_{D_S}(V)$

Looking first at Table 4.2, we can see that for the first adjustment, variance was inflated by a factor 1.81 in a direction close to the difference between $A$ and $E$, and that variance was inflated by a factor of 1.44 in a direction close to $A + E - X$. The other component had eigenvalue close to one. Table 4.3 shows that when $D_I$ is added to the adjustment, variance is inflated by a factor of 1.51 in a direction close to $A+E$. Other directions had eigenvalues close to one. Table 4.4 shows that adding the whole of $D_F$ to the adjustment actually reduces variance by a factor of 0.81 in a direction close to $A + E - X$ (other eigenvalues close to one). Table 4.5 shows the overall adjustment transformation. The overall transform is the composition of the three partial transforms (Goldstein 1991). Variance has been inflated by a factor of 1.81 in a direction close to $2A + E$, and by a factor of 1.62 in a direction close to $2A - (E + X)$. Examination of the belief transforms in this way allows interpretive analysis of the changes in belief.

| Variable | Primary eigenvector |
|---|---|
| A | -0.11 |
| E | -0.10 |
| X | 0.0 |
| Eigenvalue ratio | 1.51 |

Table 4.3: Eigenstructure of the belief transform for the mapping from $\mathrm{E}_{D_S}(V)$ to $\mathrm{E}_{D_I}(V)$

| Variable | Smallest eigenvector |
|---|---|
| A | -0.16 |
| E | -0.10 |
| X | 0.10 |
| Eigenvalue ratio | 0.81 |

Table 4.4: Eigenstructure of the belief transform for the mapping from $\mathrm{E}_{D_I}(V)$ to $\mathrm{E}_{D_F}(V)$

| Variable | Primary eigenvector | Secondary eigenvector |
|---|---|---|
| A | 0.25 | 0.10 |
| E | 0.13 | -0.05 |
| X | -0.09 | -0.06 |
| Eigenvalue ratio | 1.81 | 1.62 |

Table 4.5: Eigenstructure of the belief transform for the mapping from $\mathrm{E}(V)$ to $\mathrm{E}_{D_F}(V)$

## 4.4 Bayes linear influence diagrams for matrix objects

Figure 4.2 shows a Bayes linear influence diagram representing the adjustments and corresponding diagnostic information for the random matrices. Such diagrams are described in detail in Goldstein, Farrow, and Spiropoulos (1993) for random quantities, with a similar interpretation for random matrices, where adjusted orthogonality is determined instead by the inner-product (3.10), so that our conditional independence relation, (4.17), becomes

$$B \amalg C/D \iff \mathrm{P}[\mathrm{Tr}\{(B - \mathrm{E}_D(B))(C - \mathrm{E}_D(C))\}] = 0 \qquad (4.32)$$

This is a generalised conditional independence property, as defined in Smith (1990), and consequently, all of the usual properties of conditional independence graphs based upon such a relation will hold. Each node represents a space of covariance matrices. The outer shadings of the $V$ node represent proportions of uncertainty about $V$ resolved by projection into the various spaces. Shadings start at 3 o'clock, and progress in an anti-clockwise fashion. The full circle represents the total uncertainty about the values of the space of covariance matrices. The first outer portion shaded represents the proportion of our uncertainty resolved by the sample covariance matrix alone ($D_S$). By comparing this with the first shaded portion for the $V_I$ node, it can be seen that considerably more has been learned about the matrix object, than about the 6-dimensional space over the individual variance collection.

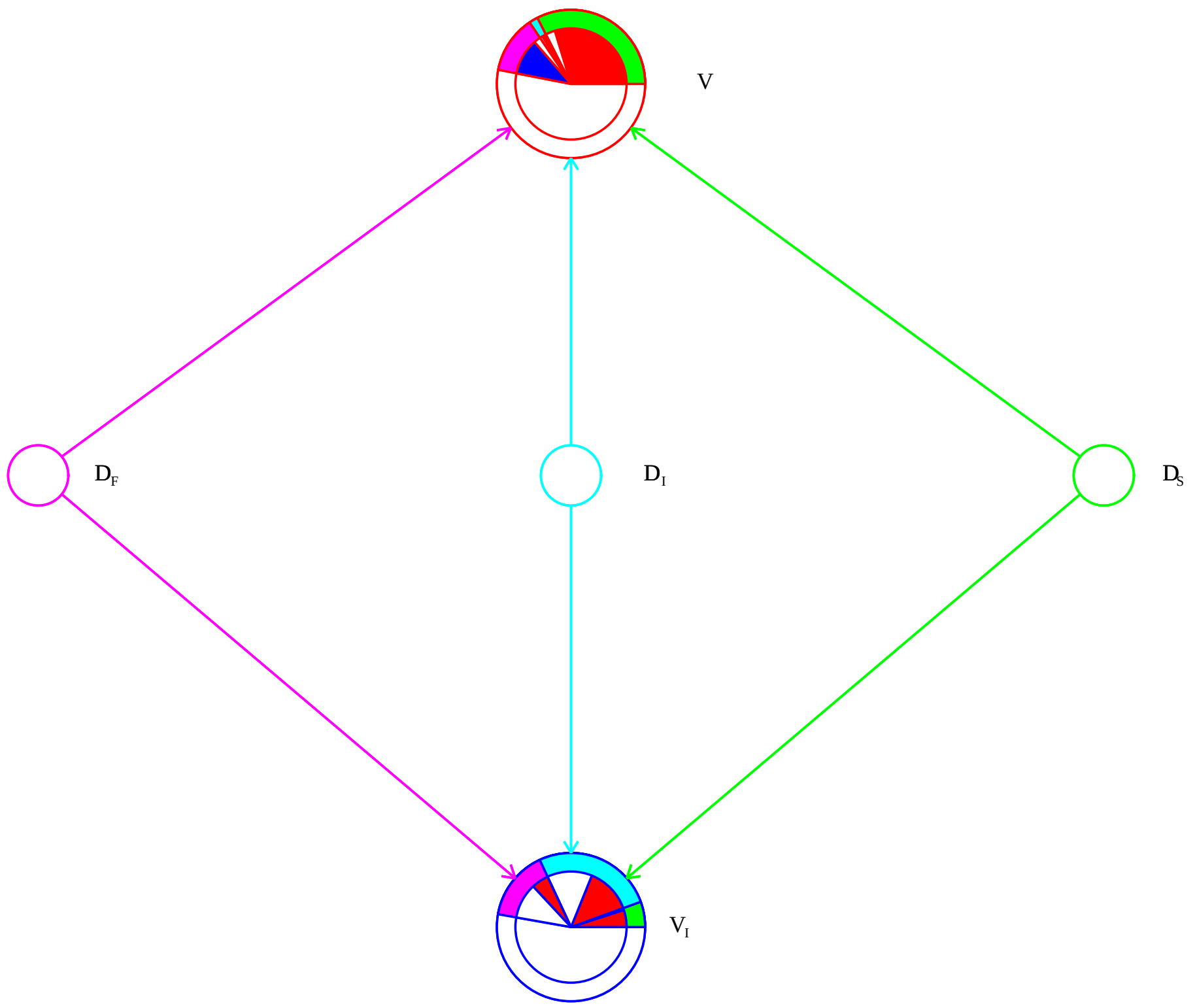

Figure 4.2: Diagnostic influence diagram summarising changes in expectation of the matrix objects

The next shading gives the *additional* information gained by using the individual collection as the projection space. This tells us a great deal more about the elements of the $V_I$ collection, but little about the matrix object as a whole. The other shading shows the additional uncertainty resolved due to including the full variance collection in our projection space. There is information to be gained by enriching our projection space, but one must balance information gained with extra effort involved. Whether or not one chooses to include the complete variance collection will depend upon the size of the problem under consideration, and upon how much the answer really matters.

It is no coincidence that the total amount of uncertainty resolved for the $V$ and $V_I$ nodes is the same in the case of fully decomposed structures. This is essentially because the $V$ node represents the *heart* of the *belief transform* for the adjustment of the $V_I$ node in this case (Goldstein 1990). This is mentioned only in passing, since it is of little relevance to the rest of the thesis. See Goldstein (1990) for a discussion of these kinds of properties of exchangeable adjustments.

Shadings in the centres of the nodes are diagnostics based on the *size* and *bearing* of the adjustments, as described in Goldstein (1988b). Diagnostics for matrix adjustments are discussed more fully in Chapter 6. For now it suffices to mention that the more red shading present, the larger the diagnostic warning. The diagram shows a lot of shading for the adjustment of the $V$ matrix, indicating a contradiction between the data and our prior beliefs.

## 4.5 Summary

Analysing matrices in a space where they live naturally not only has great aesthetic appeal, but is very powerful and illuminating in practice. Working in this space simplifies the handling of large matrices, by reducing the number of quantities involved and summarising effects over the whole covariance structure. For the same reasons, diagnostic information about adjusted beliefs is easier to interpret. Structures may be decomposed as much or as little as desired.

This approach allows learning for collections of covariance structures, and examination of their relationships. It generalises the "element by element" approach to revision, which can be viewed as taking place in a subspace of the larger space. Exchangeability representations lie at the heart of the methodology: all specifications are over observables, or quantities constructed from observables, rather than artificial model parameters, and no distributional assumptions for the data or the prior need be made.

# Chapter 5

# Covariance matrix adjustment for dynamic linear models

## 5.1 Introduction

The approach to covariance estimation is now applied to the development of a methodology for the revision of the underlying covariance structures for a dynamic linear model, free from any distributional restrictions, using Bayes linear estimators for the covariance matrices based on simple quadratic observables. This is done by constructing an inner-product space of random matrices containing both the underlying covariance matrices and observables predictive for them. Bayes linear estimates for the underlying matrices follow by orthogonal projection.

The method is illustrated using data derived from the weekly sales of six leading brands of shampoo from a medium sized cash-and-carry depot. The sales are modelled taking into account the underlying demand and competition effects, and the covariance structure over the resulting dynamic linear model is adjusted using the weekly sales data.

Covariance matrix adjustment for dynamic linear models is reviewed in West and Harrison (1989). For multivariate time series, the observational covariance matrix can be updated for a class of models known as *matrix normal models* using a simple conjugate prior approach. However, the distributional assumptions required are extremely restrictive, and it is difficult to learn about the covariance matrix for the updating of the state vector.

## 5.2 The dynamic linear model

### 5.2.1 The general model

Let $\boldsymbol{X}_1, \boldsymbol{X}_2, \ldots$ be an infinite sequence of random vectors, each of length $r$, such that $\boldsymbol{X}_t = (X_{1t}, X_{2t}, \ldots, X_{rt})^{\mathrm{T}}$. These vectors represent the observations at each time point. Suppose that we model the relationships between these vectors in the following way.

$$\boldsymbol{X}_t = F^{\mathrm{T}}\boldsymbol{\Theta}_t + \boldsymbol{\nu}_t \qquad (5.1)$$

$$\boldsymbol{\Theta}_t = G\boldsymbol{\Theta}_{t-1} + \boldsymbol{\omega}_t \qquad (5.2)$$

The prior second-order specification is as follows:

$$\mathrm{E}(\boldsymbol{\nu}_t) = \mathrm{E}(\boldsymbol{\omega}_t) = 0, \mathrm{Var}(\boldsymbol{\Theta}_0) = \Sigma, \mathrm{Var}(\boldsymbol{\nu}_t) = V, \mathrm{Var}(\boldsymbol{\omega}_t) = W, \quad \forall t \qquad (5.3)$$

$$\mathrm{Cov}(\boldsymbol{\Theta}_s, \boldsymbol{\nu}_t) = \mathrm{Cov}(\boldsymbol{\nu}_s, \boldsymbol{\omega}_t) = 0 \quad \forall s, t, \quad \mathrm{Cov}(\boldsymbol{\Theta}_s, \boldsymbol{\omega}_t) = 0 \quad \forall s < t \qquad (5.4)$$

$$\mathrm{Cov}(\boldsymbol{\omega}_s, \boldsymbol{\omega}_t) = \mathrm{Cov}(\boldsymbol{\nu}_s, \boldsymbol{\nu}_t) = 0 \quad \forall s \neq t \qquad (5.5)$$

The *state vector*, $\boldsymbol{\Theta}_t$ is $p$ dimensional, and the $p \times r$ and $p \times p$ dimensional matrices, $F$ and $G$ are assumed to be known. This is a second-order description of the (constant) multivariate time series dynamic linear model (DLM) described in West and Harrison (1989). No distributional assumptions are made for any of the components in the model. Ways to learn about $V$ and $W$ from data are now described. West and Harrison (1989, Chapter 15) give a conjugate prior solution to the problem of learning about $V$ for a class of these models known as *matrix normal models*, if one is prepared to make the necessary distributional assumptions. However methods for learning about the matrix $W$ tend primarily to be *ad hoc*. The standard method for updating $W$ is the *discount factor* approach, outlined in West and Harrison (1989), which simply inflates the $W$ matrix at each iteration, so that the prior is swamped by the data at a given rate. Such an approach fails to utilise the fact that there is information about the $W$ matrix present in the observations.

### 5.2.2 Example

As an illustration of the approach, consider a simple locally constant model for the sales of 6 leading brands of shampoo from a medium sized cash-and-carry depot. As above $\boldsymbol{X}_1, \boldsymbol{X}_2, \ldots$ is a sequence of random vectors, each of length 6, such that $\boldsymbol{X}_t = (X_{1t}, X_{2t}, \ldots, X_{6t})^{\mathrm{T}}$. The component $X_{it}$ represents the (unknown) sales of brand $i$ at time $t$, simply measured as a number of bottles. The vectors of sales are modelled as follows

$$\boldsymbol{X}_t = \boldsymbol{\Theta}_t + \boldsymbol{\nu}_t \quad \forall t \tag{5.6}$$

where

$$\boldsymbol{\Theta}_t = \boldsymbol{\Theta}_{t-1} + \boldsymbol{\omega}_t \quad \forall t \tag{5.7}$$

Prior beliefs are are given by (5.3), (5.4) and (5.5). Here it has been assumed that the process is locally constant, but with different underlying demands for each of the components of the series. This is a simple model, with no seasonal component, chosen to illustrate our methodology, and would be unrealistic if there were noticeable trends within any of the components of the series. However, for high dimensional time series with no obvious trends, it is often the case that, provided the covariance structure is appropriate, many of the interesting features of the series can be captured using just such a model. To this end covariances are introduced between components of the state vector and also for the way demand changes over time, and for the way observations vary from the underlying demand. A more detailed treatment of multivariate sales forecasting within a fully specified Bayesian framework is given by Queen, Smith, and James (1994) and Queen (1994) who consider the problem of developing a dynamic model for multivariate sales, and the development of a prior distribution with sufficient flexibility to capture the effects of market interaction. These methods are based upon the dynamic graphical model ideas discussed in Queen and Smith (1992) and Queen and Smith (1993).

The second-order DLM requires the following quantifications. Firstly, the $F$ and $G$ matrices must be specified. Then, *a priori* specifications are needed for the expectation of the initial state vector, $\boldsymbol{\mu}_0 = \mathrm{E}(\boldsymbol{\Theta}_0)$. Finally, specify the matrices $\Sigma = \mathrm{Var}(\boldsymbol{\Theta}_0),\ V = \mathrm{Var}(\boldsymbol{\nu}_t),\ W = \mathrm{Var}(\boldsymbol{\omega}_t) \forall t$.

In the example, the specification for the mean vector was

$$\mathrm{E}(\boldsymbol{\Theta}_0) = (10, 9, 9, 8, 8, 7)^{\mathrm{T}} \tag{5.8}$$

The following specifications were made for the covariance matrices, using exchangeability judgements concerning the way in which the observations vary from their means.

$$\Sigma = \begin{pmatrix} 9 & 3 & 3 & 3 & 3 & 3 \\ 3 & 9 & 3 & 3 & 3 & 3 \\ 3 & 3 & 9 & 3 & 3 & 3 \\ 3 & 3 & 3 & 9 & 3 & 3 \\ 3 & 3 & 3 & 3 & 9 & 3 \\ 3 & 3 & 3 & 3 & 3 & 9 \end{pmatrix} \tag{5.9}$$

$$W = \begin{pmatrix} 4 & 1 & 1 & 1 & 1 & 1 \\ 1 & 4 & 1 & 1 & 1 & 1 \\ 1 & 1 & 4 & 1 & 1 & 1 \\ 1 & 1 & 1 & 4 & 1 & 1 \\ 1 & 1 & 1 & 1 & 4 & 1 \\ 1 & 1 & 1 & 1 & 1 & 4 \end{pmatrix} \quad (5.10)$$

$$V = \begin{pmatrix} 36 & -4 & -4 & -4 & -4 & -4 \\ -4 & 36 & -4 & -4 & -4 & -4 \\ -4 & -4 & 36 & -4 & -4 & -4 \\ -4 & -4 & -4 & 36 & -4 & -4 \\ -4 & -4 & -4 & -4 & 36 & -4 \\ -4 & -4 & -4 & -4 & -4 & 36 \end{pmatrix} \quad (5.11)$$

For example, for the matrix, $\Sigma$, it was decided to be appropriate to associate a standard deviation of 3 with each of the variables. A correlation between variables of 1/2 was also felt appropriate. In truth, there is perhaps more symmetry in these specifications than is really appropriate, but specification is hard, and viewing variation in the sales of the various shampoos as second-order exchangeable greatly reduces the number of specifications which have to be made over the second order structure, and will allow further exchangeability modelling to simplify the fourth order specifications in later sections.

Notice however that many aspects of the underlying mechanisms have been captured by these specifications. In this model, $\mathbf{\Theta}_t$ represents the vector of demands at time $t$. From the positive correlations in $\mathrm{Var}(\mathbf{\Theta}_0)$, if the mean of one product turned out to be higher than anticipated, we would revise upwards beliefs about the means of the other products. Also, the positive correlations within $\mathrm{Var}(\boldsymbol{\omega}_t)$ indicate that there is a common component to the demands, whilst the negative correlations within $\mathrm{Var}(\boldsymbol{\nu}_t)$ indicate that brands are competing, and tend to succeed at the expense of one another.

### 5.2.3 Bayes linear analysis

With the second-order specification that has been made, sales data may be used to carry out a Bayes linear analysis which will be informative for the mean of future observations. However, learning about the covariance matrices $W = \text{Var}(\boldsymbol{\omega}_t)$ and $V = \text{Var}(\boldsymbol{\nu}_t)$ will not occur. The methods of the previous chapters will now be adapted, in order to enable such learning.

## 5.3 Quadratic products

### 5.3.1 Exchangeable decomposition of unobservable products

For the general DLM outlined in Section 5.2.1, form the quadratic products of $\boldsymbol{\omega}_t$ and $\boldsymbol{\nu}_t$, namely $\mathbf{vec}(\boldsymbol{\omega}_t{\boldsymbol{\omega}_t}^{\mathrm{T}})$ and $\mathbf{vec}(\boldsymbol{\nu}_t{\boldsymbol{\nu}_t}^{\mathrm{T}})$. We view $\mathbf{vec}(\boldsymbol{\omega}_t{\boldsymbol{\omega}_t}^{\mathrm{T}})$ and $\mathbf{vec}(\boldsymbol{\nu}_t{\boldsymbol{\nu}_t}^{\mathrm{T}})$ to be second-order exchangeable over $t$. Explicitly, second-order beliefs over the vectors of quadratic products of residuals will remain invariant under the action of an arbitrary permutation of the $t$ index (this is what I mean when describing a DLM as *constant*). As described in Section 2.2, using the second-order exchangeability representation theorem (Goldstein 1986a), an element of a second-order exchangeable collection of vectors may be represented as the sum of a mean vector, common to all elements, and a residual vector, uncorrelated with the mean vector and all other residual vectors. This representation may be applied to $\mathbf{vec}(\boldsymbol{\omega}_t{\boldsymbol{\omega}_t}^{\mathrm{T}})$, and then re-written in matrix form as

$$\boldsymbol{\omega}_t{\boldsymbol{\omega}_t}^{\mathrm{T}} = V^{\omega} + S_t^{\omega} \quad \forall t \geq 1 \tag{5.12}$$

as for (2.4), where $V^{\omega}$ and $S_t^{\omega}$ are random matrices of the same dimension as $\boldsymbol{\omega}_t\boldsymbol{\omega}_t^{\mathrm{T}}$,

$$\mathrm{P}(\mathbf{vec}(S_t^{\omega})) = \mathrm{Cov}(\mathbf{vec}(V^{\omega}), \mathbf{vec}(S_t^{\omega})) = 0,\ \forall t \tag{5.13}$$

and

$$\mathrm{Cov}(\mathbf{vec}(S_t^{\omega}), \mathbf{vec}(S_s^{\omega})) = 0,\ \forall s \neq t \tag{5.14}$$

$$\mathrm{Var}(\mathbf{vec}S_s^{\omega}) = \mathrm{Var}(\mathbf{vec}S_t^{\omega}),\ \forall s, t \tag{5.15}$$

Decomposing $\mathbf{vec}(\boldsymbol{\nu}_t\boldsymbol{\nu}_t^{\mathrm{T}})$ similarly gives

$$\boldsymbol{\nu}_t\boldsymbol{\nu}_t^{\mathrm{T}} = V^{\nu} + S_t^{\nu} \quad \forall t \geq 1 \tag{5.16}$$

with properties as for representation (5.12). Note that $\mathrm{P}(V^{\omega}) = \mathrm{P}(\boldsymbol{\omega}_t\boldsymbol{\omega}_t^{\mathrm{T}}) = \mathrm{Var}(\boldsymbol{\omega}_t)$ $= W$ and so learning about $V^{\omega}$ will allow learning about the covariance matrix for the residuals for the state, and $\mathrm{P}(V^{\nu}) = \mathrm{P}(\boldsymbol{\nu}_t\boldsymbol{\nu}_t^{\mathrm{T}}) = \mathrm{Var}(\boldsymbol{\nu}_t) = V$, and so learning about $V^{\nu}$ will allow learning about the covariance matrix for the observational residuals. Representations (5.12) and (5.16) decompose uncertainty for $\boldsymbol{\omega}_t\boldsymbol{\omega}_t^{\mathrm{T}}$ and $\boldsymbol{\nu}_t\boldsymbol{\nu}_t^{\mathrm{T}}$ into two parts. Bayes linear updating (with enough data) will eliminate the aspects of uncertainty derived from uncertainty about $V^{\omega}$ and $V^{\nu}$.

In order to conduct a Bayes linear analysis on the quadratic structure additional covariance specifications $\mathrm{Var}(\mathbf{vec}V^{\omega})$, $\mathrm{Var}(\mathbf{vec}V^{\nu})$, $\mathrm{Var}(\mathbf{vec}S_t^{\omega})$ and $\mathrm{Var}(\mathbf{vec}S_t^{\nu})$, for some $t$ are needed.

### 5.3.2 Example

In the example, the $\boldsymbol{X}_t$ vector is 6-dimensional, and so the matrices, $V^{\omega}$, $V^{\nu}$, $S^{\omega}$ and $S^{\nu}$ are $6 \times 6$-dimensional. Consequently, the matrices $\mathrm{Var}(\mathbf{vec}V^{\omega})$, $\mathrm{Var}(\mathbf{vec}V^{\nu})$, $\mathrm{Var}(\mathbf{vec}S_t^{\omega})$ and $\mathrm{Var}(\mathbf{vec}S_t^{\nu})$ are $36 \times 36$-dimensional. When referring to the compo-nents of Var($\mathbf{vec}V^\omega$), the notation $v^\omega_{ijkl}$ will be used to denote the covariance between the $(i,j)^{th}$ and $(k,l)^{th}$ elements of $V^\omega$. Similar notation is used for Var($\mathbf{vec}V^\nu$). Also $s^\omega_{ijkl}$ and $s^\nu_{ijkl}$ are used for the components of Var($\mathbf{vec}S^\omega_t$) and Var($\mathbf{vec}S^\nu_t$) respectively. The following covariance specifications were made for our example:

$$v^\omega_{iiii} = 9/4,\ \forall i, \quad v^\omega_{ijij} = 9/16,\ \forall i \neq j, \quad v^\omega_{iijj} = 1/5, \forall i \neq j, \tag{5.17}$$

$$v^\nu_{iiii} = 25,\ \forall i, \quad v^\nu_{ijij} = 1,\ \forall i \neq j, \quad v^\nu_{iijj} = 4,\ \forall i \neq j, \tag{5.18}$$

$$s^\omega_{iiii} = 30,\ \forall i, \quad s^\omega_{ijij} = 15,\ \forall i \neq j, \quad s^\nu_{iiii} = 2500,\ \forall i, \quad s^\nu_{ijij} = 1000,\ \forall i \neq j. \tag{5.19}$$

For instance, $v^\omega_{iiii}$ is the variance specification for the $(i,i)^{th}$ element of $V^\omega$, which represents the underlying variance of the $i^{th}$ element of $\boldsymbol{\omega}_t$. From (5.10), it has expectation 4. From (5.7), this value governs the rate of change of $\mathbf{\Theta}_t$. By considering the range of plausible variances for the way $\mathbf{\Theta}_t$ might change over time, it was felt reasonable that a standard deviation specification of 3/2 should be made. The other specifications were made in a similar fashion. For simplicity in this example, the specifications for $s^\omega_{ijkl}$ and $s^\nu_{ijkl}$ were made using the fourth moments of the multivariate normal distribution compatible with the given second-order structure as a guide.

Whilst considerably more specifications would be required for a full Bayes or Bayes linear analysis, (5.17), (5.18), and (5.19) are sufficient for our purposes as these are the only specifications needed for the matrix object approach to belief revision which is to be taken in the later sections.

There is a lot of symmetry in these values, greatly simplifying the specification, but once again, any non-negative covariance structure over the quadratic products is acceptable. Here assumptions of exchangeability over the variances and the covariances have been used. Many of the specifications made for the quadratic structure will be "averaged over" in the matrix object approach to covariance adjustment which shall be developed, and so there is a limit to the effort that one would wish to put into very detailed specifications at this stage, since the suggested analysis will not be overly sensitive to the individual specifications.

### 5.3.3 Observable quadratic terms

First, certain linear combinations of the observables which do not involve the state vector, $\mathbf{\Theta}_t$, will be constructed. This is useful for various reasons, and in particular because it greatly reduces the prior specification required for the analysis of the quadratic structure. In this thesis, we shall mainly be concerned with DLMs for which there exists an $r \times r$ matrix $H$, such that $HF^{\mathrm{T}} = F^{\mathrm{T}}G$. We call such DLMs *two-step invertible*. Note that a DLM will be two-step invertible if $F$ is of full rank and $r \geq p$ (as will often be the case for high-dimensional time series), and there will often be many matrices $H$ satisfying $HF^{\mathrm{T}} = F^{\mathrm{T}}G$. For example, $H = F^{\mathrm{T}}G^{\mathrm{T}}F^{\mathrm{T}^\dagger}$ (where $F^{\mathrm{T}^\dagger}$ represents any generalised inverse of $F^{\mathrm{T}}$) is a solution. Further, if $F$ is of full rank, $r < p$ and such a matrix exists, then $H = F^{\mathrm{T}}GF(F^{\mathrm{T}}F)^{-1}$, and so $H$ exists, if and only if $F^{\mathrm{T}}GF(F^{\mathrm{T}}F)^{-1}F^{\mathrm{T}} = F^{\mathrm{T}}G$. Note also that the matrix $H^2$ has the property that $H^2F^{\mathrm{T}} = F^{\mathrm{T}}G^2$. For a two-step invertible DLM, the following vectors of observables which do not involve the state vector may be constructed:

$$\boldsymbol{X}'_t = \boldsymbol{X}_t - H\boldsymbol{X}_{t-1} = F^{\mathrm{T}}\boldsymbol{\omega}_t + \boldsymbol{\nu}_t - H\boldsymbol{\nu}_{t-1} \quad \forall t \geq 2 \tag{5.20}$$

$$\boldsymbol{X}''_t = \boldsymbol{X}_t - H^2\boldsymbol{X}_{t-2} = F^{\mathrm{T}}\boldsymbol{\omega}_t + F^{\mathrm{T}}G\boldsymbol{\omega}_{t-1} + \boldsymbol{\nu}_t - H^2\boldsymbol{\nu}_{t-2} \quad \forall t \geq 3 \tag{5.21}$$

Form the matrices of quadratic products, $\boldsymbol{X}'_t\boldsymbol{X}'^{\mathrm{T}}_t$ and $\boldsymbol{X}''_t\boldsymbol{X}''^{\mathrm{T}}_t \quad \forall t$. These are predictive for $V^\omega$ and $V^\nu$.

Not all DLMs are two-step invertible, but for the constant dynamic linear model outlined in Section 5.2.1, it is always possible to construct linear combinations of the observations which do not involve the state, provided only that the constant dynamic linear model in question is *observable* (for a discussion of the very weak restriction of observability, see West and Harrison (1989, Chapter 5)). However, in general such linear combinations require more than two successive observations from the series. Consequently, for simplicity attention is restricted to the two-step invertible model. However, the approach is quite general, and may be applied similarly for any constant observable DLM, the only difference being that the covariance specification is more complicated, more quantities are involved in the adjustment, and the identifiability results are more complex, and in general, slightly weaker.

### 5.3.4 Example

For the example, $F$, $G$ and $H$ are all the identity, and so the one and two-step differences of the observables are formed:

$$\boldsymbol{X}_t^{(1)} = \boldsymbol{X}_t - \boldsymbol{X}_{t-1} = \boldsymbol{\omega}_t + \boldsymbol{\nu}_t - \boldsymbol{\nu}_{t-1} \quad \forall t \geq 2 \tag{5.22}$$

$$\boldsymbol{X}_t^{(2)} = \boldsymbol{X}_t - \boldsymbol{X}_{t-2} = \boldsymbol{\omega}_t + \boldsymbol{\omega}_{t-1} + \boldsymbol{\nu}_t - \boldsymbol{\nu}_{t-2} \quad \forall t \geq 3 \tag{5.23}$$

The quadratic products of these, $\boldsymbol{X}_t^{(1)}\boldsymbol{X}_t^{(1)\mathrm{T}}$ and $\boldsymbol{X}_t^{(2)}\boldsymbol{X}_t^{(2)\mathrm{T}}$ are then formed. These observables are predictive for $V^\omega$ and $V^\nu$ and so may be used to learn about the underlying covariance structure. All means and covariances that are required for the subsequent analysis are determined by specifications in (5.8), (5.9), (5.10), (5.11), (5.17), (5.18) and (5.19). The precise form of the covariance structure over the observables is rather complex, and given below. Recall that the operators $\otimes$ and $\star$ were defined in Definitions 1 and 2 respectively (page 28). The covariance structure over the quadratic products of the 1-step differences is determined by the following relations:

$$\mathrm{Cov}(\mathbf{vec}V^\omega, \mathbf{vec}(\boldsymbol{X}_t^{(1)}\boldsymbol{X}_t^{(1)\mathrm{T}})) = \mathrm{Var}(\mathbf{vec}V^\omega) \tag{5.24}$$

$$\mathrm{Cov}(\mathbf{vec}V^{\nu}, \mathbf{vec}(\boldsymbol{X}_t^{(1)}\boldsymbol{X}_t^{(1)\mathrm{T}})) = 2\mathrm{Var}(\mathbf{vec}V^{\nu}) \tag{5.25}$$

$$\begin{aligned}\mathrm{Cov}(\mathbf{vec}(\boldsymbol{X}_t^{(1)}\boldsymbol{X}_t^{(1)\mathrm{T}}), \mathbf{vec}(\boldsymbol{X}_t^{(1)}\boldsymbol{X}_t^{(1)\mathrm{T}})) = & \quad \mathrm{Var}(\mathbf{vec}V^{\omega}) + 4\mathrm{Var}(\mathbf{vec}V^{\nu}) \\ & + \mathrm{Var}(\mathbf{vec}S_t^{\nu}) + \mathrm{Var}(\mathbf{vec}S_{t-1}^{\nu}) \\ & + \mathrm{Var}(\mathbf{vec}S_t^{\omega}) \\ & + 2[\mathrm{E}(V^{\nu} \otimes V^{\nu}) + \mathrm{E}(V^{\nu} \star V^{\nu})] \\ & + 4[\mathrm{E}(V^{\nu}) \otimes \mathrm{E}(V^{\omega}) + \mathrm{E}(V^{\nu}) \star \mathrm{E}(V^{\omega}) \\ & + \mathrm{E}(V^{\omega}) \otimes \mathrm{E}(V^{\nu}) + \mathrm{E}(V^{\omega}) \star \mathrm{E}(V^{\nu})]\end{aligned} \tag{5.26}$$

$$\mathrm{Cov}(\mathbf{vec}(\boldsymbol{X}_t^{(1)}\boldsymbol{X}_t^{(1)\mathrm{T}}), \mathbf{vec}(\boldsymbol{X}_{t-1}^{(1)}\boldsymbol{X}_{t-1}^{(1)\ \mathrm{T}})) = 4(\mathrm{Var}(\mathbf{vec}V^{\nu}) + \mathrm{Var}(\mathbf{vec}V^{\omega})) + \mathrm{Var}(\mathbf{vec}S_{t-1}^{\nu}) \tag{5.27}$$

$$\mathrm{Cov}(\mathbf{vec}(\boldsymbol{X}_t^{(1)}\boldsymbol{X}_t^{(1)\mathrm{T}}), \mathbf{vec}(\boldsymbol{X}_{t-s}^{(1)}\boldsymbol{X}_{t-s}^{(1)\ \mathrm{T}})) = 4\mathrm{Var}(\mathbf{vec}V^{\nu}) + \mathrm{Var}(\mathbf{vec}V^{\omega}) \quad \forall t, \forall s \geq 2 \tag{5.28}$$

The covariance structure over the quadratic products of the 2-step differences are given below.

$$\mathrm{Cov}(\mathbf{vec}V^{\omega}, \mathbf{vec}(\boldsymbol{X}_t^{(2)}\boldsymbol{X}_t^{(2)\mathrm{T}})) = 2\mathrm{Var}(\mathbf{vec}V^{\omega}) \tag{5.29}$$

$$\mathrm{Cov}(\mathbf{vec}V^{\nu}, \mathbf{vec}(\boldsymbol{X}_t^{(2)}\boldsymbol{X}_t^{(2)\mathrm{T}})) = 2\mathrm{Var}(\mathbf{vec}V^{\nu}) \tag{5.30}$$

$$\begin{aligned}\mathrm{Cov}(\mathbf{vec}(\boldsymbol{X}_t^{(2)}\boldsymbol{X}_t^{(2)\mathrm{T}}), \mathbf{vec}(\boldsymbol{X}_t^{(2)}\boldsymbol{X}_t^{(2)\mathrm{T}})) = & \quad 4\mathrm{Var}(\mathbf{vec}V^{\omega}) + 4\mathrm{Var}(\mathbf{vec}V^{\nu}) \\ & + \mathrm{Var}(\mathbf{vec}S_t^{\nu}) + \mathrm{Var}(\mathbf{vec}S_{t-2}^{\nu}) \\ & + \mathrm{Var}(\mathbf{vec}S_t^{\omega}) + \mathrm{Var}(\mathbf{vec}S_{t-1}^{\omega}) \\ & + 2[\mathrm{E}(V^{\nu} \otimes V^{\nu}) + \mathrm{E}(V^{\nu} \star V^{\nu}) \\ & + \mathrm{E}(V^{\omega} \otimes V^{\omega}) + \mathrm{E}(V^{\omega} \star V^{\omega})] \\ & + 4[\mathrm{E}(V^{\nu}) \otimes \mathrm{E}(V^{\omega}) + \mathrm{E}(V^{\nu}) \star \mathrm{E}(V^{\omega}) \\ & + \mathrm{E}(V^{\omega}) \otimes \mathrm{E}(V^{\nu}) + \mathrm{E}(V^{\omega}) \star \mathrm{E}(V^{\nu})]\end{aligned} \tag{5.31}$$

$$\mathrm{Cov}(\mathbf{vec}(\boldsymbol{X}_t^{(2)}\boldsymbol{X}_t^{(2)\mathrm{T}}), \mathbf{vec}(\boldsymbol{X}_{t-1}^{(2)}\boldsymbol{X}_{t-1}^{(2)\ \mathrm{T}})) = 4[\mathrm{Var}(\mathbf{vec}V^{\nu}) + \mathrm{Var}(\mathbf{vec}V^{\omega})] + \mathrm{Var}(\mathbf{vec}S_{t-1}^{\omega}) \tag{5.32}$$

$$\mathrm{Cov}(\mathbf{vec}(\boldsymbol{X}_t^{(2)}\boldsymbol{X}_t^{(2)\mathrm{T}}), \mathbf{vec}(\boldsymbol{X}_{t-2}^{(2)}\boldsymbol{X}_{t-2}^{(2)\ \mathrm{T}})) = 4[\mathrm{Var}(\mathbf{vec}V^{\nu}) + \mathrm{Var}(\mathbf{vec}V^{\omega})] + \mathrm{Var}(\mathbf{vec}S_{t-2}^{\nu}) \tag{5.33}$$

$$\mathrm{Cov}(\mathbf{vec}(\boldsymbol{X}_t^{(2)}\boldsymbol{X}_t^{(2)\mathrm{T}}), \mathbf{vec}(\boldsymbol{X}_{t-s}^{(2)}\boldsymbol{X}_{t-s}^{(2)\ \mathrm{T}})) = 4\mathrm{Var}(\mathbf{vec}V^{\nu}) + \mathrm{Var}(\mathbf{vec}V^{\omega}) \quad \forall t, \forall s \geq 3 \tag{5.34}$$

The covariances between the one and two step differences are determined as follows:

$$\mathrm{Cov}(\mathbf{vec}(\boldsymbol{X}_t^{(1)}\boldsymbol{X}_t^{(1)\mathrm{T}}), \mathbf{vec}(\boldsymbol{X}_{t+s}^{(2)}\boldsymbol{X}_{t+s}^{(2)\,\mathrm{T}})) = 4\mathrm{Var}(\mathbf{vec}V^{\nu}) + 2\mathrm{Var}(\mathbf{vec}V^{\omega}) \quad \forall t, \forall s \geq 3 \tag{5.35}$$

$$\mathrm{Cov}(\mathbf{vec}(\boldsymbol{X}_t^{(1)}\boldsymbol{X}_t^{(1)\mathrm{T}}), \mathbf{vec}(\boldsymbol{X}_{t+2}^{(2)}\boldsymbol{X}_{t+2}^{(2)\,\mathrm{T}})) = 4\mathrm{Var}(\mathbf{vec}V^{\nu}) + 2\mathrm{Var}(\mathbf{vec}V^{\omega}) + \mathrm{Var}(\mathbf{vec}S_{t-2}) \tag{5.36}$$

$$\begin{aligned}\mathrm{Cov}(\mathbf{vec}(\boldsymbol{X}_t^{(1)}\boldsymbol{X}_t^{(1)\mathrm{T}}), \mathbf{vec}(\boldsymbol{X}_{t+1}^{(2)}\boldsymbol{X}_{t+1}^{(2)\,\mathrm{T}})) &= \quad 2\mathrm{Var}(\mathbf{vec}V^{\omega}) + 4\mathrm{Var}(\mathbf{vec}V^{\nu}) \\ &+ \quad \mathrm{Var}(\mathbf{vec}S_{t-2}^{\nu}) + \mathrm{Var}(\mathbf{vec}S_{t-1}^{\omega}) \\ &+ \quad \mathrm{E}(V^{\nu}) \otimes \mathrm{E}(V^{\omega}) + \mathrm{E}(V^{\nu}) \star \mathrm{E}(V^{\omega}) \\ &+ \quad \mathrm{E}(V^{\omega}) \otimes \mathrm{E}(V^{\nu}) + \mathrm{E}(V^{\omega}) \star \mathrm{E}(V^{\nu})\end{aligned} \tag{5.37}$$

$$\begin{aligned}\mathrm{Cov}(\mathbf{vec}(\boldsymbol{X}_t^{(1)}\boldsymbol{X}_t^{(1)\mathrm{T}}), \mathbf{vec}(\boldsymbol{X}_t^{(2)}\boldsymbol{X}_t^{(2)\mathrm{T}})) &= \quad 2\mathrm{Var}(\mathbf{vec}V^{\omega}) + 4\mathrm{Var}(\mathbf{vec}V^{\nu}) \\ &+ \quad \mathrm{Var}(\mathbf{vec}S_t^{\nu}) + \mathrm{Var}(\mathbf{vec}S_t^{\omega}) \\ &+ \quad \mathrm{E}(V^{\nu}) \otimes \mathrm{E}(V^{\omega}) + \mathrm{E}(V^{\nu}) \star \mathrm{E}(V^{\omega}) \\ &+ \quad \mathrm{E}(V^{\omega}) \otimes \mathrm{E}(V^{\nu}) + \mathrm{E}(V^{\omega}) \star \mathrm{E}(V^{\nu})\end{aligned} \tag{5.38}$$

$$\mathrm{Cov}(\mathbf{vec}(\boldsymbol{X}_t^{(1)}\boldsymbol{X}_t^{(1)\mathrm{T}}), \mathbf{vec}(\boldsymbol{X}_{t-1}^{(2)}\boldsymbol{X}_{t-1}^{(2)\,\mathrm{T}})) = 4\mathrm{Var}(\mathbf{vec}V^{\nu}) + 2\mathrm{Var}(\mathbf{vec}V^{\omega}) + \mathrm{Var}(\mathbf{vec}S_{t-1}^{\nu}) \tag{5.39}$$

$$\mathrm{Cov}(\mathbf{vec}(\boldsymbol{X}_t^{(1)}\boldsymbol{X}_t^{(1)\mathrm{T}}), \mathbf{vec}(\boldsymbol{X}_{t-s}^{(2)}\boldsymbol{X}_{t-s}^{(2)\,\mathrm{T}})) = 4\mathrm{Var}(\mathbf{vec}V^{\nu}) + 2\mathrm{Var}(\mathbf{vec}V^{\omega}) \quad \forall t, \forall s \geq 2 \tag{5.40}$$

These results are obtained by focussing on a general element of a matrix on the left hand side, and then substituting into the left hand sides the definition (5.22) and (5.23), expanding the covariances, substituting representations (5.12) and (5.16), and then simplifying the result using known orthogonalities to deduce the general element of the matrices on the right hand side. However, there are several hundred terms in some of the expansions and a computer algebra package was used to ensure the accuracy of the results. Appendix B deals with the derivation of these results, using the REDUCE computer algebra system.

## 5.4 $n$-step exchangeability

### 5.4.1 $n$-step exchangeable collections

Recall that in Section 3.4 the concept of $n$-step exchangeability was introduced. The covariance structure over $\boldsymbol{X}'_t$, $\boldsymbol{X}''_t$ and their quadratic products is *second-order $n$-step exchangeable*. In Section 5.5.1 an inner-product space of random matrices appropriate for dynamic linear models will be constructed, and collections of matrices within this space will be found which exhibit generalised second-order $n$-step exchangeability.

More formally, form collections $X^\star = \{X'_{it}X'_{jt} | \forall i, j, \forall t \geq 2\}$ and $X^{\star\star} = \{X''_{it}X''_{jt} | \forall i, j, \forall t \geq 3\}$, of the matrices $\{\boldsymbol{X}'_t\boldsymbol{X}'^{\mathrm{T}}_t | \forall t \geq 2\}$ and $\{\boldsymbol{X}''_t\boldsymbol{X}''^{\mathrm{T}}_t | \forall t \geq 3\}$ defined in Section 5.3.3, where $X'_{it}$ denotes the $i^{th}$ component of the vector $\boldsymbol{X}'_t$. Form a vector space, $\mathcal{V}$ consisting of all linear combinations of the elements of $X^\star$ and $X^{\star\star}$ and the unit constant, and define the inner-product on this space as $(X, Y) = \mathrm{P}(XY), \quad \forall X, Y \in \mathcal{V}$. We may easily check that $X^\star$ is (second order) 2-step exchangeable over $t$, and that $X^{\star\star}$ is 3-step exchangeable over $t$.

### 5.4.2 Identification of the covariance structure underlying the DLM

The $n$-step exchangeability representation theorem (Theorem 1) allows construction of models for the observable quadratic products which have been formed. The elements of the collection $\{\boldsymbol{X}'_t\boldsymbol{X}'^{\mathrm{T}}_t | \forall t \geq 2\}$ for the two-step invertible DLM, are 2-step exchangeable over $t$. Using Theorem 1, construct the representation (3.25). The identified quantities may be constructed as the Cauchy limit of the arithmetic means of the elements.

**Lemma 1** *The 2-step exchangeable collection* $\{\boldsymbol{X}'_t\boldsymbol{X}'^{\mathrm{T}}_t|\forall t \geq 2\}$ *identify the matrix*

$$M' = F^{\mathrm{T}}V^{\omega}F + V^{\nu} + HV^{\nu}H^{\mathrm{T}} \tag{5.41}$$

*and the 3-step exchangeable collection* $\{\boldsymbol{X}''_t\boldsymbol{X}''^{\mathrm{T}}_t|\forall t \geq 3\}$ *identify*

$$M'' = F^{\mathrm{T}}GV^{\omega}G^{\mathrm{T}}F + F^{\mathrm{T}}V^{\omega}F + V^{\nu} + H^2V^{\nu}H^{2\mathrm{T}} \tag{5.42}$$

*Proof*

$$\begin{aligned}
M' &= \lim_{N\to\infty}\frac{1}{N}\sum_{t=2}^{N}\boldsymbol{X}'_t\boldsymbol{X}'^{\mathrm{T}}_t && (5.43)\\
&= \lim_{N\to\infty}\frac{1}{N}\sum_{t=2}^{N}(F^{\mathrm{T}}\boldsymbol{\omega}_t + \boldsymbol{\nu}_t - H\boldsymbol{\nu}_{t-1})(F^{\mathrm{T}}\boldsymbol{\omega}_t + \boldsymbol{\nu}_t - H\boldsymbol{\nu}_{t-1})^{\mathrm{T}} && (5.44)\\
&= \lim_{N\to\infty}\frac{1}{N}\sum_{t=2}^{N}\mathbf{vec}\Big(F^{\mathrm{T}}\boldsymbol{\omega}_t{\boldsymbol{\omega}_t}^{\mathrm{T}}F + \boldsymbol{\nu}_t{\boldsymbol{\nu}_t}^{\mathrm{T}} + H\boldsymbol{\nu}_{t-1}{\boldsymbol{\nu}_{t-1}}^{\mathrm{T}}H^{\mathrm{T}}\\
&\qquad + F^{\mathrm{T}}\boldsymbol{\omega}_t{\boldsymbol{\nu}_t}^{\mathrm{T}} - F^{\mathrm{T}}\boldsymbol{\omega}_t{\boldsymbol{\nu}_{t-1}}^{\mathrm{T}}H^{\mathrm{T}} + \boldsymbol{\nu}_t{\boldsymbol{\omega}_t}^{\mathrm{T}}F\\
&\qquad - \boldsymbol{\nu}_t{\boldsymbol{\nu}_{t-1}}^{\mathrm{T}}H^{\mathrm{T}} - H\boldsymbol{\nu}_{t-1}{\boldsymbol{\omega}_t}^{\mathrm{T}}F - H\boldsymbol{\nu}_{t-1}{\boldsymbol{\nu}_t}^{\mathrm{T}}\Big) && (5.45)\\
&= \lim_{N\to\infty}\frac{1}{N}\sum_{t=2}^{N}F^{\mathrm{T}}\boldsymbol{\omega}_t{\boldsymbol{\omega}_t}^{\mathrm{T}}F + \boldsymbol{\nu}_t{\boldsymbol{\nu}_t}^{\mathrm{T}} + H\boldsymbol{\nu}_{t-1}{\boldsymbol{\nu}_{t-1}}^{\mathrm{T}}H^{\mathrm{T}} && (5.46)\\
&= F^{\mathrm{T}}V^{\omega}F + V^{\nu} + HV^{\nu}H^{\mathrm{T}} && (5.47)
\end{aligned}$$

The derivation of $M''$ is similar. □

Now since $\frac{1}{2}\left[HM'H^{\mathrm{T}} - (M'' - M')\right] = HV^{\nu}H^{\mathrm{T}}$ we deduce that for a 2-step invertible series, the collection

$$\left\{\frac{1}{2}\left[H\boldsymbol{X}'_t\boldsymbol{X}'^{\mathrm{T}}_tH^{\mathrm{T}} - (\boldsymbol{X}''_t\boldsymbol{X}''^{\mathrm{T}}_t - \boldsymbol{X}'_t\boldsymbol{X}'^{\mathrm{T}}_t)\right]\middle|\forall t \geq 3\right\} \tag{5.48}$$

identifies $HV^{\nu}H^{\mathrm{T}}$.

Now if $r \geq p$, $H$ can always be chosen to be invertible. If $r < p$, since it has been assumed that $F$ and $G$ are of full rank, so is $H$, and so $H$ is invertible. Consequently, it shall be assumed that $H$ has been chosen to be invertible. Now, since $\frac{1}{2}[M' - H^{-1}(M'' - M')H^{-1\mathrm{T}}] = V^{\nu}$ and $M' - V^{\nu} - HV^{\nu}H^{\mathrm{T}} = F^{\mathrm{T}}V^{\omega}F$ it is trivial to deduce

**Theorem 2** *For a 2-step invertible series with invertible $H$, put*

$$M_t = \frac{1}{2}\left[\boldsymbol{X}_t'\boldsymbol{X}_t'^{\mathrm{T}} - H^{-1}(\boldsymbol{X}_t''\boldsymbol{X}_t''^{\mathrm{T}} - \boldsymbol{X}_t'\boldsymbol{X}_t'^{\mathrm{T}})H^{-1\mathrm{T}}\right] \tag{5.49}$$

*(i) The collection of matrices $\{M_t | \forall t \geq 3\}$ identify $V^{\nu}$.*

*(ii) The collection of matrices $\left\{\boldsymbol{X}_t'\boldsymbol{X}_t'^{\mathrm{T}} - M_t - HM_tH^{\mathrm{T}} \middle| \forall t \geq 3\right\}$ identify $F^{\mathrm{T}}V^{\omega}F$.*

□

If $p > r$, the entire matrix $V^{\omega}$ is not identified. This would require a fuller selection of observables than we have considered here. The identified matrix, $F^{\mathrm{T}}V^{\omega}F$, is the transformation of $V^{\omega}$ with the most immediate effect on the adjustment, since it is the contribution to the uncertainty for $\boldsymbol{X}_t$ from $\boldsymbol{\Theta}_t$, given $\boldsymbol{\Theta}_{t-1}$. To see this, note that

$$\boldsymbol{X}_t = F^{\mathrm{T}}G\boldsymbol{\Theta}_{t-1} + F^{\mathrm{T}}\boldsymbol{\omega}_t + \boldsymbol{\nu}_t \tag{5.50}$$

and so

$$\mathrm{Var}(\boldsymbol{X}_t) = F^{\mathrm{T}}G\mathrm{Var}(\boldsymbol{\Theta}_{t-1})G^{\mathrm{T}}F + F^{\mathrm{T}}\mathrm{P}(V^{\omega})F + \mathrm{P}(V^{\nu}) \tag{5.51}$$

giving

$$\mathrm{Var}(\boldsymbol{X}_t|\boldsymbol{\Theta}_{t-1}) = \mathrm{P}(F^{\mathrm{T}}V^{\omega}F + V^{\nu}) \tag{5.52}$$

Also,

$$\mathrm{Var}(\boldsymbol{X}_t|\boldsymbol{\Theta}_t) = \mathrm{P}(V^{\nu}) \tag{5.53}$$

and so the additional uncertainty comes from the $F^{\mathrm{T}}V^{\omega}F$.

### 5.4.3 Example

In the example, Theorem 2 implies that the collection $\{\boldsymbol{X}_t^{(2)}\boldsymbol{X}_t^{(2)\mathrm{T}} - \boldsymbol{X}_t^{(1)}\boldsymbol{X}_t^{(1)\mathrm{T}}|\forall t \geq 3\}$ identify $V^\omega$ and that the collection $\{\boldsymbol{X}_t^{(1)}\boldsymbol{X}_t^{(1)\mathrm{T}} - \frac{1}{2}\boldsymbol{X}_t^{(2)}\boldsymbol{X}_t^{(2)\mathrm{T}}|\forall t \geq 3\}$ identify $V^\nu$.

It is worth noting that in the simple example being considered here, we have that

$$\boldsymbol{X}_t^{(2)} = \boldsymbol{X}_t^{(1)} + \boldsymbol{X}_{t-1}^{(1)} \tag{5.54}$$

and so adding the collection $\{\boldsymbol{X}_t^{(2)}\boldsymbol{X}_t^{(2)\mathrm{T}}|\forall t \geq 3\}$ to the space spanned by the $\{\boldsymbol{X}_t^{(1)}\boldsymbol{X}_t^{(1)\mathrm{T}}|\forall t \geq 2\}$ has the effect of introducing terms of the form $\boldsymbol{X}_t^{(1)}\boldsymbol{X}_{t-1}^{(1)} + \boldsymbol{X}_{t-1}^{(1)}\boldsymbol{X}_t^{(1)}$. It is also worth noting that the collection of quantities

$$\left\{-\frac{1}{2}\left(\boldsymbol{X}_t^{(1)}\boldsymbol{X}_{t-1}^{(1)}{}^{\mathrm{T}} + \boldsymbol{X}_{t-1}^{(1)}\boldsymbol{X}_t^{(1)\mathrm{T}}\right)\middle|\,\forall t \geq 3\right\} \tag{5.55}$$

identifies the matrix $V^\nu$, and that in this example, such a collection has smaller variance than the collection $\{\boldsymbol{X}_t^{(1)}\boldsymbol{X}_t^{(1)\mathrm{T}} - \frac{1}{2}\boldsymbol{X}_t^{(2)}\boldsymbol{X}_t^{(2)\mathrm{T}}|\forall t \geq 3\}$, and so represents a better collection with which to identify $V^\nu$ (since uncertainty about the underlying quantities is resolved more quickly).

By observing sales at an increasing (but finite) number of time points, one may resolve through linear fitting, as much uncertainty as is desired about the underlying covariance structure for the particular time series model being dealt with.

If all fourth order prior belief specifications have been made, a simple Bayes linear analysis can be carried out in order to learn about the underlying covariance structure by adjusting the elements of $V^\nu$, $V^\omega$ by the elements of the observable matrices $\boldsymbol{X}_t^{(1)}\boldsymbol{X}_t^{(1)\mathrm{T}}$, $\boldsymbol{X}_t^{(2)}\boldsymbol{X}_t^{(2)\mathrm{T}}$. However, for long, high-dimensional time series, the number of quantities involved in a full linear adjustment is extremely large, and so it is important to reduce the dimensionality of the problem, and preserve the inherent matrix structure. This is done by carrying out adjustments in a matrix space.

## 5.5 Matrix objects for the time series

### 5.5.1 Formation of the matrix space

To learn about $r \times r$ dimensional covariance matrices, first form the $r \times r$ constant matrix basis, by defining $C_{r(i-1)+j}$ to be the matrix with a 1 in the $(i,j)^{th}$ position, and zeros elsewhere, where $i$ and $j$ range from 1 to $r$. Call this collection $C = [C_1, \ldots, C_{r^2}]$. Define the collections of matrices

$$X_2^{\ddagger} = \{\boldsymbol{X}_2'\boldsymbol{X}_2'^{\mathrm{T}}, H\boldsymbol{X}_2'\boldsymbol{X}_2'^{\mathrm{T}}H^{\mathrm{T}}, H^{-1}\boldsymbol{X}_2'\boldsymbol{X}_2'^{\mathrm{T}}H^{-1\mathrm{T}}\} \tag{5.56}$$

$$X_t^{\ddagger} = \{\boldsymbol{X}_t'\boldsymbol{X}_t'^{\mathrm{T}}, \boldsymbol{X}_t''\boldsymbol{X}_t''^{\mathrm{T}}, H\boldsymbol{X}_t'\boldsymbol{X}_t'^{\mathrm{T}}H^{\mathrm{T}}, H^{-1}\boldsymbol{X}_t'\boldsymbol{X}_t'^{\mathrm{T}}H^{-1\mathrm{T}}, H^{-1}\boldsymbol{X}_t''\boldsymbol{X}_t''^{\mathrm{T}}H^{-1\mathrm{T}}\},\ \forall t \geq 3 \tag{5.57}$$

Following the construction given in Chapter 3, form the real vector space, $\mathcal{N}$ whose elements are linear combinations of random $r \times r$ matrices as follows.

$$\mathcal{N} = span\left\{C \cup X_2^{\ddagger} \cup X_3^{\ddagger} \cup \ldots\right\} \tag{5.58}$$

Define an inner-product on $\mathcal{N}$ via

$$(A, B) = \mathrm{P}(\mathrm{Tr}[AB^{\mathrm{T}}]), \quad \forall A, B \in \mathcal{N} \tag{5.59}$$

Complete $\mathcal{N}$ into a Hilbert space, $\mathcal{M}$. Now since the collections whose mean limit points are $HV^{\nu}H^{\mathrm{T}}$, $V^{\nu}$, and $F^{\mathrm{T}}V^{\omega}F$ are present in the space, $\mathcal{N}$, Cauchy limit points such as $HV^{\nu}H^{\mathrm{T}}$, $V^{\nu}$, and $F^{\mathrm{T}}V^{\omega}F$ are present in the completed space, $\mathcal{M}$. The inner-product on this space is determined by our beliefs about the quadratic products, since

$$(A, B) = \sum_{j=1}^{r}\sum_{k=1}^{r}[\mathrm{Cov}(A_{jk}, B_{jk}) + \mathrm{P}(A_{jk})\mathrm{P}(B_{jk})]\ \forall A, B \in \mathcal{M} \tag{5.60}$$

Bayes linear adjustment may be carried out in this space by orthogonal projection of the matrices of interest into subspaces of observable matrices. As previously noted, this matrix approach to belief adjustment is a more direct way of getting at desirable linearity properties of conditional expectations for matrices, than via the somewhat artificial constructs such as the matrix Normal, inverse Wishart and matrix $T$ distributions. The definitions and properties of such distributions are described in Dawid (1981), and their application to matrix Normal DLMs is discussed in West and Harrison (1989, Section 15.4). Essentially, the notation and distributions are chosen so that they are consistent under marginalisation (Dawid 1981, Section 2), leading to simple linear conditional and predictive distributions for matrix Normal models (Dawid 1981, Section 8). Consequently, the updating equations for a matrix Normal DLM retain a simple linear form (West and Harrison 1989, Section 15.4.4).

### 5.5.2 Example

For our example, simply construct

$$\mathcal{N} = span\{C, \boldsymbol{X}_2^{(1)}\boldsymbol{X}_2^{(1)\mathrm{T}}, \boldsymbol{X}_3^{(1)}\boldsymbol{X}_3^{(1)\mathrm{T}}, \ldots, \boldsymbol{X}_3^{(2)}\boldsymbol{X}_3^{(2)\mathrm{T}}, \boldsymbol{X}_4^{(2)}\boldsymbol{X}_4^{(2)\mathrm{T}}, \ldots\} \qquad (5.61)$$

and impose the inner-product (5.59), inducing the Hilbert space $\mathcal{M}$, which contains limit points such as $V^\nu$ and $V^\omega$. Note that in order to evaluate (5.59), the specifications needed are precisely those which were made in Section 5.3.2. The fact that many other aspects of the fourth order specifications are not necessary is very helpful, as this greatly reduces the specification burden. Often it is most straightforward to make direct primitive specifications for the matrix object inner-product. However, for simplicity here, the specifications for the matrix inner product have been built up from specifications over the scalar quadratic products, thus establishing the links between the scalar and matrix analysis, as discussed in Chapter 3.

### 5.5.3 $n$-step exchangeable matrix objects

The definition of generalised $n$-step exchangeability applies directly to matrix objects in the space $\mathcal{M}$. The collection of matrix objects $\{\boldsymbol{X}'_t\boldsymbol{X}'^{\mathrm{T}}_t|\forall t \geq 2\}$ is 2-step exchangeable in the space $\mathcal{M}$, and the collection $\{\boldsymbol{X}''_t\boldsymbol{X}''^{\mathrm{T}}_t|\forall t \geq 3\}$ is 3-step exchangeable. This leads to a restatement of Theorem 2 for matrices in the space $\mathcal{M}$. The limit points are the matrices of limit points of their elements, due to the consistency of the inner-products on the scalar and matrix spaces, as shown in Section 3.3.

**Theorem 3** *Put* $M_t = \frac{1}{2}\left[\boldsymbol{X}'_t\boldsymbol{X}'^{\mathrm{T}}_t - H^{-1}(\boldsymbol{X}''_t\boldsymbol{X}''^{\mathrm{T}}_t - \boldsymbol{X}'_t\boldsymbol{X}'^{\mathrm{T}}_t)H^{-1\mathrm{T}}\right]$.

*(i) The collection* $\{M_t|\,\forall t \geq 3\}$ *identifies* $V^\nu$ *in* $\mathcal{M}$.

*(ii) The collection* $\left\{\boldsymbol{X}'_t\boldsymbol{X}'^{\mathrm{T}}_t - M_t - HM_tH^{\mathrm{T}}\middle|\forall t \geq 3\right\}$ *identifies* $F^{\mathrm{T}}V^\omega F$ *in* $\mathcal{M}$.

□

### 5.5.4 Adjustment

Consider observing $n > 3$ time points in the series. Form the matrix space, $\mathcal{M}$, and the observable subspace $D_n \subseteq \mathcal{M}$

$$D_n = span\{C \cup X^\ddagger_2 \cup X^\ddagger_3 \cup \ldots \cup X^\ddagger_n\} \tag{5.62}$$

Then the adjusted expectation map, $\mathrm{E}_{D_n}(\cdot) : \mathcal{M} \to D_n$, is the orthogonal projection into the $D_n$ space. In particular, evaluate $\mathrm{E}_{D_n}(V^\nu)$ and $\mathrm{E}_{D_n}(F^{\mathrm{T}}V^\omega F)$, which are matrices in the $D_n$ space ($V^\nu$ and $F^{\mathrm{T}}V^\omega F$ are chosen because they are the matrices in $\mathcal{M}$ which we are most interested in).

## 5.6 Bayes linear adjustment for the example

### 5.6.1 The adjusted covariance matrices

Adjustments were carried out using 17 time points from the actual time series. The sample means for the 6 brands were 16.5, 3, 4.5, 27.5, 3.4 and 31. The matrix objects $V^\omega$ and $V^\nu$ were adjusted in the following ways:

$$\mathrm{E}(V^\omega) = \begin{pmatrix} 4 & 1 & 1 & 1 & 1 & 1 \\ 1 & 4 & 1 & 1 & 1 & 1 \\ 1 & 1 & 4 & 1 & 1 & 1 \\ 1 & 1 & 1 & 4 & 1 & 1 \\ 1 & 1 & 1 & 1 & 4 & 1 \\ 1 & 1 & 1 & 1 & 1 & 4 \end{pmatrix} \tag{5.63}$$

$$\mathrm{E}_{D_{17}}(V^\omega) = \begin{pmatrix} 4.8 & 0.9 & 1.0 & 1.0 & 0.8 & 1.5 \\ 0.9 & 3.9 & 1.2 & 0.9 & 1.1 & 0.3 \\ 1.0 & 1.2 & 4.0 & 1.1 & 1.1 & 0.7 \\ 1.0 & 0.9 & 1.1 & 6.8 & 0.7 & 0.8 \\ 0.8 & 1.1 & 1.1 & 0.7 & 3.9 & 0.8 \\ 1.5 & 0.3 & 0.7 & 0.8 & 0.8 & 4.7 \end{pmatrix} \tag{5.64}$$

$$\mathrm{E}(V^\nu) = \begin{pmatrix} 36 & -4 & -4 & -4 & -4 & -4 \\ -4 & 36 & -4 & -4 & -4 & -4 \\ -4 & -4 & 36 & -4 & -4 & -4 \\ -4 & -4 & -4 & 36 & -4 & -4 \\ -4 & -4 & -4 & -4 & 36 & -4 \\ -4 & -4 & -4 & -4 & -4 & 36 \end{pmatrix} \tag{5.65}$$

$$\mathrm{E}_{D_{17}}(V^\nu) = \begin{pmatrix} 41.8 & -5.4 & -4.4 & -8.0 & -4.7 & -2.4 \\ -5.4 & 36.7 & -3.8 & -0.2 & -3.2 & -4.1 \\ -4.4 & -3.8 & 36.1 & -4.4 & -3.5 & -7.5 \\ -8.0 & -0.2 & -4.4 & 56.6 & -5.6 & 4.8 \\ -4.7 & -3.2 & -3.5 & -5.6 & 34.9 & -4.9 \\ -2.4 & -4.1 & -7.5 & 4.8 & -4.9 & 44.0 \end{pmatrix} \tag{5.66}$$

The prior specifications were given in (5.10) and (5.11). The adjusted matrices are perturbations of the prior expectations for the matrices. Notice that the variance associated with the fourth variable has been inflated considerably in both matrices.

| Variable (Brand) | Primary eigenvector | Secondary eigenvector |
|---|---|---|
| 1 | -0.07 | 0.21 |
| 2 | -0.03 | -0.21 |
| 3 | -0.02 | -0.12 |
| 4 | 0.39 | 0.05 |
| 5 | -0.07 | -0.13 |
| 6 | -0.09 | 0.31 |
| Eigenvalue ratio | 1.88 | 1.47 |

Table 5.1: Eigenstructure of the belief transform for the $V^\omega$ adjustment

| Variable (Brand) | Primary eigenvector | Secondary eigenvector |
|---|---|---|
| 1 | 0.02 | -0.11 |
| 2 | 0.04 | 0.02 |
| 3 | 0.03 | 0.02 |
| 4 | 0.12 | 0.03 |
| 5 | 0.02 | 0.00 |
| 6 | 0.07 | -0.08 |
| Eigenvalue ratio | 1.89 | 1.24 |

Table 5.2: Eigenstructure of the belief transform for the $V^\nu$ adjustment

The sample variances for the 17 cases of the six brands considered were 167, 22, 37, 560, 18 and 427. Informally, it seems that there may indeed be more variability associated with the fourth (and last) variable.

More formally, as described in Section 4.3.2, one may analyse the eigenstructure of the belief transform implied by the adjustment. Examining Table 5.1 shows that for the adjustment of the matrix $V^\omega$, variance has been inflated by a factor of 1.88 in a direction close to the fourth brand, and by a factor of 1.47 in a direction close to the difference between the second and the sum of the first and last brands. Other components had eigenvalues close to one, and hence were of minimal interest. Table 5.2 shows that for the adjustment of the matrix $V^\nu$, variance has been inflated by a factor of 1.89 in a direction close to the fourth brand, and by a factor of 1.24 in a direction close to the difference between the first and last brands.

### 5.6.2 First order adjustment

Since the aim is to predict sales more accurately, a sensible test of the procedure is to compare the performance of the first order model, (5.6), (5.7), using both the prior and adjusted covariance matrices. Carrying out the adjustment shows that the Bayes linear diagnostic warnings (the *size* and *bearings* of the adjustments, as described in Goldstein (1988b)) are noticeably closer to their expected values when using the adjusted matrices. For the given example, most of the *size ratios* for adjustments of the first order structure were noticeably closer to one using the adjusted covariance structure to predict future values, suggesting that the adjusted matrices match more closely with the forecast performance of the model.

The improvements in forecast performance are graphically illustrated using diagnostic Bayes linear influence diagrams. Two sequences of diagrams are given in Figures 5.1, 5.2, 5.3, 5.4 and 5.5. The top diagrams represent the usual Bayes linear adjustment for a dynamic linear model using the *a priori* covariance structure, and the bottom diagrams represent the adjustment using the updated covariance structure. The shadings in the centre of the nodes are diagnostics based on the sizes of the adjustment. The larger the amount of red or blue, the stronger the diagnostic warning. Within each diagram, the upper layer of nodes represent the unobservable quantities in the models; namely $\boldsymbol{\nu}_t$ and $\boldsymbol{\omega}_t$. The lower row of nodes represent the observable quantities; namely the $\boldsymbol{X}_t$. It can be seen that the lower series of diagrams has consistently, slightly less diagnostic warning, indicating that the revised covariance structure matches more closely with the forecast performance of the model.

Of course, since the revision of the covariance matrices was carried out using the first 17 weeks worth of data, one would hope that the diagnostics would improve when the first order adjustments are carried out for those weeks, using that revised covariance structure. A purely sample based estimate of the covariance structure would probably perform even better! However, it is at least reassuring to see that the adjustment clearly hasn't made things worse. Further, if we focus attention on Figure 5.5, we see that the improved matching of the forecast performance of the model continues as first order adjustments continue on new data. Note that the diagnostics have not significantly improved using only 17 weeks of data. For these kinds of time series structures, it takes a large amount of data to learn a significant amount about the underlying variance structures, and this point is explored further in the next section. Diagnostics for scalar and matrix adjustments are discussed more fully in Chapter 6, and the calculations underlying the diagnostics for this example are discussed in Section 6.1.2.

## 5.7 Iterative adjustments

### 5.7.1 Methodological considerations

For real-time problems, data will arrive for consideration one time point at a time. Suppose we are currently at time $t$. We must consider how best to use the available data in order to make predictions for future sales. We will wish to use all of the data to revise beliefs about the covariance structure for our dynamic linear model, and then carry out first order adjustments using the revised covariance structure. However, when we receive the data for time point $t+1$, we will update beliefs for the covariance structure. Having done so, we will need to re-compute the first order adjustments for *all* time points, using the revised covariance structure. It will not be sufficient to simply add data for the last time point to the adjustment. The quadratic data is informative for the covariance structure underlying the entire first order series, and not just for the last time point.

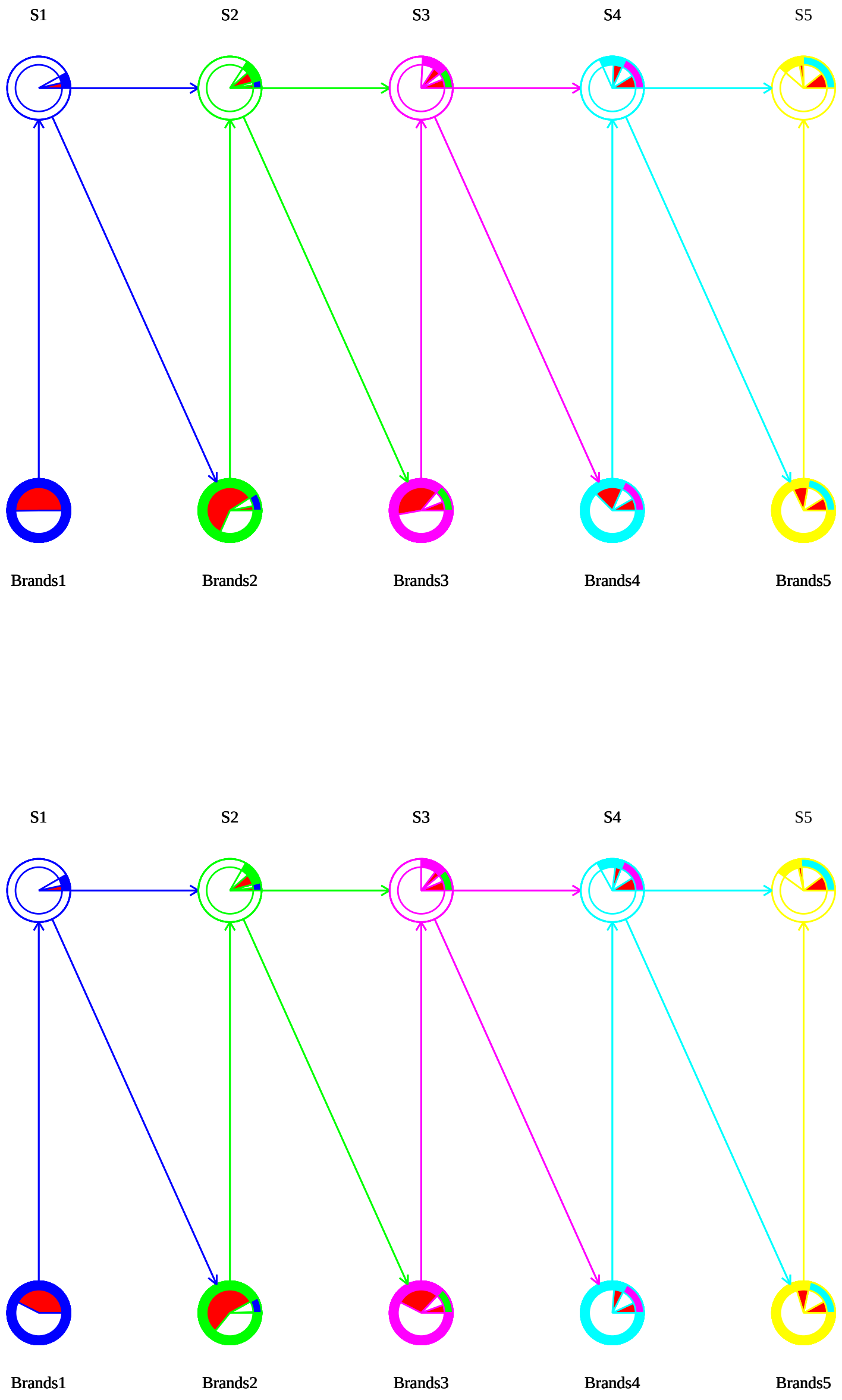

Figure 5.1: First order adjustments; weeks 1–5

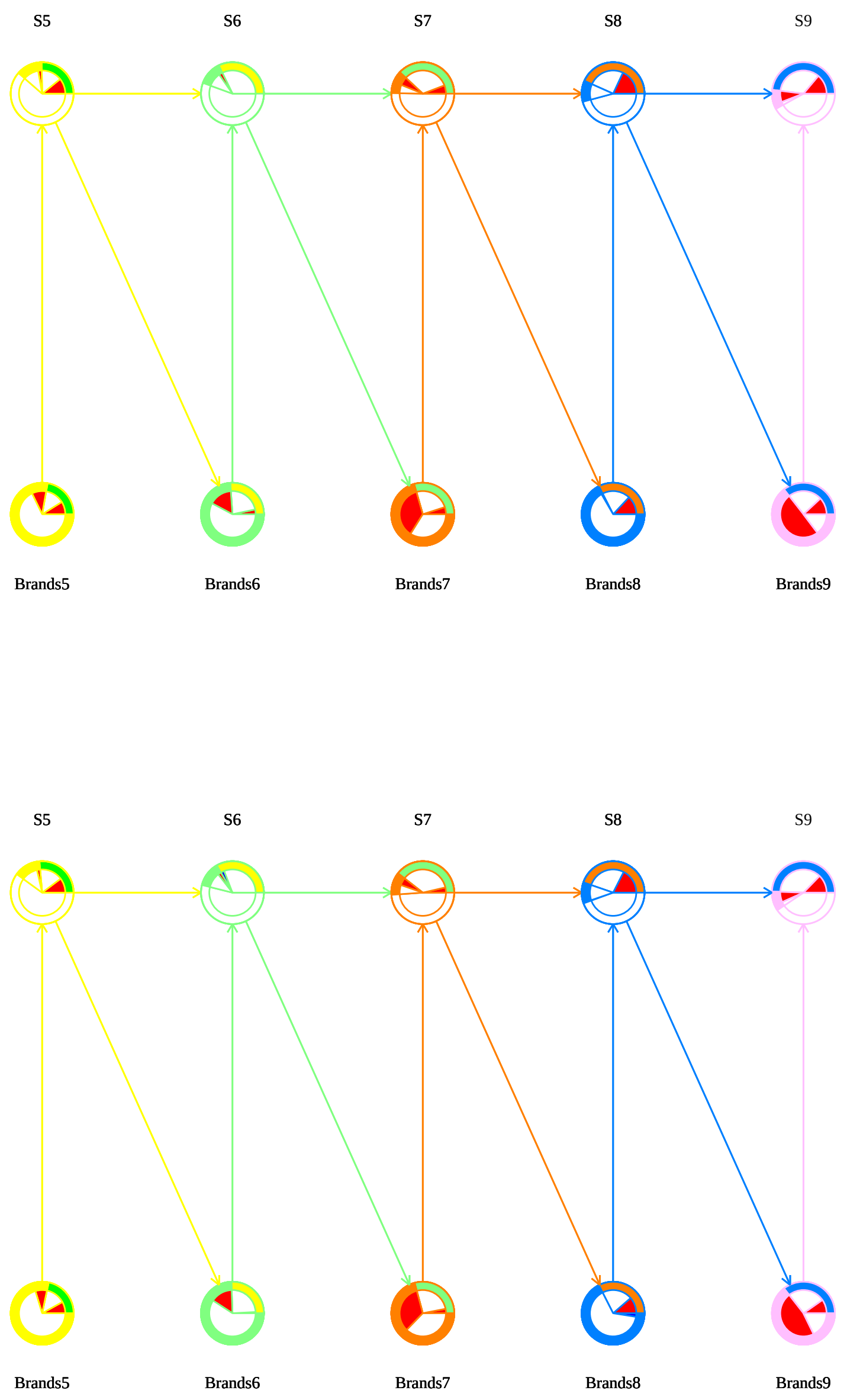

Figure 5.2: First order adjustments; weeks 5–9

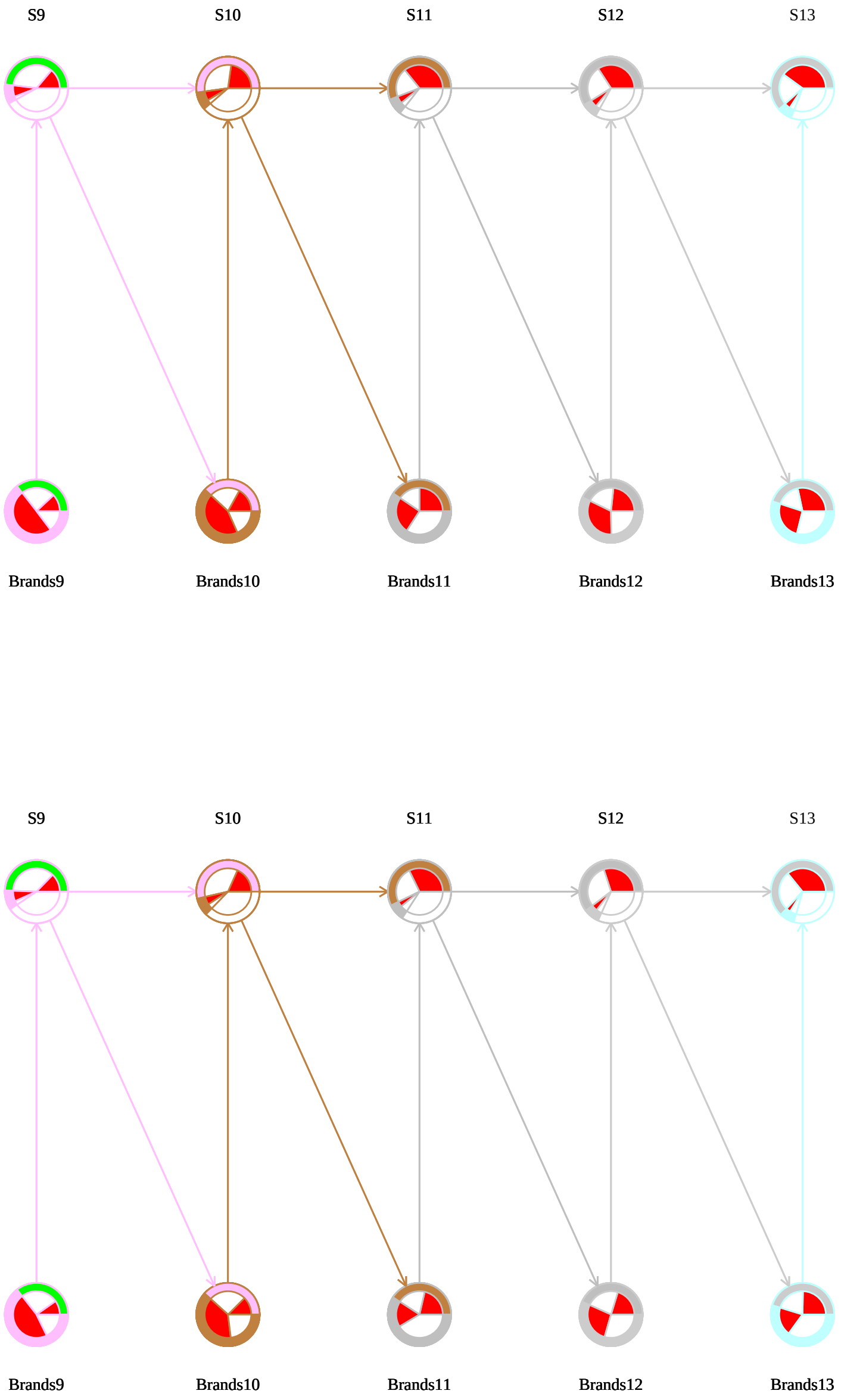

Figure 5.3: First order adjustments; weeks 9–13

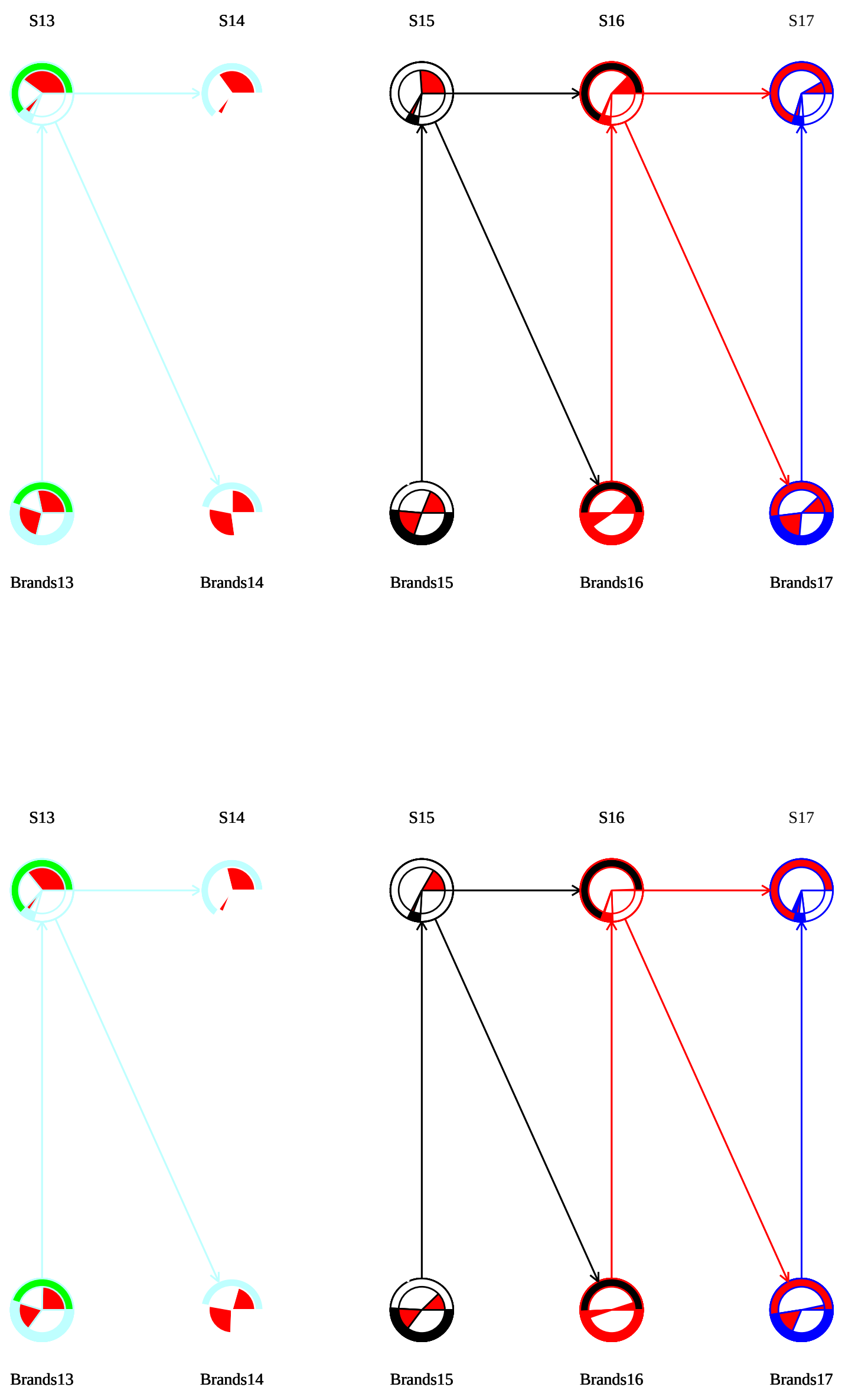

Figure 5.4: First order adjustments; weeks 13–17

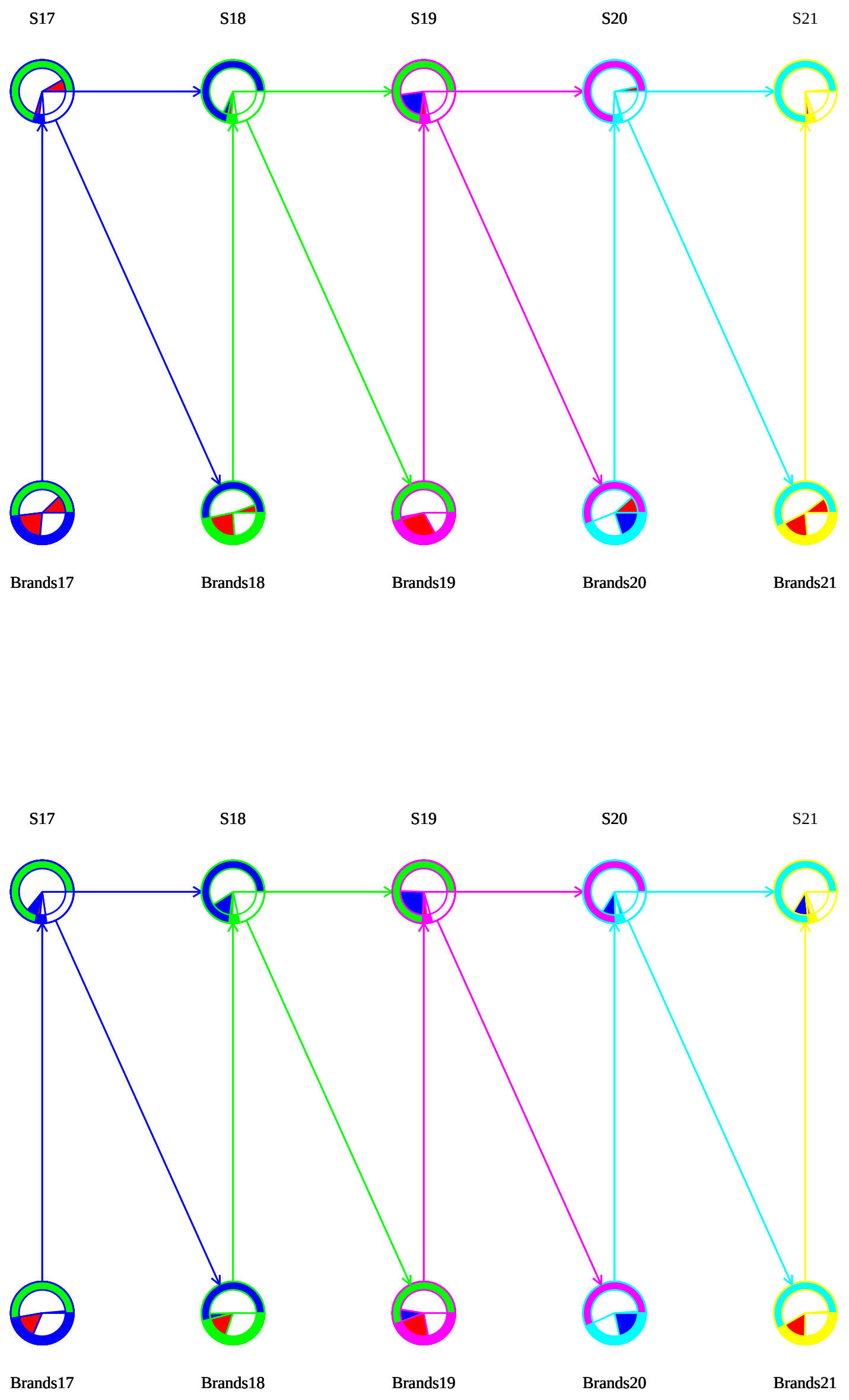

Figure 5.5: First order adjustments; weeks 17–21

### 5.7.2 Illustrative example

Consider Figure 5.6. Each small node represents 8 weeks of sales data for the top two brands of shampoo. Each node on the top row represents 16 matrix objects — one for each week for both $\boldsymbol{X}'_t\boldsymbol{X}'^{\mathrm{T}}_t$ and $\boldsymbol{X}''_t\boldsymbol{X}''^{\mathrm{T}}_t$. The large node in the middle represents the two matrix objects $V^{\omega}$ and $V^{\nu}$. The bottom rows of small nodes each represent 8 two-dimensional vectors of weekly sales. The shadings for the bottom row of nodes correspond to those for a conventional stepwise Bayes linear adjustment of the structure. The shadings for the row above take the covariance structure updates into consideration. The shadings in the centre of the nodes are diagnostic warnings, based on the *size* and *bearing* of the adjustments. It is immediately obvious that the shadings for the adjustments which take into account available quadratic data improve proportionately with time, indicating an improvement in the understanding of the forecast performance of the model as the covariance structure tends to the "true" underlying structure, as the model learns about the variability of components and correlations across them, thus improving forecast estimates, and associated standard errors. Note though, that the rate of learning in this example is very slow. Each portion of shading on the large central node represents the proportion of uncertainty resolved by eight weeks of sales data. Even at the end of the process, having used 40 weeks worth of quadratic sales data, only about one sixth of the uncertainty about the underlying covariance structure has been resolved. It is no wonder that in the last section, improvements in diagnostics using only 17 time points was marginal. Diagnostics for Bayes linear adjustments are discussed in Chapter 6.

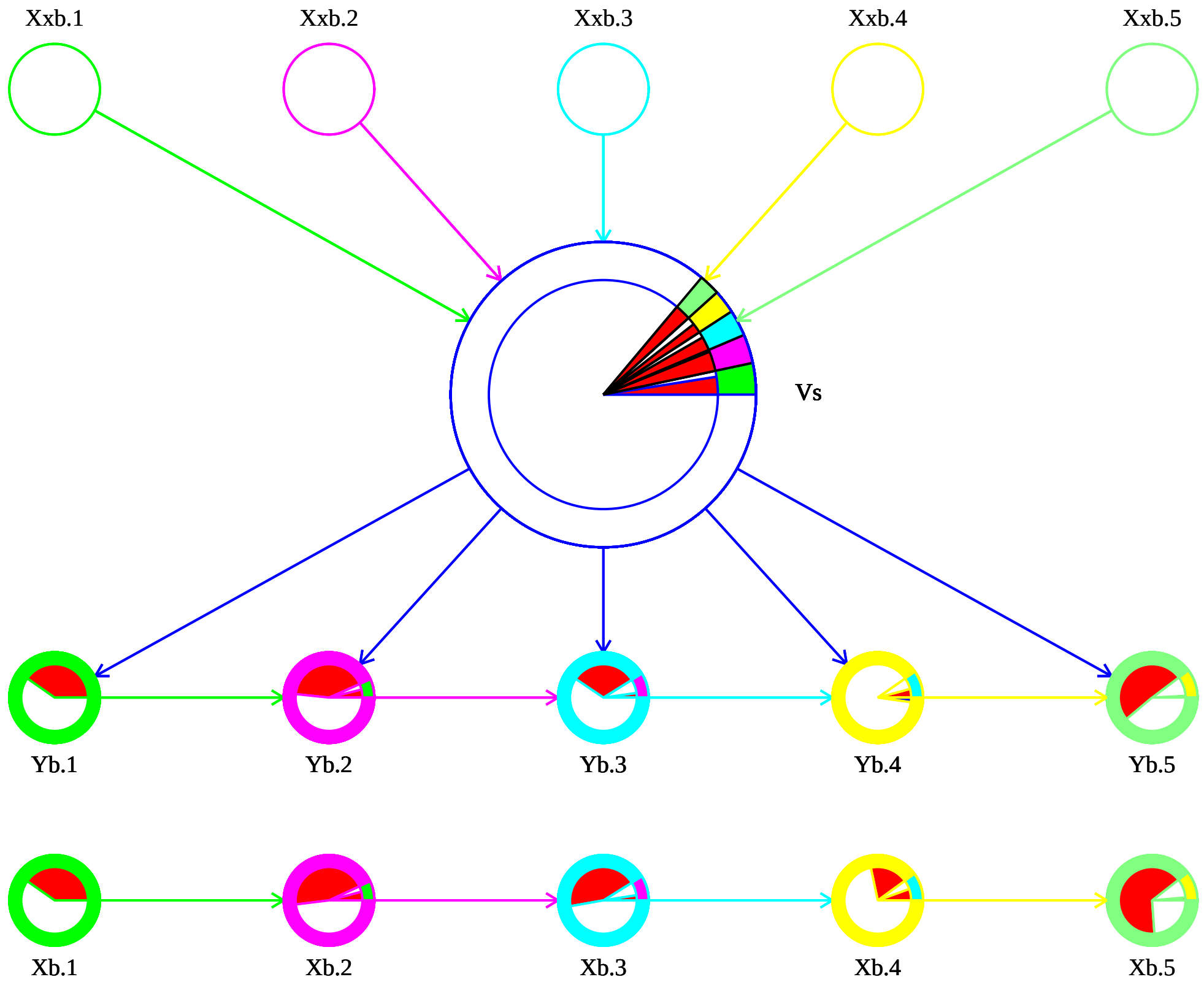

Figure 5.6: Iterative first and second order adjustments

## 5.8 Conclusions

Good forecasting requires careful updating of the covariances within the time series structure. Informally, the degree of *shrinkage* between the prior and the data is updated, and relationships between variables are properly taken into account. Since one is able to adjust the covariance matrices for both the *observational* as well as *state* residuals, it is possible to understand the *competition* and *demand* effects taking place within the series. By taking a matrix object approach, the problem is greatly simplified by reducing it's dimensionality. This is important for both simplifying belief specification and belief adjustment, and also for interpretation of the structure of the adjustment and accompanying diagnostics. There are also the general advantages of the Bayes linear approach; namely of allowing complete flexibility for the prior specifications, without placing distributional restrictions on the data or model components.

Of course, it is in principle possible to take a distributional Bayesian approach to this problem. Provided one is up to making a joint distributional statement about the components of the underlying matrices and the quadratic observables (a formidable task), one could use MCMC techniques, for example, to sample from the distribution to obtain posterior estimates. However, for non-trivial problems the problem would be sufficiently high-dimensional that assessing convergence would be extremely difficult. Also, it would be hard to assess the effect of the largely arbitrary choice of distribution made. Further, an *a priori* analysis to tackle design issues (such as how many time points are needed to reduce uncertainty about the underlying covariance matrices to one tenth of their *a priori* values) would be difficult to contemplate using such a framework. On the contrary, using the Bayes linear approach advocated in this chapter, a full *a priori* analysis is possible, allowing the handling of such design questions before receiving any data.

# Chapter 6

# Matrix adjustment diagnostics

## 6.1 Bayes linear diagnostics

### 6.1.1 The size and bearing of a scalar adjustment

Before going on to discuss general Bayes linear diagnostics, it is important to outline the usual concept of *size* and *bearing* for scalar Bayes linear adjustments. These concepts are discussed more fully in Goldstein (1988b). For the adjustment of a collection of scalar random quantities, $B = [B_1, B_2, \ldots]$, by a data space, $D$, we *a priori* form the random vector $\mathrm{E}_D(\boldsymbol{B}) = (\mathrm{E}_D(B_1), \mathrm{E}_D(B_2), \ldots)^{\mathrm{T}}$. When the data space is observed to be $D = d$, the precise value of the vector $\mathrm{E}_D(\boldsymbol{B}) = \mathrm{E}_d(\boldsymbol{B})$ becomes known. The further away $\mathrm{E}_d(\boldsymbol{B})$ is from $\mathrm{E}(\boldsymbol{B})$ relative to prior standard deviation, the more surprising our change of belief. With this in mind, the *size*, $\mathrm{Size}_d(B)$ of the adjustment is defined as follows.

$$\mathrm{Size}_d(B) = \max_{X \in span\{B\}} \frac{(\mathrm{E}_d(X) - \mathrm{E}(X))^2}{\mathrm{Var}(X)} \tag{6.1}$$

Now let $U = [U_0 = 1, U_1, U_2, \ldots]$ be an orthonormal basis for *span*$\{B\}$ with respect to the inner-product $(\cdot,\cdot)$. Then the *bearing*, $Z_d(B)$ is defined by

$$Z_d(B) = \sum_{i=1} \mathrm{E}_d(U_i)U_i \tag{6.2}$$

The bearing has the property that it is the quantity, $X$ which maximises the expression in (6.1). Hence,

$$\mathrm{Size}_d(B) = \mathrm{Var}(Z_d(B)) = \sum_{i=1} \mathrm{E}_d^2(U_i) \tag{6.3}$$

Also note that the bearing is a complete summary of the adjustment, since

$$\mathrm{E}_d(X) - \mathrm{E}(X) = \mathrm{Cov}(X, Z_d(B)) \tag{6.4}$$

The *size ratio*, $Sr_d(B)$ of an adjustment, is the magnitude of the size of the adjustment, relative to it's expected value. Hence,

$$Sr_d(B) = \frac{\mathrm{Size}_d(B)}{\mathrm{E}(\mathrm{Var}(Z_D(B)))} \tag{6.5}$$

and $\mathrm{E}(\mathrm{Var}(Z_D(B)))$ is given by the trace of the *belief transform* for the adjustment, $\mathrm{Cov}(\boldsymbol{D},\boldsymbol{B})\mathrm{Var}(\boldsymbol{B})^{-1}\mathrm{Cov}(\boldsymbol{B},\boldsymbol{D})\mathrm{Var}(\boldsymbol{D})^{-1}$ (Goldstein 1988b). A size ratio close to one indicates observations consistent with prior beliefs. A very large size ratio is due to a surprisingly large change in belief, indicative of too small specifications made for prior uncertainties. A size ratio very close to zero is due to a surprisingly small change in belief, indicative of too large specifications made for prior uncertainties. It is a simple transformation of the size ratios which is used to shade the diagnostic warnings in the centres of nodes on diagnostic Bayes linear influence diagrams.

To those readers familiar with the basic concepts of functional analysis, it should now be clear that the bearing is derived from the Riesz representation for the bounded linear functional $\mathrm{E}_d(X) - \mathrm{E}(X)$ (see Section 3.8-1 of Kreyszig (1978), for example).

Note then, that the fact that the linear function $\mathrm{E}_d(X) - \mathrm{E}(X)$ is in fact a functional (a function whose image is a subset of the real or complex numbers) is crucial to the concept of the bearing, and all of the elegant additive properties of the bearing for sequences of adjustments (Goldstein 1988b) depend upon the existence of a Riesz representation. Note also that when we are dealing with matrices, the function $\mathrm{E}_d(X) - \mathrm{E}(X)$ is most certainly *not* a functional, and so in order to make any sense of the standard diagnostic and interpretive features of the Bayes linear methodology, a clarification of precisely what is meant, in general, by terms such as the size and bearing of the adjustment, is required.

### 6.1.2 Examples

Consider Figures 5.1, 5.2, 5.3, 5.4 and 5.5. The shadings in the centre of the nodes are a non-linear transformation of the size ratios for the depicted adjustments. Red shadings represent size ratios lager than one, and blue shadings represent size ratios smaller than one. The transformation has been chosen so that a shading of more than half of the available area is "surprising". It is clear that the lower diagrams (the first order adjustments using the revised covariance structure) have slightly smaller shadings, representing size ratios closer to one, representing adjustment sizes closer to expected values. This is indicative of a covariance structure which more closely matches the forecast performance of the model.

Figure 5.6 shows the improvement over time of the forecasting performance of the model as more covariance information becomes known. The diagnostic shadings for the row of first order adjustments including covariance updating become proportionately smaller than those without covariance adjustment as more information about the covariance structure becomes known. The diagnostics for the matrix adjustments, shown in the large central node, are calculated using the subspace bearing technique discussed in the next section.

## 6.2 The subspace bearing

### 6.2.1 Definition

Here the bearing is generalised to the space of random matrices. For any given constant matrix, $G$, and projection space $D$, the bearing is defined to be the (essentially) unique random matrix, $B$, with the properties $\mathrm{E}(B) = 0$ and

$$(A - \mathrm{E}(A), B) = (\mathrm{E}_d(A), G) - (\mathrm{E}(A), G), \ \forall A \in \mathcal{M} \tag{6.6}$$

where $\mathrm{E}_d(A)$ represents the realisation of $\mathrm{E}_D(A)$ after observing $D = d$. This matrix is derived from the Riesz representation for the functional $(\mathrm{E}_d(A) - \mathrm{E}(A), G)$. Different choices of the constant matrix, $G$, give information about different projections of the adjusted expectations.

The choice of $G$ which causes diagnostics to match up exactly with those for scalar Bayes linear adjustment in the case of exclusively one-component matrices, is the choice given by the constant matrix whose elements are all 1. To see this note that the for this choice of $G$, the functional of interest, $(\mathrm{E}_d(A) - \mathrm{E}(A), G)$, becomes

$$\sum_i \sum_j \mathrm{E}_d(A_{ij}) - \mathrm{E}(A_{ij}) \tag{6.7}$$

and so for a one-component matrix, whose only non-zero element is in row $i$, column $j$, this becomes $\mathrm{E}_d(A_{ij}) - \mathrm{E}(A_{ij})$, as for the usual case of scalar adjustments. The full correspondence follows since the one-dimensional matrix subspaces are orthogonal under the matrix inner-product.

### 6.2.2 Example

Reconsider Figure 4.2 from Chapter 4. At the centre of the $V$ node, red (blue) shadings represent changes in expectation larger (smaller) than expected *a priori*. Adjusting by the sample covariance matrix, $D_S$, caused a much larger change in expectation than expected *a priori*. This is evidence of over-confidence about our ability to predict the true value of the covariance matrix, and suggests re-examination the prior specification. Notice also that adding the full variance collection, $D_F$, to the adjustment had the potential to change our expectation considerably, but in fact, hardly changed it at all. This is perhaps evidence of overestimation of the importance of the covariance terms.

### 6.2.3 A space of subspace bearings

Note that the subspace bearing is defined for a given choice of constant matrix, $G$. Consequently, for each choice of constant matrix, there is a corresponding bearing. In particular, it may be of interest to study the structure of the map, $\phi : C \longrightarrow \mathcal{M}$, which maps a given choice of constant matrix to it's corresponding bearing, and its image, $\phi(C) \subseteq \mathcal{M}$. Notice that the map $\phi$ is linear, and so an analysis of the eigenstructure of the map

$$\phi^*\phi : C \to C \tag{6.8}$$

(where $\phi^*$ denotes the Hilbert-adjoint of the $\phi$ operator) may be informative. See Section 6.3 for more details of this kind of construction and analysis.

### 6.2.4 Limitations of the subspace bearing

The above sections show that the subspace bearing can be very useful for diagnosing matrix adjustments, and in particular, for linking up with diagnostics for the component scalar adjustments. However, in general there is only a sensible link with scalar diagnostic analyses in the case where all matrices are decomposed to the one-component level. When matrices are not decomposed to the one-component level, the interpretation of the subspace bearing is much less clear. However, one may not be prepared to undertake the specification and computational burden imposed by such a decomposition of structures.

In the simple exchangeable matrices example discussed in the previous section, all specifications over the matrix components were made in order to deduce an inner-product over all sub-matrices, and so no problem arose in deducing or interpreting the bearing. However, in the dynamic linear model example of Chapter 5, only specifications sufficient to allow the deduction of the inner-product over the whole matrix objects were given. Hence in this case, the interpretation of the calculated subspace bearings are unclear. In the matrix adjustment example shown in Figure 5.6, large diagnostic warnings are present, but it is totally unclear as to whether or not these should really be considered to be a serious problem.

Moreover, in general one will often wish to make primitive specifications for the matrix inner-product, without making any recourse to scalar components, and in this case it is clear that obtaining beliefs over the scalar subspaces is not possible. Also, the methodology underlying the subspace bearing seems contrived, and the resulting definitions of sizes and bearings of adjustments do not seem to fit well with an intuitive concept of size and bearing for an adjustment.

The above motivates the development of a more general concept of bearing, not tied the the concept of the Riesz representation for bounded linear functionals on Hilbert space.

## 6.3 General Bayes linear diagnostics

### 6.3.1 Observation operators

**Definition 4** *Consider the Hilbert space $\mathcal{H}$ with Hilbert subspaces, $C$ and $D$ such that $C \subseteq D \subseteq \mathcal{H}$. $C$ represents the subspace of known quantities, and $D$ represents the subspace of observable quantities. Define the operator, $\mathrm{E}_D(\cdot) : \mathcal{H} \to D$ to be the orthogonal projection into the space $D$. Then, for a given realisation $d$ of the space, $D$, define the observation operator, $F_d$ as follows:*

$$F_d(\cdot) : D \to C \tag{6.9}$$

*where for every $X \in D$, $F_d(X)$ is the realisation of $X$. Note that the observation operator, $F_d$ is linear. Also, the $C$ space will, in general, be chosen so that $F_d$ is necessarily bounded. Next define the observed adjusted expectation operator, $\mathrm{E}_d$, via*

$$\mathrm{E}_d(\cdot) : \mathcal{H} \to C \tag{6.10}$$

*where*

$$\mathrm{E}_d(X) = F_d \mathrm{E}_D(X), \quad \forall X \in \mathcal{H} \tag{6.11}$$

*Note that $\mathrm{E}_d$ is bounded and linear.*

□

### 6.3.2 General diagnostic measures

The bounded linear operator, $\mathrm{E}_d : \mathcal{H} \to C$ contains all the information about the observed adjustment. If we are interested in the effect of the observations on some particular Hilbert subspace, $B \subseteq \mathcal{H}$, such that $B \perp C^*$, we can look at the restriction of $\mathrm{E}_d$ to $B$,

$$\mathrm{E}_d|_B : B \to C \tag{6.12}$$

We may compare this with the restriction of the $\mathrm{E}_D$ projection to $B$,

$$\mathrm{E}_D|_B : B \to D \tag{6.13}$$

in order to get an impression of the magnitude of the changes, compared to what we expected *a priori*. The structure of the $\mathrm{E}_d|_B$ operator is revealed by examining the eigenstructure of the operator,

$$\mathrm{E}_d|_B^* \mathrm{E}_d|_B : B \to B \tag{6.14}$$

where $\mathrm{E}_d|_B^*$ denotes the Hilbert-adjoint of the operator $\mathrm{E}_d|_B$. The Hilbert-adjoint of an operator is defined in Kreyszig (1978, Section 3.9). This is compared to the eigenstructure of the operator,

$$\mathrm{E}_D|_B^* \mathrm{E}_D|_B : B \to B \tag{6.15}$$

The *belief transform for $B$ adjusted by $D$*,

$$\mathrm{E}_B|_D \mathrm{E}_D|_B : B \to B \tag{6.16}$$

---

*We are interested in changes of expectation. It is simpler to choose to work with a space $B$ such that $B \perp C$ than to allow general $B$, and look at the operator $\mathrm{E}_d - \mathrm{E}$. It doesn't make any difference to the analysis however, since if $B \perp C$, all elements of $B$ have zero expectation.

is defined and discussed in Goldstein (1981). The eigenstructure of this operator contains all information about expected changes in belief. It is worth pointing out that the operator $\mathrm{E}_D|_B^*\mathrm{E}_D|_B$ is simply the belief transform $\mathrm{E}_B|_D\mathrm{E}_D|_B$, since $\mathrm{E}_D|_B^* = \mathrm{E}_B|_D$. To see this, note that

$$\mathrm{E}_D|_B^* : D \rightarrow B \tag{6.17}$$

is defined by the property

$$(\mathrm{E}_D|_B(b), d) = (b, \mathrm{E}_D|_B^*(d)), \quad \forall b \in B, d \in D \tag{6.18}$$

and that

$$\mathrm{E}_B|_D : D \rightarrow B \tag{6.19}$$

has the property

$$(b, \mathrm{E}_B|_D(d)) = (b, d) = (\mathrm{E}_D|_B(b), d), \quad \forall b \in B, d \in D \tag{6.20}$$

Properties of projection operators are discussed in Kreyszig (1978, Section 9.5).

For the special case of scalar adjustments ($\dim(C) = 1$), it is clear that the operator $\mathrm{E}_d|_B^*\mathrm{E}_d|_B$ has rank 1, and so has only one non-trivial eigen-pair. The eigenvalue of this operator is the *size* of the adjustment (note that the size of a scalar adjustment is defined to be the square of the induced norm of the $\mathrm{E}_d|_B$ operator, that the norm of an operator is equal to that of it's adjoint, and that the induced norm of the $\mathrm{E}_d|_B^*\mathrm{E}_d|_B$ operator is equal to it's single eigenvalue), and it's corresponding eigenvector is known as the *bearing* of the adjustment. The properties of the eigenvalues and eigenvectors of the $\mathrm{E}_d|_B^*\mathrm{E}_d|_B$ operator are discussed more fully for the general case later in this section.

For general adjustments, the operator $\mathrm{E}_d|_B^*\mathrm{E}_d|_B$ may have more than one non-trivial eigen-pair, and so a generalisation of the concepts of size and bearing are required.

One may generalise the concept of size in a number of ways. For example, one could define the size of the adjustment to be the maximum eigenvalue of the operator $\mathrm{E}_d|_B^*\mathrm{E}_d|_B$, which would lead to the (perhaps desirable) property

$$\begin{aligned}\mathrm{Size}_d(B) &= \sup_{X \in B+C, \mathrm{E}(X)=0, (X,X)=1} (\mathrm{E}_d(X), \mathrm{E}_d(X)) && (6.21)\\ &= \sup_{X \in B+C, \mathrm{E}(X)=0, (X,X)=1} \|\mathrm{E}_d(X)\|^2 && (6.22)\end{aligned}$$

However, one of the key properties of the size of the adjustment is the way it's expected values naturally sum over sequences of adjustments. If the size was defined as described above, such a property would be lost. Explicitly, it is the case that for orthogonal subspaces, $D_1$ and $D_2$,

$$\mathrm{E}_{D_1 \oplus D_2}|_B = \mathrm{E}_{D_1}|_B \oplus \mathrm{E}_{D_2}|_B \tag{6.23}$$

Consequently,

$$\begin{aligned}\mathrm{E}_{D_1 \oplus D_2}|_B^*\mathrm{E}_{D_1 \oplus D_2}|_B &= (\mathrm{E}_{D_1}|_B \oplus \mathrm{E}_{D_2}|_B)^*(\mathrm{E}_{D_1}|_B \oplus \mathrm{E}_{D_2}|_B) && (6.24)\\ &= (\mathrm{E}_{D_1}|_B^* \oplus \mathrm{E}_{D_2}|_B^*)(\mathrm{E}_{D_1}|_B \oplus \mathrm{E}_{D_2}|_B) && (6.25)\\ &= \mathrm{E}_{D_1}|_B^*\mathrm{E}_{D_1}|_B + \mathrm{E}_{D_2}|_B^*\mathrm{E}_{D_2}|_B && (6.26)\end{aligned}$$

In particular,

$$\mathrm{Tr}(\mathrm{E}_{D_1 \oplus D_2}|_B^*\mathrm{E}_{D_1 \oplus D_2}|_)) = \mathrm{Tr}(\mathrm{E}_{D_1}|_B^*\mathrm{E}_{D_1}|_B) + \mathrm{Tr}(\mathrm{E}_{D_2}|_B^*\mathrm{E}_{D_2}|_B) \tag{6.27}$$

and so the expected value of the trace of the $\mathrm{E}_d|^*_B\mathrm{E}_d|_B$ operator sums over orthogonal sequential adjustments. To see that $\mathrm{Tr}(\mathrm{E}_D|^*_B\mathrm{E}_D|_B)$ is the expected value of $\mathrm{Tr}(\mathrm{E}_d|^*_B\mathrm{E}_d|_B)$, note that $\mathrm{P}(\mathrm{Tr}(\mathrm{E}_d|^*_B\mathrm{E}_d|_B)) = \mathrm{Tr}(\mathrm{E}_D|^*_B\mathrm{P}(F_d^*F_d)\mathrm{E}_D|_B) = \mathrm{Tr}(\mathrm{E}_D|^*_B\mathrm{E}_D|_B)$.

Of course, for scalar adjustments, $\mathrm{Tr}(\mathrm{E}_d|^*_B\mathrm{E}_d|_B)$ is just the single eigenvalue of $\mathrm{E}_d|^*_B\mathrm{E}_d|_B$, and so it corresponds precisely with the usual definition of size. In general however, $\mathrm{Tr}(\mathrm{E}_d|^*_B\mathrm{E}_d|_B)$ is the sum of the eigenvalues of $\mathrm{E}_d|^*_B\mathrm{E}_d|_B$. This is useful for exactly the same reason that it's expected value, $\mathrm{Tr}(\mathrm{E}_D|^*_B\mathrm{E}_D|_B)$ (the trace of the belief transform) is useful for summarising *a priori* analysis of belief structures. Of course, just as it is of great value to examine in detail the full eigenstructure of the belief transform, $\mathrm{E}_D|^*_B\mathrm{E}_D|_B$, it is always desirable to examine the full eigenstructure of $\mathrm{E}_d|^*_B\mathrm{E}_d|_B$ in order to fully understand the observed changes in belief. To conclude, the following definitions are made.

**Definition 5** *The* size *of the adjustment of $B$ given $d$,* $\mathrm{Size}_d(B)$*, is given by*

$$\mathrm{Size}_d(B) = \mathrm{Tr}(\mathrm{E}_d|^*_B\mathrm{E}_d|_B) \tag{6.28}$$

*This may be compared with it's a priori expected value,* $\mathrm{Size}_D(B)$ *given by*

$$\mathrm{Size}_D(B) = \mathrm{Tr}(\mathrm{E}_D|^*_B\mathrm{E}_D|_B) \tag{6.29}$$

*The set of non-trivial eigenvectors of* $\mathrm{E}_d|^*_B\mathrm{E}_d|_B$ *are known as the* bearings *of the adjustment. The eigenvalue corresponding to a particular bearing, is known as the* size *of that bearing.*

□

The bearings of the adjustment correspond to the elements of $B$ whose observed expectations are different to their *a priori* expectations. The corresponding sizes give an indication of the magnitude of the changes. To see this, let $\{Z_1, Z_2, \ldots, Z_r\}$ be an orthonormal set of non-trivial eigenvectors of $\mathrm{E}_d|_B^* \mathrm{E}_d|_B$ with corresponding ordered eigenvalues $\{\lambda_1, \lambda_2, \ldots, \lambda_r\}$. Now for any $A \in B$, write

$$A = \alpha_1 Z_1 + \alpha_2 Z_2 + \cdots + \alpha_r Z_r + \bar{A} \tag{6.30}$$

where $\bar{A} \perp span\{Z_1, Z_2, \ldots, Z_r\}$. Consequently,

$$\begin{aligned} \mathrm{E}_d(A) &= \mathrm{E}_d(\alpha_1 Z_1 + \alpha_2 Z_2 + \cdots + \alpha_r Z_r + \bar{A}) && (6.31) \\ &= \alpha_1 \mathrm{E}_d(Z_1) + \alpha_2 \mathrm{E}_d(Z_2) + \cdots + \alpha_r \mathrm{E}_d(Z_r) + \mathrm{E}_d(\bar{A}) && (6.32) \\ &= \alpha_1 \mathrm{E}_d(Z_1) + \alpha_2 \mathrm{E}_d(Z_2) + \cdots + \alpha_r \mathrm{E}_d(Z_r) && (6.33) \\ & && (6.34) \end{aligned}$$

and so the bearings define the directions within the $B$ space whose observed adjusted expectations differ from their *a priori* expected values. Note that when $dim(C) = 1$, there is only one bearing, which is just the usual scalar bearing for the adjustment.

In practice, for finite dimensional problems, a matrix representation of the operator $\mathrm{E}_D|_B$ with respect to an orthonormal bases on $B$ and $D$ is formed, which may then be transposed to give a matrix representation of $\mathrm{E}_D|_B^*$. The two matrices can then be multiplied together, and the eigenstructure of the resulting matrix may be analysed to give the bearings, and their corresponding sizes.

Explicitly, let $\{B_1, B_2, \ldots, B_m\}$ be an orthonormal basis for $B$, and let $\{C_1, C_2, \ldots, C_n\}$ be an orthonormal basis for $C$. Now, for each $B_i$, evaluate $\mathrm{E}_d(B_i)$ with respect to the basis on $C$. Then, for each $i$ we have

$$\mathrm{E}_d(B_i) = \alpha_{1i} C_1 + \alpha_{2i} C_2 + \cdots + \alpha_{ni} C_n \tag{6.35}$$

and so we deduce that with respect to the given bases, the operator $\mathrm{E}_d|_B$ is represented by the $n \times m$ matrix

$$\begin{pmatrix} \alpha_{11} & \alpha_{12} & \cdots & \alpha_{1m} \\ \alpha_{21} & \alpha_{22} & \cdots & \alpha_{2m} \\ \vdots & \vdots & \ddots & \vdots \\ \alpha_{n1} & \alpha_{n2} & \cdots & \alpha_{nm} \end{pmatrix} \tag{6.36}$$

Consequently, the operator $\mathrm{E}_d|_B^*$ is represented by the $m \times n$ matrix

$$\begin{pmatrix} \alpha_{11} & \alpha_{21} & \cdots & \alpha_{n1} \\ \alpha_{12} & \alpha_{22} & \cdots & \alpha_{n2} \\ \vdots & \vdots & \ddots & \vdots \\ \alpha_{1m} & \alpha_{2m} & \cdots & \alpha_{mn} \end{pmatrix} \tag{6.37}$$

The $m \times m$ matrix representation for the $\mathrm{E}_d|_B^* \mathrm{E}_d|_B$ operator may then be obtained by multiplying together the matrix representations for $\mathrm{E}_d|_B^*$ and $\mathrm{E}_d|_B$. The eigenstructure of the resulting matrix may then be analysed in order to understand the structure of the $\mathrm{E}_d|_B^* \mathrm{E}_d|_B$ operator.

Note that the theory developed in this section applies to a general Bayes linear adjustment where the random objects have a multi-dimensional constant space, and not just to the matrix space example developed in this thesis.

### 6.3.3 Example

In order to illustrate the definitions above with a concrete example, the simplest possible case will be used. Consider the adjustment of the matrix $V$ by the space $D_S$ of Chapter 4. Here, to simplify notation, $D$ will be used to denote the space $C + D_S$. The expression for the adjusted expectation turned out to be

$$\mathrm{E}_D(V) = \frac{2}{3}\mathrm{P}(V) + \frac{1}{3}S \tag{6.38}$$

This is (4.7) with $\alpha = 1/3$. The adjustment space of interest is the space

$$B = span\{V - \mathrm{E}(V)\} \tag{6.39}$$

First consider the evaluation of $\mathrm{Size}_D(B)$. Put

$$X = \frac{1}{23.618}[V - \mathrm{E}(V)] \tag{6.40}$$

This is an orthonormal basis for $B$. Put

$$Y = \frac{1}{40.58}[S - \mathrm{E}(V)] \tag{6.41}$$

and note that this has norm 1, and forms an orthonormal basis for the subspace of $D$ being projected onto. Now, since

$$\mathrm{E}_D|_B(V - \mathrm{E}(V)) = \frac{1}{3}[S - \mathrm{E}(V)] \tag{6.42}$$

we trivially deduce that

$$\mathrm{E}_D|_B(X) = 0.573Y \tag{6.43}$$

and so on $span\{X, Y\}$, the operator $\mathrm{E}_D|_B$ has matrix representation (0.573). Consequently, $\mathrm{E}_D|_B^*\mathrm{E}_D|_B$ has matrix representation (0.328), and so

$$\mathrm{Size}_D(B) = 0.328 \tag{6.44}$$

Now evaluate $\mathrm{Size}_d(B)$.

$$\|s - \mathrm{E}(V)\| = \left\| \begin{pmatrix} 0.3 & 9.01 & 9 \\ 9.01 & 122.04 & 107.69 \\ 9 & 107.69 & 158.26 \end{pmatrix} \right\| \tag{6.45}$$

$$= \quad 51.75 \tag{6.46}$$

where $s$ is the observed value of $S$. So put

$$Z = \frac{s - \mathrm{E}(V)}{51.75} \tag{6.47}$$

and note that this is an orthonormal basis for the subspace of $C$ being projected onto. Then

$$\mathrm{E}_d|_B(X) = 0.731Z \tag{6.48}$$

and so on $span\{X, Z\}$, the operator $\mathrm{E}_d|_B$ has matrix representation $(0.731)$. Consequently, $\mathrm{E}_d|_B^*\mathrm{E}_d|_B$ has matrix representation $(0.534)$, and so

$$\mathrm{Size}_d(B) = 0.534 \tag{6.49}$$

which gives a *size ratio* of 1.63. This tells us that the changes in belief were very slightly larger than expected *a priori*, but should not be considered a serious diagnostic warning.

### 6.3.4 Summary

The work of this section gives a completely unified framework for both the *a priori* and *a posteriori* analysis of totally general Bayes linear statistical problems. *A priori*, the eigenstructure of the operator $\mathrm{E}_D|_B^*\mathrm{E}_D|_B$ (the belief transform) is analysed in order to understand expected changes in belief. *A posteriori*, analysis of the eigenstructure of the operator $\mathrm{E}_d|_B^*\mathrm{E}_d|_B$ (the *observed belief transform*) may then be analysed in order to understand the observed changes in belief. The approach advocated is seen to generalise the diagnostic methodology used for scalar adjustments, and the generalised concept of *size* remains additive over sequences of orthogonal adjustments. It is also the case that analysis of the eigenstructure of the operator $F_d^* F_d$ provides direct information on the discrepancy between prior specification and observations, though this is not obvious from the current discussion, and will be discussed further elsewhere.

## 6.4 Negative eigenvalues in adjusted matrices

All the matrices which have been considered in this thesis have been non-negative definite (NND). However, in general, a matrix which is revised in an unconstrained manner may in certain situations turn out not to be NND. Such adjusted matrices are described as *incoherent*. In general, negative eigenvalues in an adjusted matrix act as a diagnostic warning of a possible contradiction between prior belief specifications and the data, or indeed of inappropriate choice of model or projection space. However, if after careful reflection, it is decided that the prior specifications made were proper and appropriate, given the available information at the time, there is a "quick fix" for the problem which may under certain circumstances, be worth considering.

Given a matrix which has been revised in an unconstrained manner, and has negative eigenvalues, one can construct a sure-loser argument which shows that the matrix formed by diagonalising the matrix, setting negative terms to zero, and then un-diagonalising (namely, the projection of the matrix into the subspace of coherent alternatives), necessarily has smaller loss associated with it than the original adjusted matrix, and therefore should be preferred to the original adjusted matrix.

Consider a random $n \times n$ matrix $A$, with elements $a_{ij}$. A specification for $\mathrm{P}(A)$ is made by specifying the matrix $X$ with elements $x_{ij}$ to minimise the loss†

$$L = \sum_{i=1}^{n} \sum_{j=1}^{n} (a_{ij} - x_{ij})^2 = \|A - X\|_F^2 \tag{6.50}$$

†Allowing the more general loss, $L = \sum_{i=1}^{n} \sum_{j=1}^{n} K_{ij}(a_{ij} - x_{ij})^2$ complicates the argument slightly, but does not alter the conclusion.

Now, suppose that the specification made is incoherent (*ie.* there is a negative eigenvalue in the matrix $X$). We wish to construct a matrix $\widehat{X}$ which is coherent, and necessarily has smaller loss associated with it than $X$. Write the orthogonal decomposition of $X$ as

$$X = \Sigma X^{\star} \Sigma^{\mathrm{T}} \tag{6.51}$$

Since the loss is rotationally invariant, we may transform to the orthogonal coordinate system implied by the matrix $\Sigma$. Write $A^{\star} = \Sigma^{\mathrm{T}} A \Sigma$. Then

$$L = \|A^{\star} - X^{\star}\|_F^2 \tag{6.52}$$

where the non-diagonal elements of $A^{\star}$ have zero prevision. All of the non-diagonal elements of $X^{\star}$ have been chosen to be zero. Since the matrix we are considering is incoherent, some of the diagonal elements of $X^{\star}$ have been specified to be negative. However, since we know that the corresponding element of $A^{\star}$ is at least zero, the matrix $\widehat{X}^{\star}$ which is formed by setting the negative elements of $X^{\star}$ to zero, necessarily has smaller loss associated with it than $X^{\star}$. Therefore, one would be a sure-looser not to prefer $\widehat{X}^{\star}$ to $X^{\star}$. Back in our original coordinate system, $\widehat{X} = \Sigma \widehat{X}^{\star} \Sigma^{\mathrm{T}}$ is necessarily prefered to $X$.

However, zero eigenvalues imply the existence of known combinations of variables, which is likely to contradict available data. For this reason, I feel that *a posteriori* correction of the matrix is undesirable, and that the constrained projection spaces discussed in the next chapter are a more promising way to go about ensuring that adjusted matrices are NND.

# Chapter 7

# Alternative approaches and further work

## 7.1 Projections into non-negative spaces

### 7.1.1 Motivation

Consider the example of learning about the covariance matrix for a collection of exchangeable random vectors given in Chapter 4. In this simplest case, projection into the space spanned by the prior expectation for the matrix and the sample covariance matrix was used. Both of these matrices are non-negative definite (NND), and the coefficients for the projection are necessarily greater than zero, whatever the belief specifications. In this case, the adjusted expectation is necessarily a positive combination of NND matrices, and hence NND. This is clearly a desirable state of affairs, since an adjusted matrix which is not NND is incoherent. In the other examples, the adjusted matrices are not constrained to be NND.

This issue is not specific to covariance matrix adjustment. For example, just consider the fitting of a strictly positive quantity on an unconstrained predictor. The adjusted value for that quantity is not necessarily positive. Usually however, such a negative revision would be regarded as evidence of a contradiction between prior beliefs and observations or evidence of an inappropriate model, or projection space. Some Bayes linear theory needs to be developed for the consideration of non-linearly constrained problems.

For the specific problem of constrained matrix adjustments, a few techniques due to special properties and decompositions of matrices are worth considering.

### 7.1.2 Eigenspace projections

Consider once more, the example of Chapter 4.

$$\boldsymbol{R}_k {\boldsymbol{R}_k}^{\mathrm{T}} = V + U_k \tag{7.1}$$

Also consider a sample covariance matrix,

$$S = \frac{1}{n-1} \sum_{w=1}^{n} (\boldsymbol{X}_w - \bar{\boldsymbol{X}})(\boldsymbol{X}_w - \bar{\boldsymbol{X}})^{\mathrm{T}} \tag{7.2}$$

predictive for $V$. Write

$$S = \Sigma^{\mathrm{T}} \Theta \Sigma \tag{7.3}$$

where $\Sigma$ is orthonormal, and $\Theta = diag(\theta_1, \theta_2, \ldots, \theta_r)$. Note that *a priori*, $\Sigma$ and $\Theta$ are both random. Now let $\Theta_i$ be the matrix with $\theta_i$ in the $(i,i)^{th}$ position, and zeros elsewhere. Then

$$\Theta = \Theta_1 + \Theta_2 + \cdots + \Theta_r \tag{7.4}$$

Now define

$$S_i = \Sigma^{\mathrm{T}} \Theta_i \Sigma \tag{7.5}$$

Note that the $S_i$ are observable, necessarily NND, and that

$$S = S_1 + S_2 + \cdots + S_r \tag{7.6}$$

Consequently, instead of projecting into the space

$$span\{C, S\} \tag{7.7}$$

one may project into the much richer space

$$span\{C, S_1, S_2, \ldots, S_r\} \tag{7.8}$$

allowing resolution of a much greater proportion of uncertainty, whilst retaining a necessarily NND adjustment (provided that all coefficients are necessarily positive).

Unfortunately in general, it is not possible to deduce the beliefs over the $S_i$ from the usual belief specifications made. However, when a better understanding of primitive specifications for matrices is obtained, this may turn out not be an insurmountable problem. In fact, all that is really needed in order to carry out such an analysis is a primitive specification of belief for the eigenstructure of the matrices, rather than the elements of the matrices. Note also that certain modelling assumptions, such as the assumption of second-order exchangeability, may fix the eigenstructure, giving beliefs over the eigenstructure from beliefs over the scalars directly.

### 7.1.3 Choelesky projections

Reconsider the same example of learning about $V$ from a predictive $S$. Form the Choelesky decomposition of $S$,

$$S = \Lambda\Lambda^{\mathrm{T}} \tag{7.9}$$

where $\Lambda$ is lower triangular and essentially unique. The Choelesky triangle is a useful parameterisation of the covariance matrix, because it has the property that there is a natural one to one correspondence between it and symmetric positive definite matrices. Unfortunately, the Choelesky triangle of a sum of matrices is not the sum of the Choelesky triangles, making it unclear exactly how one should decompose the triangle into a sensible projection space.

### 7.1.4 Logarithmic transformations

Another useful re-parameterisation of the covariance matrix is afforded by the matrix logarithm. The matrix logarithm is defined to be the inverse of the matrix exponential function, which is defined as follows:

$$exp(A) = I + A + \frac{A^2}{2!} + \frac{A^3}{3!} + \cdots \qquad (7.10)$$

It is easy to see that *exp* and *log* are rotationally invariant. Explicitly, if $A = \Sigma\Theta\Sigma^{\mathrm{T}}$ is the eigen-decomposition of $A$, and $\Theta = diag\{\theta_i\}$, we have

$$log(A) = \Sigma log(\Theta)\Sigma^{\mathrm{T}} = \Sigma \; diag\{log(\theta_i)\}\Sigma^{\mathrm{T}} \qquad (7.11)$$

As for the Choelesky triangle, there is a correspondence between it and the positive definite matrices. There is the same problem as with the Choelesky triangles, in that the logarithm of a sum is not the sum of the logarithms. However, the sum of logarithms is the logarithm of the product, and hence if some sort of multiplicative model was felt appropriate, there may indeed be a sensible way to form a projection space of logarithmically transformed matrices. Of course, this will be informative for the logarithm of the covariance matrix, which is not something necessarily of direct interest.

Leonard and Hsu (1992) focus their attention on the logarithm of the covariance matrix for exactly these reasons. However, their assumption of joint multivariate normality for the distribution of the elements of the logarithm of the covariance matrix seems a little speculative. Of course, since they make inferences about the full joint distribution of the elements of the logarithm of the covariance matrix, they have no problem translating those inferences into statements about the untransformed matrix. However, the usual computational problems associated with the sampling methodology that they use make the approach impractical for large problems.

More generally, some Bayes linear theory for multiplicative models and logarithmic transforms for scalars may prove an easier problem to tackle in the first instance, and may shed some light on the matrix version.

### 7.1.5 Summary

In conclusion, I think that some progress on the problem of ensuring NND belief revisions is possible, and will be of great value. However, I feel that it would be more appropriate to examine related problems from a scalar perspective in the first instance.

## 7.2 Restricted estimates

A related problem to that of restricting estimates to be NND, is that of handling general restrictions on the form of the covariance matrix. For example, it is easy to imagine that a particular linear combination of variables is known, and hence that there is a known eigenvector, with corresponding eigenvalue known to be zero in the covariance matrix. It would consequently be desirable to preserve such a structure in any adjusted matrices.

Also, two variables may be known to be independent for logical reasons and this would correspond to a zero in the covariance matrix which should be preserved.

Also, there may be a particular conditional independence present which is known to be the case for logical reasons. This will correspond to a zero in the inverse of the covariance matrix, which is required to be preserved. For example, if belief specification has been based upon a graphical model, there may be some arcs which one feels should not be introduced.

I imagine that the best way of handling such restrictions would be to choose a projection space which makes preserving such properties most natural. For example, an eigen-decomposition may be most appropriate for a known linear combination. The obvious element-wise decomposition demonstrated in Chapter 4 is easily modified to handle a known orthogonality — decompose all matrices to the one component level, and then neglect to include sample matrices corresponding to the known orthogonality in the projection space. In a similar vein, some kind of Choelesky or inverse matrix decomposition may prove to be the way to handle conditional orthogonalities.

Again, there is clearly a great deal of important work to be done in this area, but progress looks quite possible.

## 7.3 Diagnostics

Matrix adjustment diagnostics still require attention. In the last chapter, some theory regarding general Bayes linear diagnostics was developed, but there is still work to be done. In general, without the concept of a unique bearing derived from a Riesz functional representation, the Bayes linear concept of the *path correlation* (Goldstein 1988b) no longer makes a lot of sense. It is still possible to define a path correlation as a ratio of sizes of the various partial adjustments, but interpretation is now much less clear.

# Chapter 8

# Summary and conclusions

## 8.1 Summary

Quantifying relationships between variables is of fundamental importance in subjective statistical inference. However, there are many difficulties associated even with learning about covariances. It is often difficult to make prior covariance specifications, and usually even harder to make the statements about the uncertainty in these covariance statements which are required in order to learn about the covariance statements from data. Further, a covariance structure is more than just a collection of random quantities, so we should aim to analyse such structures in a space where they live naturally. In this thesis, such an approach was developed and illustrated, based around a geometric representation for variance matrices and exploiting second-order exchangeability specifications for them.

All authors who have considered the problem of covariance matrix revision seem to have come to the conclusion that it is such a difficult problem that they are prepared to make whatever distributional assumptions necessary in order to make the analysis as simple as possible. The distributional assumptions they make are usually such that expectations and conditional expectations have desirable linearity properties, which simplify the problem. In this thesis, no such distributional assumptions were made, but exactly those sorts of linearity properties were imposed and exploited.

Specifications made for the scalar components of random matrices can be used as a basis for a Bayes linear analysis of the covariance structures. However, for large matrices, the number of quantities involved in the adjustments will be prohibitively large. It is therefore very desirable to consider matrices in a space where they may be treated as a single object, greatly reducing the specification burden.

There is a common form of symmetry which often arises amongst ordered vectors of random quantities. It is essentially just a slightly weaker concept than that of (second-order) exchangeability. The covariance structure is invariant under arbitrary translations and reflections of the ordering, and the auto-correlation function becomes constant after some distance, $n$. Ordered vectors with this property were called, *second-order $n$-step exchangeable*. This same symmetry also occurs, under the same sorts of circumstances, for collections of random matrices in a random matrix inner-product space. Hence, a concept of $n$-step exchangeability which was sufficiently general that it was also valid for spaces of matrices was developed, and a representation theorem analagous to that for second-order exchangeability was derived. The representation theorem provides a simple way of decomposing variation for $n$-step exchangeable quantities into a part which is identifiable by the data, and a residual part, for which data is uninformative via linear fitting.

Just as there are many advantages to making expectation primitive, and specifying expectations directly, so there are with matrix inner-products. A scheme for elicitation based upon graphical modelling of the relationships between matrices, and specification of uncertainty and uncertainty reduction could be used in a way very similar to that often used for random scalars. In this way, the specification burden will be vastly reduced. Given a problem involving just a few (possibly large) matrices, all that will be required in order to carry out a basic analysis is a specification for the inner-product between every pair of matrices, rather than between every pair of scalars of which they are comprised.

Analysing matrices in a space where they live naturally not only has great aesthetic appeal, but is very powerful and illuminating in practice. Working in this space simplifies the handling of large matrices, by reducing the number of quantities involved and summarising effects over the whole covariance structure. For the same reasons, diagnostic information about adjusted beliefs is easier to interpret. Structures may be decomposed as much or as little as is desired.

This approach allows us to learn about collections of covariance structures, and examine their relationships. It generalises the "element by element" approach to revision, which can be viewed as taking place in a subspace of the larger space. Exchangeability representations lie at the heart of the methodology: all of our specifications are over observables, or quantities constructed from observables, rather than artificial model parameters, and no distributional assumptions for the data or the prior need be made.

This approach to covariance estimation was applied to the development of a methodology for the revision of the underlying covariance structures for a dynamic linear model, using Bayes linear estimators for the covariance matrices based on simple quadratic observables. This was done by constructing an inner-product space of random matrices containing both the underlying covariance matrices and observables predictive for them. Bayes linear estimates for the underlying matrices followed by orthogonal projection.

Good forecasting requires careful updating of the covariances within the time series structure. Informally, the degree of *shrinkage* between the prior and the data requires updating, and relationships between variables must be properly taken into account. Taking a matrix object approach greatly simplifies the problem by reducing it's dimensionality. This is important for both simplifying belief specification and belief adjustment, and also for interpretation of the structure of the adjustment and accompanying diagnostics. There are also the general advantages of the Bayes linear approach; namely of allowing complete flexibility for the prior specifications, without placing distributional restrictions on the data or model components.

General *a priori* and *a posteriori* analysis of Bayes linear statistical problems has been shown to be possible within a single framework, via analysis of the eigenstructure of the belief transform and the observed belief transform. This will allow practical object-oriented implementations of the Bayes linear methodology which are not restricted to a particular type of random entity, making analysis of complex problems with unusual objects possible using standard procedures.

Some of the work from this thesis is beginning to appear in the literature. In Wilkinson and Goldstein (1995a) we briefly describe a matrix inner-product, decompositions of covariance matrices, and methods for learning about a covariance matrix for exchangeable random vectors. In Wilkinson and Goldstein (1995b) we discuss applications to covariance matrix revision for multivariate time series dynamic linear models, and the $n$-step exchangeability representation theorem.

## 8.2 Conclusions

Genuine subjective revision of belief for covariance matrices is a very difficult, but important problem. The methodology detailed in this thesis represents a useful contribution towards understanding the problem, and important methodological advances for carrying out revision in certain situations. However, I do not claim to have all of the answers — there are several important outstanding questions which still need addressing, and some of these are outlined in Chapter 7. The other important func-tion of this thesis was to demonstrate the power and flexibility of the Bayes linear methodology. It is impressive that the theory coped so well with the introduction of non-scalar quantities. No modification or re-interpretation of the theory was required in order to consider linear spaces of random matrices, and to carry out adjustments. Further, it is of considerable interest that the Bayes linear diagnostic theory required a generalisation before being applicable to spaces of matrices. The development of the work for this thesis has highlighted a necessity to re-evaluate the theory surrounding general Bayes linear adjustment diagnostics. The general theory for Bayes linear diagnostics presented in Section 6.3 is of considerable importance and interest independently of the matrix adjustment theory developed in this thesis.

On completing this thesis, I feel more strongly than ever that the Bayes linear approach to subjective statistical inference is currently the best and most natural approach. It is however, equally clear that the theory is still in it's infancy, and that much work needs to be done. I feel that the approach to covariance estimation contained in this thesis captures very well the problems associated with covariance estimation, namely, the difficulties of belief specification, simplification, and ensuring sensible coherent revisions. By stripping away arbitrary distributional assumptions, the inherent problems and difficulties, often obscured by distributional and computational issues, are revealed. Covariance estimation is a problem which I feel will be with us for some time to come.

# Appendix A

# Covariances between sample covariances

This appendix is concerned with the derivation of equations (2.7), (2.8) and (2.9). Recall that we had exchangeable vectors, $\boldsymbol{X}_k$ with exchangeable decomposition

$$\boldsymbol{X}_k = \boldsymbol{M} + \boldsymbol{R}_k \tag{A.1}$$

(2.1), and that the quadratic products of the residuals were considered exchangeable, so that

$$\boldsymbol{R}_k \boldsymbol{R}_k{}^{\mathrm{T}} = V + U_k \tag{A.2}$$

(2.4). The formation of a sequence of sample covariance matrices, each based upon $n$ observations of the series, was considered. Now, from (2.6), we have

$$S_q = \frac{1}{n-1} \sum_{w=1}^{n} (\boldsymbol{R}_w^q - \bar{\boldsymbol{R}}^q)(\boldsymbol{R}_w^q - \bar{\boldsymbol{R}}^q)^{\mathrm{T}} \tag{A.3}$$

using an obvious extension of notation. Now using the notation $R_{ik}^q$ for the $i^{th}$ element of $\boldsymbol{R}_k^q$, and the notation $V_{ij}$ for the $(i,j)^{th}$ element of $V$, we have the following simple results.

**Lemma 2**

$$\begin{aligned}
\mathrm{Cov}(V_{ij}, R^q_{lk}R^q_{mk}) &= \mathrm{Cov}(V_{ij}, V_{lm}) && \text{(A.4)}\\
\mathrm{Cov}(V_{ij}, R^q_{lk}R^q_{m\cdot}) &= \frac{1}{n}\mathrm{Cov}(V_{ij}, V_{lm}) && \text{(A.5)}\\
\mathrm{Cov}(V_{ij}, R^q_{l\cdot}R^q_{mk}) &= \frac{1}{n}\mathrm{Cov}(V_{ij}, V_{lm}) && \text{(A.6)}\\
\mathrm{Cov}(V_{ij}, R^q_{l\cdot}R^q_{m\cdot}) &= \frac{1}{n}\mathrm{Cov}(V_{ij}, V_{lm}) && \text{(A.7)}
\end{aligned}$$

*$\forall i, j, k, l, m, q$, where $\cdot$ denotes the sample mean of the $n$ cases. Also,*

$$\begin{aligned}
\mathrm{Cov}(R^q_{ik}R^q_{jk}, R^q_{lk}R^q_{mk}) &= \mathrm{Cov}(V_{ij}, V_{lm}) + \mathrm{Cov}(U^q_{ijk}, U^q_{lmk}) && \text{(A.8)}\\
\mathrm{Cov}(R^q_{ik}R^q_{jk}, R^q_{lk}R^q_{m\cdot}) &= \frac{1}{n}\mathrm{Cov}(V_{ij}, V_{lm}) + \frac{1}{n}\mathrm{Cov}(U^q_{ijk}, U^q_{lmk}) && \text{(A.9)}\\
\mathrm{Cov}(R^q_{ik}R^q_{jk}, R^q_{l\cdot}R^q_{m\cdot}) &= \frac{1}{n}\mathrm{Cov}(V_{ij}, V_{lm}) + \frac{1}{n^2}\mathrm{Cov}(U^q_{ijk}, U^q_{lmk}) && \text{(A.10)}\\
\mathrm{Cov}(R^q_{ik}R^q_{j\cdot}, R^q_{lk}R^q_{m\cdot}) &= \frac{1}{n^2}\mathrm{Cov}(V_{ij}, V_{lm}) + \frac{1}{n^2}\mathrm{Cov}(U^q_{ijk}, U^q_{lmk}) && \text{(A.11)}\\
\mathrm{Cov}(R^q_{ik}R^q_{j\cdot}, R^q_{l\cdot}R^q_{m\cdot}) &= \frac{1}{n^2}\mathrm{Cov}(V_{ij}, V_{lm}) + \frac{1}{n^3}\mathrm{Cov}(U^q_{ijk}, U^q_{lmk}) && \text{(A.12)}\\
\mathrm{Cov}(R^q_{i\cdot}R^q_{j\cdot}, R^q_{l\cdot}R^q_{m\cdot}) &= \frac{1}{n^2}\mathrm{Cov}(V_{ij}, V_{lm}) + \frac{1}{n^3}\mathrm{Cov}(U^q_{ijk}, U^q_{lmk}) && \text{(A.13)}
\end{aligned}$$

*$\forall i, j, k, l, m, q$. Also,*

$$\begin{aligned}
\mathrm{Cov}(R^q_{ik}R^q_{jk}, R^q_{lp}R^q_{mp}) &= \mathrm{Cov}(V_{ij}, V_{lm}) && \text{(A.14)}\\
\mathrm{Cov}(R^q_{ik}R^q_{jk}, R^q_{ln}R^q_{m\cdot}) &= \frac{1}{n}\mathrm{Cov}(V_{ij}, V_{lm}) && \text{(A.15)}\\
\mathrm{Cov}(R^q_{ik}R^q_{jk}, R^q_{l\cdot}R^q_{m\cdot}) &= \frac{1}{n}\mathrm{Cov}(V_{ij}, V_{lm}) + \frac{1}{n^2}\mathrm{Cov}(U^q_{ijk}, U^q_{lmk}) && \text{(A.16)}\\
\mathrm{Cov}(R^q_{ik}R^q_{j\cdot}, R^q_{lp}R^q_{m\cdot}) &= \frac{1}{n^2}\mathrm{Cov}(V_{ij}, V_{lm}) && \text{(A.17)}\\
\mathrm{Cov}(R^q_{ik}R^q_{j\cdot}, R^q_{l\cdot}R^q_{m\cdot}) &= \frac{1}{n^2}\mathrm{Cov}(V_{ij}, V_{lm}) + \frac{1}{n^3}\mathrm{Cov}(U^q_{ijk}, U^q_{lmk}) && \text{(A.18)}\\
\mathrm{Cov}(R^q_{i\cdot}R^q_{j\cdot}, R^q_{l\cdot}R^q_{m\cdot}) &= \frac{1}{n^2}\mathrm{Cov}(V_{ij}, V_{lm}) + \frac{1}{n^3}\mathrm{Cov}(U^q_{ijk}, U^q_{lmk}) && \text{(A.19)}
\end{aligned}$$

$\forall i, j, l, m, q, \forall k \neq p.$

□

Now, using the notation $s_{ij}^q$ for the $(i, j)^{th}$ element of $S_q$, we have the following results.

**Theorem 4**

$$\text{Cov}(V_{ij}, s_{lm}^q) = \text{Cov}(V_{ij}, V_{lm}) \tag{A.20}$$

$$\text{Cov}(s_{ij}^q, s_{lm}^p) = \text{Cov}(V_{ij}, V_{lm}) \tag{A.21}$$

$$\text{Cov}(s_{ij}^q, s_{lm}^q) = \text{Cov}(V_{ij}, V_{lm}) + \frac{\text{Cov}(U_{ij}^q, U_{lm}^q)}{n} \tag{A.22}$$

*Proof*

$$\begin{aligned}
\text{Cov}(V_{ij}, s_{lm}^q) &= \frac{1}{n-1} \sum_{k=1}^{n} \text{Cov}(V_{ij}, (R_{lk}^q - R_{l\cdot}^q)(R_{mk}^q - R_{m\cdot}^q)) && \text{(A.23)} \\
&= \frac{1}{n-1} \sum_{k=1}^{n} \text{Cov}(V_{ij}, R_{lk}^q R_{mk}^q - R_{lk}^q R_{m\cdot}^q - R_{l\cdot}^q R_{mk}^q + R_{l\cdot}^q R_{m\cdot}^q) && \text{(A.24)} \\
&= \frac{1}{n-1} \sum_{k=1}^{n} \left( \frac{n-1}{n} \right) \text{Cov}(V_{ij}, V_{lm}) && \text{(A.25)} \\
&= \text{Cov}(V_{ij}, V_{lm}) && \text{(A.26)}
\end{aligned}$$

(A.25) follows using equations (A.4) to (A.7). This gives (A.20). (A.21) and (A.22) can be derived similarly. □

Vectorising (A.20), (A.21) and (A.22) gives (2.7), (2.8) and (2.9).

# Appendix B

# Using REDUCE to assist DLM quadratic covariance calculations

The covariance calculations for the quadratic products of one and two step differences, given in Section 5.3.4, were calculated using the REDUCE computer algebra system, described in Rayna (1987). Also, the precise form of the matrix inner product for the example discussed in Chapter 5 was deduced using the same REDUCE script. The following script was used to do all of the calculations required.

```
% reduce program
% for covariance calculations

operator cov,ex,x,xx,a,r,v,s,va,sa;

for all j,k
        let cov(j,k)=ex(j*k)-ex(j)*ex(k);
for all j
        let ex(-j)=-ex(j);
for all j,k
        let ex(j+k)=ex(j)+ex(k);
for all j
        let ex(2*j)=2*ex(j);
for all j
        let ex(4*j)=4*ex(j);
for all j
        let ex(j/2)=ex(j)/2;
for all j
        let ex(j/4)=ex(j)/4;

% model
for all j,t1
        let x(j,t1)=a(j,t1)+r(j,t1)-r(j,t1-1);
```

```
for all j,t1
        let xx(j,t1)=a(j,t1)+a(j,t1-1)+r(j,t1)-r(j,t1-2);
for all j,t1
        let ex(a(j,t1))=0;
for all j,t1
        let ex(r(j,t1))=0;
for all j,k,t1,t2
        let ex(a(j,t2)*r(k,t1))=0;
for all j,k,t1
        let a(j,t1)*a(k,t1)=va(j,k)+sa(j,k,t1);
for all j,k,t1
        let ex(sa(j,k,t1))=0;
for all j,k,l,m,t1
        let ex(sa(j,k,t1)*va(l,m))=0;
for all j,k,t1
        let r(j,t1)*r(k,t1)=v(j,k)+s(j,k,t1);
for all j,k,t1
        let ex(s(j,k,t1))=0;
for all j,k,l,m,t1
        let ex(s(j,k,t1)*v(l,m))=0;

for all j,k,l,m
        let ex(v(j,k)*va(l,m))=ex(va(l,m))*ex(v(j,k));
for all j,k,l,m,t1
        let ex(va(l,m)*s(j,k,t1))=0;
for all j,k,l,m,t1
        let ex(sa(l,m,t1)*v(j,k))=0;
for all j,k,l,m,t1,t2
        let ex(sa(l,m,t2)*s(j,k,t1))=0;

for all j,k,t1,t2 such that t1 neq t2
        let ex(r(j,t1)*r(k,t2))=0;
for all j,k,t1,t2 such that t1 neq t2
        let ex(a(j,t1)*a(k,t2))=0;
for all j,k,l,m,t1,t2 such that t1 neq t2
        let ex(s(j,k,t1)*s(l,m,t2))=0;
for all j,k,l,m,t1,t2 such that t1 neq t2
        let ex(sa(j,k,t1)*sa(l,m,t2))=0;

% simplifications
for all j,k,l,m,t1,t2
        let ex(a(k,t1)*r(j,t2)*va(l,m))=0;
for all j,k,l,m,t1,t2
        let ex(a(m,t1)*r(j,t2)*v(k,l))=0;
for all j,k,l,m,t1,t2,t3
        let ex(a(m,t1)*r(j,t2)*s(k,l,t3))=0;
for all j,k,l,m,t1,t2,t3
        let ex(a(k,t1)*r(j,t2)*sa(l,m,t3))=0;

for all j,k,l,m,t1,t2 such that t1 neq t2
        let ex(r(j,t1)*r(k,t2)*v(l,m))=0;
for all j,k,l,m,t1,t2,t3 such that t1 neq t2
        let ex(r(j,t1)*r(k,t2)*s(l,m,t3))=0;
for all j,k,l,m,t1,t2 such that t1 neq t2
        let ex(r(j,t1)*r(k,t2)*va(l,m))=0;
for all j,k,l,m,t1,t2,t3 such that t1 neq t2
        let ex(r(j,t1)*r(k,t2)*sa(l,m,t3))=0;
for all j,k,l,m,t1,t2,t3 such that t1 neq t2
        let ex(a(l,t1)*a(m,t2)*s(j,k,t3))=0;
for all j,k,l,m,t1,t2,t3 such that t1 neq t2
        let ex(a(l,t1)*a(m,t2)*sa(j,k,t3))=0;
for all j,k,l,m,t1,t2 such that t1 neq t2
        let ex(a(l,t1)*a(m,t2)*v(j,k))=0;
for all j,k,l,m,t1,t2 such that t1 neq t2
        let ex(a(l,t1)*a(m,t2)*va(j,k))=0;

for all j,k,l,m,t1,t2,t3,t4
        such that t1 neq t2 and t1 neq t3 and t2 neq t3
```

```
        let ex(a(j,t1)*a(k,t2)*a(l,t3)*r(m,t4))=0;

for all j,k,l,m,t1,t2,t3,t4
        such that t1 neq t2 and t3 neq t4
        let ex(a(l,t1)*a(m,t2)*r(j,t3)*r(k,t4))=0;

for all j,k,l,m,t1,t2,t3,t4
        such that t2 neq t3 and t3 neq t4 and t2 neq t4
        let ex(a(m,t1)*r(j,t2)*r(k,t3)*r(l,t4))=0;

for all j,k,l,m,t1,t2,t3,t4
        such that t1 neq t2 and t1 neq t3 and t1 neq t4
        and t2 neq t3 and t2 neq t4 and t3 neq t4
        let ex(r(j,t1)*r(k,t2)*r(l,t3)*r(m,t4))=0;

for all j,k,l,m,t1,t2,t3,t4
        such that t1 neq t2 and t1 neq t3 and t1 neq t4
        and t2 neq t3 and t2 neq t4 and t3 neq t4
        let ex(a(j,t1)*a(k,t2)*a(l,t3)*a(m,t4))=0;

% expressions

% covariances for the one step diffs
l1:=cov(va(j,k),x(l,t1)*x(m,t1));
l2:=cov(v(j,k),x(l,t1)*x(m,t1));
l3:=cov(x(j,t1)*x(k,t1),x(l,t1)*x(m,t1));
l4:=cov(x(j,t1)*x(k,t1),x(l,t1-1)*x(m,t1-1));
l5:=cov(x(j,t1)*x(k,t1),x(l,t1-2)*x(m,t1-2));
l6:=cov(x(j,t1)*x(k,t1),x(l,t1-3)*x(m,t1-3));

% covariances for the 2-step diffs
l11:=cov(va(j,k),xx(l,t1)*xx(m,t1));
l12:=cov(v(j,k),xx(l,t1)*xx(m,t1));
l13:=cov(xx(j,t1)*xx(k,t1),xx(l,t1)*xx(m,t1));
l14:=cov(xx(j,t1)*xx(k,t1),xx(l,t1-1)*xx(m,t1-1));
l15:=cov(xx(j,t1)*xx(k,t1),xx(l,t1-2)*xx(m,t1-2));
l16:=cov(xx(j,t1)*xx(k,t1),xx(l,t1-3)*xx(m,t1-3));
l17:=cov(xx(j,t1)*xx(k,t1),xx(l,t1-4)*xx(m,t1-4));

% covariances between the one and 2 step diffs
l23:=cov(x(j,t1)*x(k,t1),xx(l,t1)*xx(m,t1));
l24:=cov(x(j,t1)*x(k,t1),xx(l,t1-1)*xx(m,t1-1));
l25:=cov(x(j,t1)*x(k,t1),xx(l,t1-2)*xx(m,t1-2));
l26:=cov(x(j,t1)*x(k,t1),xx(l,t1-3)*xx(m,t1-3));
l27:=cov(x(j,t1)*x(k,t1),xx(l,t1-4)*xx(m,t1-4));

l34:=cov(xx(j,t1)*xx(k,t1),x(l,t1-1)*x(m,t1-1));
l35:=cov(xx(j,t1)*xx(k,t1),x(l,t1-2)*x(m,t1-2));
l36:=cov(xx(j,t1)*xx(k,t1),x(l,t1-3)*x(m,t1-3));
l37:=cov(xx(j,t1)*xx(k,t1),x(l,t1-4)*x(m,t1-4));

% actual number substitutions

for all j
        let ex(va(j,j))=4;
for all j,k such that j neq k
        let ex(va(j,k))=1;
for all j
        let ex(v(j,j))=36;
for all j,k such that j neq k
        let ex(v(j,k))=-4;

for all j
        let ex(va(j,j)^2)= (1.5)^2 + (4)^2;
for all j,k such that j neq k
        let ex(va(j,k)^2)= (0.75)^2 + (1)^2;
for all j
        let ex(v(j,j)^2)= (5)^2 + (36)^2;
```

```
for all j,k such that j neq k
        let ex(v(j,k)^2)= (1)^2 + (-4)^2;

for all j,k such that j neq k
        let ex(va(j,j)*va(k,k))=0.2*1 + (4)^2;
for all j,k such that j neq k
        let ex(v(j,j)*v(k,k))= 4 + (36)^2;

for all j,t1
        let ex(s(j,j,t1)^2)=(2500);
for all j,k,t1 such that j neq k
        let ex(s(j,k,t1)^2)=(1000);
for all j,t1
        let ex(sa(j,j,t1)^2)=(30);
for all j,k,t1 such that j neq k
        let ex(sa(j,k,t1)^2)=(15);

% index ordering
for all j,k such that j neq k and ordp(j,k)
        let v(k,j)=v(j,k);
for all j,k such that j neq k and ordp(j,k)
        let va(k,j)=va(j,k);
for all j,k,t1 such that j neq k and ordp(j,k)
        let s(k,j,t1)=s(j,k,t1);
for all j,k,t1 such that j neq k and ordp(j,k)
        let sa(k,j,t1)=sa(j,k,t1);

% just want diagonal terms
let l=j;
let m=k;

% expressions
ld1:=l1;
ld2:=l2;
ld3:=l3;
ld4:=l4;
ld5:=l5;
ld6:=l6;

ld11:=l11;
ld12:=l12;
ld13:=l13;
ld14:=l14;
ld15:=l15;
ld16:=l16;
ld17:=l17;

ld23:=l23;
ld24:=l24;
ld25:=l25;
ld26:=l26;
ld27:=l27;

ld34:=l34;
ld35:=l35;
ld36:=l36;
ld37:=l37;

let k=j;

ls1:=l1;
ls2:=l2;
ls3:=l3;
ls4:=l4;
ls5:=l5;
ls6:=l6;

ls11:=l11;
```

```
ls12:=l12;
ls13:=l13;
ls14:=l14;
ls15:=l15;
ls16:=l16;
ls17:=l17;

ls23:=l23;
ls24:=l24;
ls25:=l25;
ls26:=l26;
ls27:=l27;

ls34:=l34;
ls35:=l35;
ls36:=l36;
ls37:=l37;

% matrix expressions

on bigfloat,numval;
precision 20;

n:=6;
c01:=n*(5)^2 + n*(n-1)*(1)^2;
c02:=n*(1.5)^2 + n*(n-1)*(0.75)^2;
c03:=0;

c04:=n*ls3 + n*(n-1)*ld3;
c05:=n*ls4 + n*(n-1)*ld4;
c06:=n*ls5 + n*(n-1)*ld5;

c07:=n*ls2 + n*(n-1)*ld2;
c08:=n*ls1 + n*(n-1)*ld1;

c09:=n*ls13 + n*(n-1)*ld13;
c10:=n*ls14 + n*(n-1)*ld14;
c11:=n*ls15 + n*(n-1)*ld15;
c12:=n*ls16 + n*(n-1)*ld16;

c13:=n*ls12 + n*(n-1)*ld12;
c14:=n*ls11 + n*(n-1)*ld11;

c15:=n*ls23 + n*(n-1)*ld23;
c16:=n*ls34 + n*(n-1)*ld34;
c17:=n*ls24 + n*(n-1)*ld24;
c18:=n*ls35 + n*(n-1)*ld35;
c19:=n*ls25 + n*(n-1)*ld25;
c20:=n*ls26 + n*(n-1)*ld26;

% now do expectations

cc01:=4;     % ex(va(i,j)) for i=j
cc02:=1;     % ex(va(i,j)) for i neq j
cc03:=36;    % ex(v(i,j)) for i=j
cc04:=-4;    % ex(v(i,j)) for i neq j

cc05:=cc01 + 2*cc03;
cc06:=cc02 + 2*cc04;
cc07:=2*(cc01 + cc03);
cc08:=2*(cc02 + cc04);

ccc1:=n*cc01 + n*(n-1)*cc02; % ex(va)
ccc2:=n*cc03 + n*(n-1)*cc04; % ex(v)
ccc3:=n*cc05 + n*(n-1)*cc06; % ex(x')
ccc4:=n*cc07 + n*(n-1)*cc08; % ex(x'')

out "datavec.bd";
```

```
write "@covvec";
write c01;
write c02;
write c03;
write c04;
write c05;
write c06;
write c07;
write c08;
write c09;
write c10;
write c11;
write c12;
write c13;
write c14;
write c15;
write c16;
write c17;
write c18;
write c19;
write c20;
write "@meanvec";
write ccc1;
write ccc2;
write ccc3;
write ccc4;
shut "datavec.bd";
system "rcp datavec.bd @gauss:bd";

bye;
```

The first part of the script defines the properties of expectation and covariance. The next part sets up the model. Following this, simplifications in the form of known orthogonalities are given. The next part of the script produces the algebraic form of the covariances which we are interested in. The remainder of the script substitutes in the belief specifications made for the example, in order to deduce the covariances over the quadratic products, and then for the matrix inner product. Note that the covariances for the matrix inner-product are the *constant-adjusted* versions, as these are what is required by [B/D]. Running this script produced the following output.

```
Using directory /usr/local/reduce/current as Reduce root
REDUCE 3.4.1, 15-Jul-92 ...

1: 1: 1: 1: 1: 1: 1: 1: 1:
2: 2: 2:
3: 3:
4: 4:
5: 5:
6: 6:
7: 7:
8: 8:
9: 9: 9: 9:
10: 10:
11: 11:
12: 12:
13: 13:
14: 14:
15: 15:
16: 16:
17: 17:
```

```
18: 18:
19: 19:
20: 20: 20:
21: 21:
22: 22:
23: 23:
24: 24: 24:
25: 25:
26: 26:
27: 27:
28: 28: 28: 28:
29: 29:
30: 30:
31: 31:
32: 32: 32:
33: 33:
34: 34:
35: 35:
36: 36:
37: 37:
38: 38:
39: 39:
40: 40: 40: 40:
41: 41: 41: 41:
42: 42: 42: 42:
43: 43: 43: 43: 43:
44: 44: 44: 44: 44:
45: 45: 45: 45: 45:
L1 := EX(VA(J,K)*VA(M,L)) - EX(VA(J,K))*EX(VA(M,L))

46:
L2 := 2*( - EX(V(J,K))*EX(V(M,L)) + EX(V(J,K)*V(M,L)))

47:
L3 :=  - 4*EX(V(K,J))*EX(V(M,L)) + 2*EX(V(L,J))*EX(VA(M,K))

       + 2*EX(V(L,K))*EX(VA(M,J)) + 2*EX(V(M,J))*EX(VA(L,K))

       + 2*EX(V(M,K))*EX(VA(L,J)) + EX(S(K,J,T1 - 1)*S(M,L,T1 - 1))

       + EX(S(K,J,T1)*S(M,L,T1)) + 4*EX(V(K,J)*V(M,L))

       + 2*EX(V(L,J)*V(M,K)) + 2*EX(V(L,K)*V(M,J))

       + EX(VA(K,J)*VA(M,L)) + EX(SA(K,J,T1)*SA(M,L,T1))

       - EX(VA(K,J))*EX(VA(M,L))

48:
L4 :=  - 4*EX(V(K,J))*EX(V(M,L)) + EX(S(K,J,T1 - 1)*S(M,L,T1 - 1))

       + 4*EX(V(K,J)*V(M,L)) + EX(VA(K,J)*VA(M,L))

       - EX(VA(K,J))*EX(VA(M,L))

49:
L5 :=  - 4*EX(V(K,J))*EX(V(M,L)) + 4*EX(V(K,J)*V(M,L))

       + EX(VA(K,J)*VA(M,L)) - EX(VA(K,J))*EX(VA(M,L))

50:
L6 :=  - 4*EX(V(K,J))*EX(V(M,L)) + 4*EX(V(K,J)*V(M,L))

       + EX(VA(K,J)*VA(M,L)) - EX(VA(K,J))*EX(VA(M,L))

51: 51: 51:
L11 := 2*(EX(VA(J,K)*VA(M,L)) - EX(VA(J,K))*EX(VA(M,L)))
```

```
52:
L12 := 2*( - EX(V(J,K))*EX(V(M,L)) + EX(V(J,K)*V(M,L)))

53:
L13 :=  - 4*EX(V(K,J))*EX(V(M,L)) + 4*EX(V(L,J))*EX(VA(M,K))

        + 4*EX(V(L,K))*EX(VA(M,J)) + 4*EX(V(M,J))*EX(VA(L,K))

        + 4*EX(V(M,K))*EX(VA(L,J)) + EX(S(K,J,T1 - 2)*S(M,L,T1 - 2))

        + EX(S(K,J,T1)*S(M,L,T1)) + 4*EX(V(K,J)*V(M,L))

        + 2*EX(V(L,J)*V(M,K)) + 2*EX(V(L,K)*V(M,J))

        + 4*EX(VA(K,J)*VA(M,L)) + 2*EX(VA(L,J)*VA(M,K))

        + 2*EX(VA(L,K)*VA(M,J)) + EX(SA(K,J,T1 - 1)*SA(M,L,T1 - 1))

        + EX(SA(K,J,T1)*SA(M,L,T1)) - 4*EX(VA(K,J))*EX(VA(M,L))

54:
L14 :=  - 4*EX(V(K,J))*EX(V(M,L)) + 4*EX(V(K,J)*V(M,L))

        + 4*EX(VA(K,J)*VA(M,L)) + EX(SA(K,J,T1 - 1)*SA(M,L,T1 - 1))

        - 4*EX(VA(K,J))*EX(VA(M,L))

55:
L15 :=  - 4*EX(V(K,J))*EX(V(M,L)) + EX(S(K,J,T1 - 2)*S(M,L,T1 - 2))

        + 4*EX(V(K,J)*V(M,L)) + 4*EX(VA(K,J)*VA(M,L))

        - 4*EX(VA(K,J))*EX(VA(M,L))

56:
L16 := 4*( - EX(V(K,J))*EX(V(M,L)) + EX(V(K,J)*V(M,L))

           + EX(VA(K,J)*VA(M,L)) - EX(VA(K,J))*EX(VA(M,L)))

57:
L17 := 4*( - EX(V(K,J))*EX(V(M,L)) + EX(V(K,J)*V(M,L))

           + EX(VA(K,J)*VA(M,L)) - EX(VA(K,J))*EX(VA(M,L)))

58: 58: 58:
L23 :=  - 4*EX(V(K,J))*EX(V(M,L)) + EX(V(L,J))*EX(VA(M,K))

        + EX(V(L,K))*EX(VA(M,J)) + EX(V(M,J))*EX(VA(L,K))

        + EX(V(M,K))*EX(VA(L,J)) + EX(S(K,J,T1)*S(M,L,T1))

        + 4*EX(V(K,J)*V(M,L)) + 2*EX(VA(K,J)*VA(M,L))

        + EX(SA(K,J,T1)*SA(M,L,T1)) - 2*EX(VA(K,J))*EX(VA(M,L))

59:
L24 :=  - 4*EX(V(K,J))*EX(V(M,L)) + EX(S(K,J,T1 - 1)*S(M,L,T1 - 1))

        + 4*EX(V(K,J)*V(M,L)) + 2*EX(VA(K,J)*VA(M,L))

        - 2*EX(VA(K,J))*EX(VA(M,L))

60:
L25 := 2*( - 2*EX(V(K,J))*EX(V(M,L)) + 2*EX(V(K,J)*V(M,L))

           + EX(VA(K,J)*VA(M,L)) - EX(VA(K,J))*EX(VA(M,L)))

61:
```

```
L26 := 2*( - 2*EX(V(K,J))*EX(V(M,L)) + 2*EX(V(K,J)*V(M,L))

           + EX(VA(K,J)*VA(M,L)) - EX(VA(K,J))*EX(VA(M,L)))

62:
L27 := 2*( - 2*EX(V(K,J))*EX(V(M,L)) + 2*EX(V(K,J)*V(M,L))

           + EX(VA(K,J)*VA(M,L)) - EX(VA(K,J))*EX(VA(M,L)))

63: 63:
L34 :=  - 4*EX(V(K,J))*EX(V(M,L)) + EX(V(L,J))*EX(VA(M,K))

        + EX(V(L,K))*EX(VA(M,J)) + EX(V(M,J))*EX(VA(L,K))

        + EX(V(M,K))*EX(VA(L,J)) + EX(S(K,J,T1 - 2)*S(M,L,T1 - 2))

        + 4*EX(V(K,J)*V(M,L)) + 2*EX(VA(K,J)*VA(M,L))

        + EX(SA(K,J,T1 - 1)*SA(M,L,T1 - 1))

        - 2*EX(VA(K,J))*EX(VA(M,L))

64:
L35 :=  - 4*EX(V(K,J))*EX(V(M,L)) + EX(S(K,J,T1 - 2)*S(M,L,T1 - 2))

        + 4*EX(V(K,J)*V(M,L)) + 2*EX(VA(K,J)*VA(M,L))

        - 2*EX(VA(K,J))*EX(VA(M,L))

65:
L36 := 2*( - 2*EX(V(K,J))*EX(V(M,L)) + 2*EX(V(K,J)*V(M,L))

           + EX(VA(K,J)*VA(M,L)) - EX(VA(K,J))*EX(VA(M,L)))

66:
L37 := 2*( - 2*EX(V(K,J))*EX(V(M,L)) + 2*EX(V(K,J)*V(M,L))

           + EX(VA(K,J)*VA(M,L)) - EX(VA(K,J))*EX(VA(M,L)))

67: 67: 67: 67: 67:
68: 68:
69: 69:
70: 70:
71: 71: 71:
72: 72:
73: 73:
74: 74:
75: 75: 75:
76: 76:
77: 77: 77:
78: 78:
79: 79:
80: 80:
81: 81: 81: 81:
82: 82:
83: 83:
84: 84:
85: 85: 85:
86:
87: 87: 87:
        9
LD1 := ----
        16

88:
LD2 := 2

89:
```

```
        83417
LD3 := -------
         16

90:
        16073
LD4 := -------
         16

91:
        73
LD5 := ----
        16

92:
        73
LD6 := ----
        16

93: 93:
         9
LD11 := ---
         8

94:
LD12 := 2

95:
         233031
LD13 := --------
           40

96:
         85
LD14 := ----
         4

97:
         4025
LD15 := ------
          4

98:
         25
LD16 := ----
         4

99:
         25
LD17 := ----
         4

100: 100:
         10401
LD23 := -------
           8

101:
         8041
LD24 := ------
          8

102:
         41
LD25 := ----
         8

103:
```

```
         41
LD26 := ----
         8

104:
         41
LD27 := ----
         8

105: 105:
         10401
LD34 := -------
           8

106:
         8041
LD35 := ------
          8

107:
         41
LD36 := ----
         8

108:
         41
LD37 := ----
         8

109: 109:
110: 110:
        9
LS1 := ---
        4

111:
LS2 := 50

112:
        46273
LS3 := -------
          4

113:
        10409
LS4 := -------
          4

114:
        409
LS5 := -----
         4

115:
        409
LS6 := -----
         4

116: 116:
         9
LS11 := ---
         2

117:
LS12 := 50

118:
LS13 := 12830
```

```
119:
LS14 := 139

120:
LS15 := 2609

121:
LS16 := 109

122:
LS17 := 109

123: 123:
         6421
LS23 := ------
          2

124:
         5209
LS24 := ------
          2

125:
         209
LS25 := -----
          2

126:
         209
LS26 := -----
          2

127:
         209
LS27 := -----
          2

128: 128:
         6421
LS34 := ------
          2

129:
         5209
LS35 := ------
          2

130:
         209
LS36 := -----
          2

131:
         209
LS37 := -----
          2

132: 132: 132: 132:
*** Please use ROUNDED instead

133:
12

134: 134:
N := 6

135:
```

```
C01 := 180

136:
C02 := 30.375

137:
C03 := 0

138: 138:
C04 := 225816.375

139:
C05 := 45750.375

140:
C06 := 750.375

141: 141:
C07 := 360

142:
C08 := 30.375

143: 143:
C09 := 251753.25

144:
C10 := 1471.5

145:
C11 := 45841.5

146:
C12 := 841.5

147: 147:
C13 := 360

148:
C14 := 60.75

149: 149:
C15 := 58266.75

150:
C16 := 58266.75

151:
C17 := 45780.75

152:
C18 := 45780.75

153:
C19 := 780.75

154:
C20 := 780.75

155: 155: 155: 155:
CC01 := 4

156:
CC02 := 1

157:
CC03 := 36
```

```
158:
CC04 := -4

159: 159:
CC05 := 76

160:
CC06 := -7

161:
CC07 := 80

162:
CC08 := -6

163: 163:
CCC1 := 54

164:
CCC2 := 96

165:
CCC3 := 246

166:
CCC4 := 300

167: 167: 168: 169: 170: 171: 172: 173: 174: 175: 176: 177: 178:
 179: 180: 181: 182: 183: 184: 185: 186: 187: 188: 189: 190: 191:
 192: 193: 194:
195:
0

196: 196:
Quitting
```

First, the algebraic form of the covariances are given. The REDUCE operators `v(·)` and `s(·)` correspond to the matrices $V^{\nu}_{\cdot}$ and $S^{\nu}_{\cdot}$ respectively, and the operators `va(·)` and `sa(·)` correspond to the matrices $V^{\omega}_{\cdot}$ and $S^{\omega}_{\cdot}$ respectively. We may re-write the derived expressions in more conventional notation as follows.

$$\mathrm{Cov}(V^{\omega}_{jk}, X^{(1)}_{lt}X^{(1)}_{mt}) = \mathrm{Cov}(V^{\omega}_{jk}, V^{\omega}_{ml}) \tag{B.1}$$

$$\mathrm{Cov}(V^{\nu}_{jk}, X^{(1)}_{lt}X^{(1)}_{mt}) = 2\mathrm{Cov}(V^{\nu}_{jk}, V^{\nu}_{ml}) \tag{B.2}$$

$$\begin{aligned} \mathrm{Cov}(X^{(1)}_{jt}X^{(1)}_{kt}, X^{(1)}_{lt}X^{(1)}_{mt}) = & \quad \mathrm{Cov}(V^{\omega}_{kj}, V^{\omega}_{ml}) + 4\mathrm{Cov}(V^{\nu}_{kj}, V^{\nu}_{ml}) \\ & + \mathrm{Cov}(S^{\nu}_{kjt}, S^{\nu}_{mlt}) + \mathrm{Cov}(S^{\nu}_{kj(t-1)}, S^{\nu}_{ml(t-1)}) \\ & + \mathrm{Cov}(S^{\omega}_{kjt}, S^{\omega}_{mlt}) + 2[\mathrm{E}(V^{\nu}_{lj}V^{\nu}_{mk}) + \mathrm{E}(V^{\nu}_{lk}V^{\nu}_{mj})] \\ & + 4[\mathrm{E}(V^{\nu}_{lj})\mathrm{E}(V^{\omega}_{mk}) + \mathrm{E}(V^{\nu}_{lk})\mathrm{E}(V^{\omega}_{mj}) \\ & + \mathrm{E}(V^{\nu}_{mj})\mathrm{E}(V^{\omega}_{lk}) + \mathrm{E}(V^{\nu}_{mk})\mathrm{E}(V^{\omega}_{lj})] \end{aligned} \tag{B.3}$$

$$\mathrm{Cov}(X_{jt}^{(1)}X_{kt}^{(1)},X_{l(t-1)}^{(1)}X_{m(t-1)}^{(1)}) = 4(\mathrm{Cov}(V_{kj}^{\nu},V_{ml}^{\nu}) + \mathrm{Cov}(V_{kj}^{\omega},V_{ml}^{\omega})) + \mathrm{Cov}(S_{kj(t-1)}^{\nu},S_{ml(t-1)}^{\nu}) \tag{B.4}$$

$$\mathrm{Cov}(X_{jt}^{(1)}X_{kt}^{(1)},X_{l(t-s)}^{(1)}X_{m(t-s)}^{(1)}) = 4\mathrm{Cov}(V_{kj}^{\nu},V_{ml}^{\nu}) + \mathrm{Cov}(V_{kj}^{\omega},V_{ml}^{\omega}) \quad \forall i,j,k,l,t,s \geq 2 \tag{B.5}$$

$$\mathrm{Cov}(V_{jk}^{\omega},X_{lt}^{(2)}X_{mt}^{(2)}) = 2\mathrm{Cov}(V_{jk}^{\omega},V_{ml}^{\omega}) \tag{B.6}$$

$$\mathrm{Cov}(V_{jk}^{\nu},X_{lt}^{(2)}X_{mt}^{(2)}) = 2\mathrm{Cov}(V_{jk}^{\nu},V_{ml}^{\nu}) \tag{B.7}$$

$$\begin{aligned}\mathrm{Cov}(X_{jt}^{(2)}X_{kt}^{(2)},X_{lt}^{(2)}X_{mt}^{(2)}) = & \ 4\mathrm{Cov}(V_{kj}^{\omega},V_{ml}^{\omega}) + 4\mathrm{Cov}(V_{kj}^{\nu},V_{ml}^{\nu}) \\ & + \mathrm{Cov}(S_{kjt}^{\nu},S_{mlt}^{\nu}) + \mathrm{Cov}(S_{kj(t-2)}^{\nu},S_{ml(t-2)}^{\nu}) \\ & + \mathrm{Cov}(S_{kjt}^{\omega},S_{mlt}^{\omega}) + \mathrm{Cov}(S_{kj(t-1)}^{\omega},S_{ml(t-1)}^{\omega}) \\ & + 2[\mathrm{E}(V_{lj}^{\nu}V_{mk}^{\nu}) + \mathrm{E}(V_{lk}^{\nu}V_{mj}^{\nu}) + \mathrm{E}(V_{lj}^{\omega}V_{mk}^{\omega}) + \mathrm{E}(V_{lk}^{\omega}V_{mj}^{\omega})] \\ & + 4[\mathrm{E}(V_{lj}^{\nu})\mathrm{E}(V_{mk}^{\omega}) + \mathrm{E}(V_{lk}^{\nu})\mathrm{E}(V_{mj}^{\omega}) \\ & + \mathrm{E}(V_{mj}^{\nu})\mathrm{E}(V_{lk}^{\omega}) + \mathrm{E}(V_{mk}^{\nu})\mathrm{E}(V_{lj}^{\omega})]\end{aligned} \tag{B.8}$$

$$\mathrm{Cov}(X_{jt}^{(2)}X_{kt}^{(2)},X_{l(t-1)}^{(2)}X_{m(t-1)}^{(2)}) = 4(\mathrm{Cov}(V_{kj}^{\nu},V_{ml}^{\nu}) + \mathrm{Cov}(V_{kj}^{\omega},V_{ml}^{\omega})) + \mathrm{Cov}(S_{kj(t-1)}^{\omega},S_{ml(t-1)}^{\omega}) \tag{B.9}$$

$$\mathrm{Cov}(X_{jt}^{(2)}X_{kt}^{(2)},X_{l(t-2)}^{(2)}X_{m(t-2)}^{(2)}) = 4(\mathrm{Cov}(V_{kj}^{\nu},V_{ml}^{\nu}) + \mathrm{Cov}(V_{kj}^{\omega},V_{ml}^{\omega})) + \mathrm{Cov}(S_{kj(t-2)}^{\nu},S_{ml(t-2)}^{\nu}) \tag{B.10}$$

$$\mathrm{Cov}(X_{jt}^{(2)}X_{kt}^{(2)},X_{l(t-s)}^{(2)}X_{m(t-s)}^{(2)}) = 4\mathrm{Cov}(V_{kj}^{\nu},V_{ml}^{\nu}) + \mathrm{Cov}(V_{kj}^{\omega},V_{ml}^{\omega}) \quad \forall i,j,k,l,t,s \geq 3 \tag{B.11}$$

$$\mathrm{Cov}(X_{jt}^{(1)}X_{kt}^{(1)},X_{l(t+s)}^{(2)}X_{m(t+s)}^{(2)}) = 4\mathrm{Cov}(V_{kj}^{\nu},V_{ml}^{\nu}) + 2\mathrm{Cov}(V_{kj}^{\omega},V_{ml}^{\omega}) \quad \forall i,j,k,l,t,s \geq 3 \tag{B.12}$$

$$\mathrm{Cov}(X_{jt}^{(1)}X_{kt}^{(1)},X_{l(t+2)}^{(2)}X_{m(t+2)}^{(2)}) = 4\mathrm{Cov}(V_{kj}^{\nu},V_{ml}^{\nu}) + 2\mathrm{Cov}(V_{kj}^{\omega},V_{ml}^{\omega}) + \mathrm{Cov}(S_{kj(t-2)},S_{ml(t-2)}) \tag{B.13}$$

$$\begin{aligned}\mathrm{Cov}(X_{jt}^{(1)}X_{kt}^{(1)},X_{l(t+1)}^{(2)}X_{m(t+1)}^{(2)}) = & \ 2\mathrm{Cov}(V_{kj}^{\omega},V_{ml}^{\omega}) + 4\mathrm{Cov}(V_{kj}^{\nu},V_{ml}^{\nu}) \\ & + \mathrm{Cov}(S_{kj(t-2)}^{\nu},S_{ml(t-2)}^{\nu}) + \mathrm{Cov}(S_{kj(t-1)}^{\omega},S_{ml(t-1)}^{\omega}) \\ & + \mathrm{E}(V_{lj}^{\nu})\mathrm{E}(V_{mk}^{\omega}) + \mathrm{E}(V_{lk}^{\nu})\mathrm{E}(V_{mj}^{\omega}) \\ & + \mathrm{E}(V_{mj}^{\nu})\mathrm{E}(V_{lk}^{\omega}) + \mathrm{E}(V_{mk}^{\nu})\mathrm{E}(V_{lj}^{\omega})\end{aligned} \tag{B.14}$$

$$\begin{aligned}\mathrm{Cov}(X_{jt}^{(1)}X_{kt}^{(1)},X_{lt}^{(2)}X_{mt}^{(2)}) = & \ 2\mathrm{Cov}(V_{kj}^{\omega},V_{ml}^{\omega}) + 4\mathrm{Cov}(V_{kj}^{\nu},V_{ml}^{\nu}) \\ & + \mathrm{Cov}(S_{kjt}^{\nu},S_{mlt}^{\nu}) + \mathrm{Cov}(S_{kjt}^{\omega},S_{mlt}^{\omega}) \\ & + \mathrm{E}(V_{lj}^{\nu})\mathrm{E}(V_{mk}^{\omega}) + \mathrm{E}(V_{lk}^{\nu})\mathrm{E}(V_{mj}^{\omega}) \\ & + \mathrm{E}(V_{mj}^{\nu})\mathrm{E}(V_{lk}^{\omega}) + \mathrm{E}(V_{mk}^{\nu})\mathrm{E}(V_{lj}^{\omega})\end{aligned} \tag{B.15}$$

$$\mathrm{Cov}(X_{jt}^{(1)}X_{kt}^{(1)},X_{l(t-1)}^{(2)}X_{m(t-1)}^{(2)}) = 4\mathrm{Cov}(V_{kj}^{\nu},V_{ml}^{\nu}) + 2\mathrm{Cov}(V_{kj}^{\omega},V_{ml}^{\omega}) + \mathrm{Cov}(S_{kj(t-1)}^{\nu},S_{ml(t-1)}^{\nu}) \tag{B.16}$$

$$\mathrm{Cov}(X^{(1)}_{jt}X^{(1)}_{kt}, X^{(2)}_{l(t-s)}X^{(2)}_{m(t-s)}) = 4\mathrm{Cov}(V^{\nu}_{kj}, V^{\nu}_{ml}) + 2\mathrm{Cov}(V^{\omega}_{kj}, V^{\omega}_{ml}) \quad \forall i, j, k, l, t, s \geq 2 \tag{B.17}$$

Vectorising these equations gives the formulae from Section 5.3.4. The rest of the output is used as [B/D] input for the example adjustments.

# Bibliography

Bayes, T. (1763). An essay towards solving a problem in the doctrine of chances. *Phil. Trans. 53*, 370–418.

Bayes, T. (1958). Thomas Bayes' essay towards solving a problem in the doctrine of chances. *Biometrika 45*, 293–315.

Brown, P. J., N. D. Le, and J. V. Zidek (1994). Inference for a covariance matrix. In P. Freeman and A. F. M. Smith (Eds.), *Aspects of Uncertainty: A Tribute to D.V. Lindley*, pp. 77–92. Wiley.

Chen, C. F. (1979). Bayesian inference for a normal dispersion matrix. *J. Roy. Statist. Soc. B:41*, 235–248.

Dawid, A. P. (1981). Some matrix-variate distribution theory: Notational considerations and a Bayesian application. *Biometrika 68*, 265–274.

de Finetti, B. (1974). *Theory of probability, vol. 1.* Wiley.

de Finetti, B. (1975). *Theory of probability, vol. 2.* Wiley.

Dickey, J. M., D. V. Lindley, and S. J. Press (1985). Bayesian estimation of the dispersion matrix of a multivariate normal distribution. *Comm. Statist. Theory Methods 14*, 1019–1034.

Evans, I. G. (1965). Bayesian estimation of parameters of a multivariate normal distribution. *J. Roy. Statist. Soc. B:27*, 279–283.

Farrow, M. and M. Goldstein (1993). Bayes linear methods for grouped multivariate repeated measurement studies with application to crossover trials. *Biometrika 80*(1), 39–59.

Farrow, M. and M. Goldstein (1995). Diagnostic monitoring of Bayes linear prediction systems. In J.-M. Bernardo et al. (Eds.), *Bayesian Statistics 5, to appear.*

Garthwaite, P. H. and J. M. Dickey (1992). Elicitation of prior distributions for variable selection problems in regression. *Ann. Statist. 20*, 1697–1719.

Goldstein, M. (1979). The variance modified linear Bayes estimator. *J. R. Statist. Soc. B:41*, 96–100.

Goldstein, M. (1981). Revising previsions: a geometric interpretation. *J. R. Statist. Soc. B:43*, 105–130.

Goldstein, M. (1983). General variance modifications for linear Bayes estimators. *J. Amer. Statist. Ass. 78*, 616–618.

Goldstein, M. (1986a). Exchangeable belief structures. *J. Amer. Statist. Ass. 81*, 971–976.

Goldstein, M. (1986b). Separating beliefs. In P. Goel and A. Zellner (Eds.), *Bayesian Inference and Decision Techniques.* Amsterdam: North Holland.

Goldstein, M. (1988a). Adjusting belief structures. *J. R. Statist. Soc. B:50*, 133–154.

Goldstein, M. (1988b). The data trajectory. In J.-M. Bernardo et al. (Eds.), *Bayesian Statistics 3*, pp. 189–209. Oxford University Press.

Goldstein, M. (1990). Influence and belief adjustment. In J. Q. Smith and R. M. Oliver (Eds.), *Influence Diagrams, Belief Nets and Decision Analysis.* Chichester: Wiley.

Goldstein, M. (1991). Belief transforms and the comparison of hypotheses. *The Annals of Statistics 19*, 2067–2089.

Goldstein, M. (1994). Revising exchangeable beliefs: subjectivist foundations for the inductive argument. In P. Freeman and A. F. M. Smith (Eds.), *Aspects of Uncertainty: A Tribute to D. V. Lindley*. Wiley.

Goldstein, M. (1995). Bayes linear methods I - Adjusting beliefs: concepts and properties. Technical Report 1995/1, Department of Mathematical Sciences, University of Durham.

Goldstein, M. (1996). Prior inferences for posterior judgements. In M. L. D. Chiara et al. (Eds.), *Proceedings of the tenth international congress of Logic, Methodology and Philosophy of Science*. Kluwer.

Goldstein, M., M. Farrow, and T. Spiropoulos (1993). Prediction under the influence: Bayes linear influence diagrams for prediction in a large brewery. *The Statistician 42*(2), 445–459.

Goldstein, M. and A. O'Hagan (1995). Bayes linear methods for systems of expert posterior assessments. *J. R. Statist. Soc., to appear*.

Goldstein, M. and D. A. Wooff (1994). Robustness measures for Bayes linear analyses. *Journal of Statistical Planning and Inference 40*(2-3), 261–277.

Goldstein, M. and D. A. Wooff (1995a). Adjusting exchangeable beliefs. *In submission*.

Goldstein, M. and D. A. Wooff (1995b). Bayes linear computation: concepts, implementation and programming environment. *Statistics and Computing, to appear*.

Haff, L. R. (1980). Empirical bayes estimation of the multivariate normal covariance matrix. *Ann. Statist. 8*, 586–597.

Hartigan, J. A. (1969). Linear Bayesian methods. *J. Roy. Statist. Soc. B:31*, 446–454.

Henderson, H. V. and S. R. Searle (1979). Vec and vech operators for matrices, with some uses in jacobians and multivariate statistics. *Can. J. Statist. 7*, 65–81.

Henderson, H. V. and S. R. Searle (1981). The vec-permutation matrix, the vec operator and kronecker products: A review. *Linear and Multilinear Algebra 9*, 271–288.

Kreyszig, E. (1978). *Introductory functional analysis with applications.* Wiley.

Leonard, T. and J. S. J. Hsu (1992). Bayesian inference for a covariance matrix. *Ann. Statist. 20*, 1669–1696.

Queen, C. M. (1994). Using the multiregression dynamic-model to forecast brand sales in a competitive product market. *Statistician 43:1*, 87–98.

Queen, C. M. and J. Q. Smith (1992). Dynamic graphical models. In J. Bernardo et al. (Eds.), *Bayesian Statistics 4*, pp. 741–751. Oxford University Press.

Queen, C. M. and J. Q. Smith (1993). Multiregression dynamic-models. *J. Roy. Statist. Soc. B:55:4*, 849–870.

Queen, C. M., J. Q. Smith, and D. James (1994). Bayesian forecasts in markets with overlapping structures. *International journal of forecasting 10*, 209–233.

Ramsey, F. P. (1931). Truth and probability. In R. B. Braithwaite (Ed.), *The foundations of mathematics*, London. Kegan Paul, Trench, Trubner and Co.

Ramsey, F. P. (1990). Truth and probability. In D. H. Mellor (Ed.), *Philosophical papers.* Cambridge University Press.

Rayna, G. (1987). *REDUCE: Software for algebraic computation.* Springer-Verlag.

Rohde, C. A. and G. M. Tallis (1969). Exact first- and second-order moments of estimates of components of covariance. *Biometrika 56*, 517–525.

Savage, L. J. (1954). *The foundations of statistics.* New York: Dover.

Savage, L. J. (1971). Elicitation of personal probabilities and expectations. *J. Amer. Statist. Ass. 66*, 783–801.

Searle, S. R. (1971). *Linear models.* Wiley.

Searle, S. R. (1982). *Matrix Algebra Useful For Statistics.* New York: Wiley.

Smith, C. A. B. (1961). Consistency in statistical inference (with discussion). *J. Roy. Statist. Soc. B:23*, 1–37.

Smith, J. Q. (1990). Statistical principles on graphs. In J. Smith and R. Oliver (Eds.), *Influence Diagrams, Belief Nets and Decision Analysis.* Chichester: Wiley.

Stone, M. (1963). Robustness of non-ideal decision procedures. *J. Amer. Statist. Ass. 58*, 480–486.

Walley, P. (1991). *Statistical reasoning with imprecise probabilities.* Chapman and Hall.

West, M. and P. J. Harrison (1989). *Bayesian forecasting and dynamic models.* New York: Springer.

Whittaker, J. (1989). *Graphical models in applied multivariate statistics.* Chichester: Wiley.

Wilkinson, D. J. and M. Goldstein (1995a). Bayes linear adjustment for variance matrices. In J.-M. Bernardo et al. (Eds.), *Bayesian Statistics 5, to appear.*

Wilkinson, D. J. and M. Goldstein (1995b). Bayes linear covariance matrix adjustment for multivariate dynamic linear models. *Bayesian analysis e-print*,

bayes–an/9506002.

Wooff, D. A. (1992). [B/D] works. In J.-M. Bernardo et al. (Eds.), *Bayesian Statistics 4*, pp. 851–859. Oxford University Press.

Wooff, D. A. (1995a). Bayes linear methods II - An example with an introduction to [B/D]. Technical Report 1995/2, Department of Mathematical Sciences, University of Durham.

Wooff, D. A. (1995b). [B/D] Manual. Technical Report 1995/4, Department of Mathematical Sciences, University of Durham.

Wooff, D. A. and M. Goldstein (1995). Bayes linear methods III - analysing Bayes linear influence diagrams and Exchangeability in [B/D]. Technical Report 1995/3, Department of Mathematical Sciences, University of Durham.